\documentclass{ws-ijmpa}
\usepackage{color}

\usepackage{amsmath}
\begin{document}

\markboth{F. Henry-Couannier}{The Dark Side of Gravity}

\catchline{}{}{}{}{}

\title{ THE DARK SIDE OF GRAVITY}

\author{FREDERIC HENRY-COUANNIER}
\address{CPPM, 163 Avenue De Luminy, Marseille 13009 France.\\
henry@cppm.in2p3.fr}


\maketitle

\maketitle


\begin{abstract}
\color{red}
Section 20 revisited
\color{black}

Adopting a non geometrical point of view, we are led to an alternative theory of the order two and symmetric gravitational 
tensor field of GR. Considering our space-time to be flat and non dynamical,
this field can no more be interpreted as its metric. 
The true metric is a globally Minkowskian background, 
a context which justifies a genuine rehabilitation 
 of the global discrete space-time symmetries involved in the structure of the Lorentz group along with their 
'problematic' representations: the negative energy and tachyonic ones. 
It turns out that about this background our gravitational field must appear in two discrete symmetry reversal conjugate forms, 
different but non independent symmetric order two tensors which must be treated on the same footing in
 our actions. The discrete symmetry reversal invariant actions, equations and their conjugate 
solutions are obtained. The conjugate form generates a Dark Side of Gravity. Indeed, matter living there can only interact antigravitationally 
with matter on our side. We show that stability is granted. In this Dark Gravity theory (DG), the 
new Schwarzschild solution in vacuum only starts to differ from that of General Relativity at the Post-Post-Newtonian order. DG does not violate 
WEP and though it violates SEP, the theory passes all present SEP no violation tests except 
the Pioneer anomalous blue-shift one, an SEP violating effect which is a natural outcome. 
No horizon (no Black Hole) arises in the Schwarzschild solution.
A discontinuity of the gravitational field seems to be propagating in the solar system. Depending on its present position, new gravitomagnetic effects well
 within the accuracy of the Gravity Probe B experiment might be observed in place of the GR frame dragging one.
 A spatially flat universe accelerated expansion phase is obtained without resorting to inflation nor a cosmological constant and the Big-Bang 
singularity is avoided. A helicity zero wave solution is obtained leading to the 
same binary pulsar decay rate as observed and predicted by GR. The context is also promising to help us elucidate several outstanding cosmological
 enigmas such as the flat galactic rotation curves. 

\end{abstract}



\section{Introduction}

In special relativity, it is a well known and obvious result that the energy, being the time component
of a four-vector must flip its sign under time reversal. It was also soon recognised however 
that negative energy objects lead to intolerable instabilities in our theories and should be discarded. 
In a quantum framework this is easily done thanks to the complex conjugation involved in the 
antiunitary time reversal operator of our Quantum Field Theories which avoids 
the generation of negative energies under time reversal. In Ref.~\refcite{fhc1} we nevertheless 
wanted to come back on the less conventional mathematical option i.e. that the time reversal operator 
be unitary which not surprisingly reintroduces negative energy fields. 
We then noticed that, since there is no 
way to reverse a scalar field Hamiltonian sign under time reversal, it is impossible to reach 
a coherent description of unitary time reversal at least in a nongravitational framework.

We here propose a modified theory of gravity on flat space-time which extends, clarifies and explores 
in much more details some of our ideas already published in Ref.~\refcite{fhc3} 
to solve the issue at a classical level and show that it leads to a knew understanding of
 time reversal and by the way 'the problematic' negative energy and tachyonic representations 
of the Lorentz group. The reader is referred to Ref.~\refcite{fhc1} for our detailed investigation of negative energies and time 
reversal restricted to non gravitational QFT. In the next section, we only remind the main conclusions 
of this study. An interesting analysis can also be found in Ref.~\refcite{Chen}. 

\section { Negative Energies and Motivations for an Alternative Theory of Gravity} 
Let us gather the main information we learned from our investigation in Ref.~\refcite{fhc1} of negative energies in Relativistic 
QFT indicating that the correct theoretical framework for handling them should be a modified GR.

\begin{itemize}
\item TheoreticaI Motivations

In second quantization, all relativistic field equations admit genuine negative energy field solutions 
creating and annihilating negative energy quanta. Unitary time reversal links these fields to the positive energy ones.
 With the unitary choice, usual for all other symmetries in physics, reversing time does not mean
going backward intime anymore.
Positive and negative energy fields vacuum divergences we encounter after second quantization are 
unsurprisingly found to be exactly opposite. The negative energy fields action must be maximized.
However, there is no way to reach a coherent theory involving negative energies in a non-gravitational
framework. Indeed, if positive and negative energy scalar fields are time reversal conjugate, also must 
be their Hamiltonian densities and actions. But for a scalar field, the Hamiltonian density is simply a 
sum of squarred terms so there is obviously no way to make it flip its sign under time reversal, at least 
in a non gravitational context. However, when gravity comes into the game, $g_{\mu\nu}$ terms are expected
 in the Hamiltonian expression and their behaviour under time reversal might lead us to a solution 
to this issue.
But this solution certainly cannot be found in GR because it is a geometric theory of space-time: global 
space-time symmetries and their associated Noether currents are lost and in particular, time reversal
 is no more the well defined global symmetry it was in flat space-time. Therefore we must find a new 
theory of gravitation on a flat space-time background (we shall name it Dark Gravity or DG) in order to hopefully 
rehabilitate negative energy objects. 
 
\item  Phenomenological Motivations

In a standalone mirror negative energy world which fields remain non coupled to our world positive energy fields, stability is insured
 and the behaviour of matter and radiation is as usual. Hence, it is just a matter of convention to define each one as a positive
 or negative energy world as was already noticed by Linde (Ref.~\refcite{Lin1} \refcite{Lin2}). Otherwise, if the two worlds are allowed to interact, both signs of the energy 
being involved in this interaction, a promising new phenomenology can be expected. Indeed, many outstanding enigmas seem to be 
suggesting that repelling gravity might play an important role in physics: flat galactic rotation curves, 
the flatness and acceleration (Ref.~\refcite{SNLS}) of the universe, its voids...But negative energy particles never manifested themselves 
up to now in our detectors, also suggesting that a barrier is at work preventing the two worlds to interact except through gravity. 
The DG theory framework will involve by construction such a barrier.

The concordant cosmological SM is of course successfull in its confrontation with as many precision observables as are the CMB, LSS, BAO and 
primordial elements abundances but this is at the price of simplicity since many added components and ideas of different kinds
 along with their free parameters enter into the game. 
These are dark energy, dark matter, inflation, some of them remaining badly understood and introducing new issues such as the fine tuning and 
coincidence problems.
The impression that this construction is at least effective if not epicyclic has therefore recently motivated the research for alternative theories 
of gravity such as MOND (Ref.~\refcite{MOND}) and its relativistic extension TeVeS (Ref.~\refcite{TVS}). Part of our motivation for DG is similar.

\item  A Modified GR to Circumvent the Main Issues

A trivial cancellation between positive and negative vacuum divergences is not acceptable since the Casimir effect shows
 evidence for vacuum fluctuations. But in DG we might hopefully get cancellations of vacuum gravitational effects only.

 Also, a generic catastrophic instability issue arises whenever positive and negative energy fields 
are allowed to interact. Restricting the stability issue to the modified gravity of DG, we shall show why 
 this disastrous scenario is avoided. 

Finally, even neglecting the former instability issue, allowing both positive and negative energy virtual 
bosons to propagate an interaction simply makes it vanish. Because electromagnetic, weak and strong interactions propagated by positive 
energy bosons will be isolated from the ones propagated by negative energy bosons, each being restricted to a different side of our gravitational 
barrier, the vanishing interaction issue remains a concern only for gravity in DG. 
However, the gravity interaction also will be treated very differently in DG than it is in GR so that this unpleasant feature, 
expected when gravitons of both energy signs can participate to the interaction, will also be avoided here.

\end{itemize}
\section{Theories of Gravity on Flat space-time}

Understanding negative energies as explained in the previous section is a strong motivation for looking for a flat spacetime theory of gravity 
but constructing a viable quantum theory of gravity is an even stronger one. Indeed, in the seventies, due to the increasing difficulties in trying to reach
a coherent theory of quantum gravity, many theorists were led to the 
conviction that we might have to come back to a flat space-time theory of gravity and there were many attempts along this line. 
After all, geometric considerations play no role in the derivation of GR which only rests on the general covariance requirement and the equivalence principle: covariantization which introduces $g_{\mu\nu}$ (and its derivatives) in place of $\eta_{\mu\nu}$ is just required to get equations valid in any general coordinate system and the equivalence principle is needed to physically interpret $g_{\mu\nu}$ not only as a pseudoforce field but as a real interaction field: the field of gravity for which the Einstein Hilbert action is the simplest one to add in order to make it dynamical.

This is all we need to get a complete GR and it's only afterwards that we are tempted to interpret $g_{\mu\nu}$ as the genuine metric of space-time when we realize that this gravitational field affects the 
space and time intervals we measure and  has, as any order two symmetric tensor field, the required mathematical properties for that. However, following this geometrical viewpoint clearly hinders the construction of any field theory with more than one symmetric order two tensor field  without postulating the existence of extradimensions because two such fields describing different geometries could not at the same time be the metric of our four dimensional space-time. Such limitation is very unnatural from the usual field theoretic approach where we felt free to introduce as many fields of any type as we wanted and comes from the fact that the geometrical interpretation of GR gives the $g_{\mu\nu}$ field a very special status which were not necessary at all in the derivation of GR from first principles. This gap between the so interpreted $g_{\mu\nu}$ and the other fields is also probably one of the main obstacles to unification.

Fortunately, it still remains possible to adopt a non geometrical viewpoint for GR i.e. we could still consider that the gravitational field has nothing to 
do with
 deformations of space-time itself which remains flat with true metric the globally Minkowskian $\eta_{\mu\nu}$.
Just as in optical geometry, where we do not understand the curved trajectory of a light ray in a varying index medium 
as the effect of a deformed space-time but as the consequence of the interaction of light 
with the fields involved in this medium, just in the same way do we consider all the 
effects of gravity on our observables as due to mere interactions of our clocks and rods 
with the field $g_{\mu\nu}$, as any other field in a theory with flat space time background 
 $\eta_{\mu\nu}$. 

However, as has been often pointed out, because
 $\eta_{\mu\nu}$ is absent from GR equations, there is no way to observationally distinguish the geometrical from the non geometrical point of view so 
that this issue remains one of metaphysics in the case of GR. 
But, taking serious the flat space-time metric $\eta_{\mu\nu}$ makes it possible and even necessary to build different kinds of alternative theories of gravity.
The so called multimetric theories exploit the fact that $g_{\mu\nu}$, if it can still be a metric topologically defined on our space-time
 is not considered as the metric of this space-time, the geometry defined by this metric having nothing to do with the 
geometrical properties of space-time which true metric is $\eta_{\mu\nu}$. Then we are free to introduce as many fields of the $g_{\mu\nu}$ type as we want on a manifold without having to 
make appeal to extra-manifolds or extra-dimensions. Following this line of thinking, Petit (Ref.~\refcite{petit2} with references therein) has built 
a bimetric theory which associated phenomenology is very similar to the one we shall derive in DG. See also a review in
Ref.~\refcite{dam} and
 references therein. S. Hossenfelder (Ref.~\refcite{Hos}) has also recently initiated a study of anti-gravitation following probably the 
closest approach to the one followed here as far as we know.
In Rosen's (Ref.~\refcite{Ros}) approach, the true flat space-time metric $\eta_{\mu\nu}$ was explicitely 
introduced in the action. 
This theory is reviewed by C.M Will (Ref.~\refcite{Will}) and found to be ruled out by
 tests of the Strong Equivalence Principle (SEP) due to its background dependence. Because generically all background dependent theories are in trouble 
with the very constraining tests of SEP, the flat space-time (or background dependent) approach was progressively given up. DG is also a background dependent theory 
and it also violates SEP, though most such
violations remain too small to be detectable thanks to fortunate compensations of the largest 
SEP violating terms in our field equations, as we shall show. 

In the following sections, we argue that not only the building of a theory involving both $g_{\mu\nu}$ and $\eta_{\mu\nu}$ is possible without conflicting
 with present observational Post-Newtonian constraints but moreover that this possibility is crucially based on satisfying a new essential 
symmetry that went unnoticed in the 70's attempts and that our Dark Gravity theory is the minimal one that respects this symmetry.
\section{Conjugate Worlds Gravitational Coupling }

Now let us settle down the basic principles of DG.
As usual for any order two tensor, symmetric and covariant gravitational field ${g}_{\mu \nu}$
we define its inverse, the contravariant $\left[{g}^{-1}\right]^{\mu \nu }$. But since ${g}_{\mu \nu}$ is not understood to be 
the metric of space-time, the latter is a contravariant tensor $\tilde 
{g}^{\mu \nu }$ but not the ${g}^{\mu \nu }$ tensor one would obtain by raising 
${g}_{\mu \nu}$ indices using the true metric $\eta_{\mu\nu}$.  Though the Ricci scalar one builds from ${g}_{\mu \nu}$ and $\left[{g}^{-1}\right]^{\mu \nu }$  is of 
course left unchanged by the pure renaming 
of $\left[g^{-1}\right]^{\mu\nu}$ into  $\tilde{g}^{\mu\nu}=\eta^{\mu\rho}\eta^{\nu\sigma}\tilde{g}_{\rho\sigma}$, this rewriting reveals the existence 
of a second form of the field, still an order two symmetric covariant tensor $\tilde{g}_{\mu\nu}$ different but not independent from $g_{\mu\nu}$: 
this is $\tilde {g}_{\mu \nu } =\eta _{\mu \rho } \eta _{\nu \sigma } \left[ 
{g^{-1}} \right]^{\rho \sigma }=\left[ {\eta ^{\mu \rho }\eta ^{\nu \sigma 
}g_{\rho \sigma } } \right]^{-1}$. Thus the Janus gravitational field, like the Janus god, has two faces, $g_{\mu\nu}$ and $\tilde {g}_{\mu \nu}$ linked by 
the above manifestly covariant and background dependent relation. 
As the following simple picture (where moving vertically inverses the tensors, and diagonally raises or lowers
their indices with $\eta_{\mu\nu}$) makes it obvious, the two forms play perfectly equivalent roles relative to
the background metric $\eta_{\mu\nu}$ so should be treated on the same footing
 in our actions if we dont want to artificially destroy the basic symmetry of the picture under their permutation.
\[
\begin{array}{l}
 \left[ {g^{-1}} \right]^{\mu \nu }\hspace{0.5cm}\left[ {\tilde {g}^{-1}} 
\right]^{\mu \nu } \\ 
 \\ 
\\\\
 g_{\mu \nu } \hspace{1.5cm} \tilde {g}_{\mu \nu } \\ 
 \end{array}
\]
Symmetrizing the roles of $g_{\mu\nu}$ and $\tilde{g}_{\mu\nu}$ will be performed by simply adding to the usual GR action, the similar action built from
 $\tilde{g}_{\mu\nu}$ and its inverse. Indeed, it is possible to have a connection compatible with $\tilde{g}_{\mu\nu}$ just in the same way as we had one 
compatible with $g_{\mu\nu}$ and exploit 
the mathematical apparatus of GR to constrain the general coordinate scalar built out of this form and its derivatives, another Ricci scalar. 
The theory that follows from symmetrizing the roles of $g_{\mu\nu}$ and $\tilde{g}_{\mu\nu}$ is DG which turns out to be essentially the other option of 
a binary choice that must be done at the level of the conceptual fondations of a 
covariant theory of a symmetric order two tensor field: either the space-time is curved with metric $g_{\mu\nu}$ and we get GR, or it is flat with background metric 
$\eta_{\mu\nu}$ and we get DG! GR actually just corresponds to the special case where g and $\eta$ identify in which case, taking the inverse is 
equivalent to twice raising or lowering the indices with the space-time metric $g_{\mu\nu}=\eta_{\mu\nu}$. At last, even if we did not consider from the 
begining that $\eta_{\mu\nu}$ is the true metric of space-time, in presence of this background the gravitational field $g_{\mu\nu}$, automatically acquires its "inverse form", $\tilde{g}_{\mu\nu}$ with a priori perfectly symmetric role to play. If both are gravitational fields, i.e. if both are as usual minimally coupled to a distinct class of matter and radiation fields (F fields propagating along $g_{\mu\nu}$ geodesics and
$\tilde{F}$ fields propagating along $\tilde{g}_{\mu\nu}$ geodesics) , hence 
determine the space and time intervals measured by clocks and rods made from these fields, $g_{\mu\nu}$ could a priori be considered as the genuine metric of space-time by the observer living in $g_{\mu\nu}$ while the observer living in $\tilde{g}_{\mu\nu}$ would certainly favour this latter form, even more if he knows nothing about the existence of $g_{\mu\nu}$. The two forms are incompatible in the sense that 
these in general describe a different geometry so only one of the two might be the genuine metric of spacetime but since none should be preferred it only remains $\eta_{\mu\nu}$ to, may be, play the central role of this true space-time metric. 

Then, we have to distinguish between those fields coupling to $\tilde {g}_{\mu \nu }$ and the others, coupling to $g_{\mu \nu }$.
 These matter and radiation fields of two mirror standard models respectively living in the two different forms of the gravity field never meet since they do not couple 
to each other. However, the relation between the conjugate forms, $\tilde{g}_{\mu\nu}=\eta_{\mu\rho}\eta_{\nu\sigma}[g^{-1}]^{\rho\sigma}$, 
indirectly and invisibly connects F and $\tilde{F}$ and allows these fields 
to interact antigravitationally as we shall check later since 
a positive mass induced curvature of one form translates 
into a negative mass induced curvature of the conjugate form. It is already clear why the gravitational barrier involved in DG explains 
the non detection of these "negative mass" particles and isolates the stability issue in the gravitational sector: these live on the other side of gravity preventing them 
from interacting weakly, strongly or electromagnetically with our side particles.

 But DG not only has two sectors, two mirror Standard Models of particles and Fields  F and $\tilde{F}$ living in g and $\tilde{g}$ respectively, one of which implying the existence of irremediably dark repelling matter from 
our side point of view. The introduction of an additional 
Einstein Hilbert action for the conjugate form $\tilde{g}$ also modifies the geometric side of the Einstein equation after exploiting 
$\tilde{g}_{\mu\nu}=\eta_{\mu\rho}\eta_{\nu\sigma}[g^{-1}]^{\rho\sigma}$ in order to eliminate $\tilde {g}_{\mu \nu }$.
But this will have mainly PostPostNewtonian consequences, only non negligible in the cosmological and strong gravity domains.
\footnote{
It was shown by Straumann in Ref.~\refcite{Strau}  that the building of a spin two $h_{\mu \nu }$ field theory over flat space-time automatically leads 
to identify a metric field 
and that its action turns out to be nothing but the Einstein-Hilbert action of GR. It follows that GR is the only theory of such spin two field perturbation 
over flat space-time, an argument which has been used to deny the validity of DG. 
The obvious fact that our action is eventually not the GR action after eliminating $\tilde {g}_{\mu \nu }$ simply tells us that DG is in no way the spin two
 theory over flat space-time Straumann started from. The deep reason why it is not is the gauge invariance requirement of the Straumann spin two theory, 
which is not meaningful in DG
since it is not the general covariant theory of a single field, but of two separate fields  $g_{\mu \nu }$ and the non dynamical $\eta_{\mu \nu }$. Thus DG is not invariant 
under an infinitesimal general coordinate transformation that only applies to $g_{\mu \nu }$ and not separately to $\eta_{\mu \nu }$, such general coordinate invariance
 condition being the one that usually translates into the gauge invariance requirement 
for $h_{\mu \nu }$ entering in $g_{\mu \nu } = \eta_{\mu \nu } +h_{\mu \nu }$
(the invariance of $\eta_{\mu \nu }$ under infinitesimal transformations was recognised as an important drawback 
of the perturbative treatment even in the context of GR: Ref.~\refcite{rov}).  By the way, having lost the usual gauge invariance of gravity in DG also
 puts under question a pilar of quantum field theory: gauge invariance as a building tool 
toward a theory of the electromagnetic or any other field.}

\section{Global space-time discrete symmetries }

Since we are working on a globally flat non dynamical space-time we will be free to use the flat space-time tools
to build the Noether currents associated with global symmetries and obtain genuine energy-momentum tensors not to 
be confused with
the GR energy-momentum pseudotensors.  This was also Rosen's initial motivation for prefering flat space-time and many others followed him in this way 
(see Ref.~\refcite{Pitt} ~\refcite{Pitt2} and references therein).
 We can also rehabilitate global space and time reversal symmetries which are now well defined discrete symmetries along
 with their associated negative energy and tachyonic representations. We anticipate that we will have the good surprise to notice after solving our equations that the two faces 
of our Janus field solutions will be related by obvious discrete space-time symmetries. For instance, the $g_{\mu\nu}(t)$ and $\tilde{g}_{\mu\nu}(t)$ cosmological 
solutions will be found to be conjugate under time reversal in the coordinate system where 
$\eta_{\mu\nu}=diag(-1,+1,+1,+1)$. 

We can already try to deeper understand the meaning of $g_{\mu\nu}$ and $\tilde{g}_{\mu\nu}$ being linked by a global space-time discrete symmetry.  
From any reference GR textbook such as Ref.~\refcite{Wein} we learned that in a locally inertial coordinate system   
$ \xi ^\alpha$, the propertime $d\tau$ satisfies
\begin{equation}
d\tau^2=\eta_{\alpha \beta}d\xi^\alpha d\xi^\beta
\end{equation}
and of course we are free to perform the replacement $d\xi^\alpha=\frac{\partial \xi^\alpha}{\partial x^\mu}dx^\mu$ to get the propertime
\begin{equation}
d\tau^2=g_{\mu \nu}(x^\mu)dx^\mu dx^\nu \hspace{2cm} g_{\mu \nu } \equiv \eta _{\alpha \beta } \frac{\partial \xi ^\alpha }{\partial 
x^\mu }\frac{\partial \xi ^\beta }{\partial x^\nu }
\label{eqgdef}
\end{equation}
expressed in any general coordinate system $x^\mu$.
In these expressions the inertial coordinates are understood to be functions of the general chosen coordinates:   $\xi ^\alpha = \xi ^\alpha(x^\mu)$ as well as $g_{\mu \nu}=g_{\mu \nu}(x^\mu)$.
But we could as well have chosen another coordinate system $x^{\prime \mu}$ to write 
\begin{equation}
d\tau^2=g^{\prime}_{\mu \nu}(x^{\prime \mu})dx^{\prime \mu} dx^{\prime \nu} \hspace{2cm} g^{\prime}_{\mu \nu } =\eta _{\alpha \beta } \frac{\partial \xi ^\alpha }{\partial 
x^{\prime \mu }}\frac{\partial \xi ^\beta }{\partial x^{\prime \nu} }
\end{equation}
where of course the inertial coordinates $\xi ^\alpha$ are still the same as in (\ref{eqgdef}) though we can now consider them as the result of a new set of functions $\xi^{\prime \alpha}(x^{\prime \mu})$ acting on the $x^{\prime \mu}$ :  $\xi^{\prime \alpha}(x^{\prime \mu}) = \xi ^\alpha(x^\mu)$.
 The change of coordinate system from $x^\mu$ to $x^{\prime \mu}$ might appear to be just an arbitrary reparametrization without any physical content whatever it is, even for instance the transformation from $x^\mu$ to $x_T^\mu$ that would reverse $x^0$ and let invariant the other three coordinates.
 However this is not the case: not all transformations of the coordinates have the same status even in GR where we know that, though general covariance by itself does not favour Lorentz transformations over any other ones such as for instance Galilean transformations, the equations once expressed in the inertial frame do so. Indeed, after replacing $g_{\mu \nu}$ by $\eta _{\mu \nu }$, Lorentz invariance is then demanded which is a strong constraint (By the way, it makes more sense to consider here $\eta _{\mu \nu }$, the local value taken by $g_{\mu \nu }$, as a mere matrix $diag(-1,+1,+1,+1)$ rather than a Lorentz tensor because then the invariance of $d\tau$ and covariance of the equations under a transformation between two sets of inertial coordinates is a real constraint on how these should transform: only Lorentz Transformations can satisfy it).
 Similarly, we know that thanks to our fixed background metric, we are offered privileged coordinates in which the theory is globally Lorentz invariant and in particular we have a  
well defined time coordinate $x^0$ to reverse.
As explained above this T transformation lets the inertial coordinates unchanged: $\xi_T ^\alpha(x_T^\mu) = \xi ^\alpha(x^\mu)$.
In the context of GR, this time reversal would just be a general coordinate transformation as any other one, but in DG we can and actually have to give it a special role to play by promoting the time reversal conjugate field $\xi_T ^\alpha(x^\mu) = \xi ^\alpha(x_T^\mu)$ revealed by the previous equality, hence the new time reversal conjugate field $g_{\mu \nu}(x_T^\mu)$ different from $g_{\mu \nu}(x^\mu)$.

We can give it a special role even more since,
\begin{itemize}
\item  Time reversal as well as Parity are the basic elements of the improper Lorentz group. No smooth deformation can result in T or P starting from any parameterized element of the group that does not explicitely involve P or T or both. So we find natural to have in our theory another field such as $g_{\mu \nu}(x_T^\mu)$ that no general coordinate transformation can transform into $g_{\mu \nu}(x^\mu)$. Only a discrete jump can connect them.

\item We already have a new order two tensor field available: the inverse form of $g_{\mu \nu}(x^\mu)$ which we have called $\tilde{g}_{\mu \nu}(x^\mu)$. Let us anticipate the spectacular result and bright confirmation of this analysis that we shall obtain by solving all our cosmological equations:
\begin{equation}
g_{\mu\nu}(x_{T}^\mu) = \tilde{g}_{\mu\nu}(x^\mu)
\end{equation}
Time reversal invariance is therefore only insured provided we add to the GR actions the ones built in the same way from $\tilde{g}_{\mu\nu}$. 
\end{itemize}
The propertime measured by clocks made of $\tilde{F}$ fields will be

\begin{equation}
d\tilde{\tau}^2=\tilde{g}_{\mu \nu}(x^\mu)dx^\mu dx^\nu
\end{equation}
In order to derive the fundamental equations of DG, solve them and check that we have a complete and acceptable theory (coherent and in good agreement with all 
observational data) we shall proceed in several steps: In step one, space-time permutation 
symmetries and discrete symmetries allow us 
to freeze various degrees of freedom and a priori identify only two allowed forms for the field in a privileged 
coordinate system where $\eta_{\mu \nu }$ reads $diag (-1,+1,+1,+1)$:

\begin{equation}g_{\mu \nu } = diag(B,A,A,A) \Longrightarrow  \tilde {g}_{\mu \nu }=diag(B^{-1},A^{-1},A^{-1},A^{-1}) \end{equation}
with either $B=-A$ or $B=-A^{-1}$. 
In step two we build the action for the gravitational field, sum of usual $I_{G}$ and the conjugate one $\tilde {I}_{G}$ built in the same way 
from $\tilde {g}_{\mu \nu }$. The conjugate actions are separately general coordinate scalars and adding the two pieces is necessary to 
obtain a discrete symmetry reversal invariant total action. In step three, we vary together the unfrozen conjugate matrix elements, 
eliminate $\tilde {g}_{\mu \nu }$ thanks to the relation $\tilde{g}_{\mu\nu}=\eta_{\mu\rho}\eta_{\nu\sigma}[g^{-1}]^{\rho\sigma}$ 
and apply the extremum action principle to finally obtain our modified Einstein equation.  In step four, we solve it and discuss our conjugate solutions in terms 
of discrete space-time symmetries. 

\section{Space/Time Permutation and Isotropy}

We carry out the program in four steps outlined at the end of the previous section,
 starting from step 1: in this section we will justify from global symmetry principles of 
flat space-time the only two allowed forms, $B=-A$ or $B=-A^{-1}$ in:
\begin{equation}
g_{\mu \nu } = diag(B,A,A,A) \Longrightarrow  \tilde {g}_{\mu \nu }=diag(B^{-1},A^{-1},A^{-1},A^{-1})
\end{equation}
 in a privileged coordinate system where $\eta_{\mu \nu } = diag (-1,+1,+1,+1)$. 
 But first we need to investigate the special case of the isotropic metric. 
 
 \subsection{The isotropic metric}

 A reference GR text book such as Ref.~\refcite{Wein} (page 176 and 335) reminds us the most general isotropic metric (rotationally form invariant  in the sense that the transformed metric is the same function of the rotated arguments as the old metric was of its arguments) in x,y,z,t coordinates:
 \begin {equation}
 M = \begin{pmatrix} B&xE&yE&zE\\ xE&A-x^2D&0&0 \\ yE&0&A-y^2D&0 \\ zE&0&0&A-z^2D \end{pmatrix}
\end {equation}
where B, A, E and D are allowed to be functions of time and/or $r = \sqrt{x^2+y^2+z^2}$ alone. However in DG the two forms of the metric must of course share the same isometries hence the inverse of this matrix $\tilde{M}=M^{-1}$ must also be of the same form
\begin {equation}
 \tilde{M} = \begin{pmatrix} \tilde{B}&x\tilde{E}&y\tilde{E}&z\tilde{E}\\ x\tilde{E}&\tilde{A}-x^2\tilde{D}&0&0 \\ y\tilde{E}&0&\tilde{A}-y^2\tilde{D}&0 \\ z\tilde{E}&0&0&\tilde{A}-z^2\tilde{D} \end{pmatrix}
\end {equation}
where $\tilde{B}$, $\tilde{A}$, $\tilde{E}$ and $\tilde{D}$ can only be functions of time and/or r. 
A straightforward inspection of the elements after matrix inversion of M shows that D=0 (thus $\tilde{D}=0$) is a necessary condition for having $\tilde{B}$ only r 
and/or t dependent and then that E=0 (thus $\tilde{E}=0$) is also necessary for having $\tilde{A}$ depending on r 
and/or t alone.
This is a striking result as compared to the GR case that allowed many possible forms for the isotropic metric, corresponding to as many different choices of coordinate systems at will. At the contrary here the only allowed form are the diagonal: 
\begin{equation}
g_{\mu \nu } = diag(B,A,A,A) \Longrightarrow  \tilde {g}_{\mu \nu }=diag(B^{-1},A^{-1},A^{-1},A^{-1})
\end{equation}
with B=B(r,t) and A=A(r,t) as the coordinate system where the background $\eta_{\mu \nu } = diag (-1,+1,+1,+1)$ fixes the two forms of the metric in DG much more efficiently than the requirement of rotational invariance alone in GR.

The Standard form ($g_{\mu \nu } = diag (B,A,r^{2},r^{2}sin^{2}(\theta)$) usually used to derive the Schwarzschild 
solution in GR is forbidden in true polar spherical coordinates where $\eta_{\mu \nu } = diag (-1,+1,r^{2},r^{2}sin^{2}(\theta))$. Instead the isotropic form ($g_{\mu \nu } = diag (B,A,Ar^{2},Ar^{2}sin^{2}(\theta)$) is imposed by this choice of coordinates!
But let's come back to the cartesian coordinates. Actually we realize that the particular form $\eta_{\mu \nu } = diag (-1,+1,+1,+1)$ of the background metric has fixed the form M of the metric in a way which is equivalent to having imposed from the begining another constraint added to the required rotational form invariance of M: the invariance under permutation $X_{ij}$ of any spatial indices i,j of the metric elements implying from the begining 
\begin {equation}
 M = \begin{pmatrix} B&E&E&E\\ E&A&C&C \\ E&C&A&C \\ E&C&C&A \end{pmatrix}
\end {equation}
Because then rotational form invariance implies not only C=0 but also E=0.

Our simple isotropic metric should not make us forget that as in GR we must be able to solve our equations and get solutions for any extended, in general not isotropic source distribution, which is a specially hard task in GR because the theory is so non linear. But if nature prefers simplicity, we can still hope that DG will facilitate the game in two possible (and non exclusive) ways:
\begin{itemize}
\item
First way: May be any extended distribution of matter and radiation should not any more be considered as a continuous and dynamical medium when the question is asked how it becomes a source of the gravitational field. A possibility is that at a microscopic scale, the quantum mechanical collapse implies the existence of a network of fixed points in vacuum where the quanta are actually created and annihilated, and may be, these are the privileged locations from where gravity is sourced and from nowhere else. Each of this point would be a particular center of isotropy and our isotropic metric would be suitable to represent the gravitational field having its roots there. This way will be explored and worked out in details later. 

\item
Second way: May be could we imagine alternatively that there are special sharply demarkated field configurations such as massive atoms or other more extended bound or not systems which center of gravity is the origin of a particular comoving privileged frame in which we have to work to compute the gravitational field of this system and have the right to apply our isotropic metric even if the system is not isotropic. This rests on the idea that the isotropic form of the metric is not actually imposed by the symmetries of the source but rather by fundamental principles, such as the isotropy of spacetime itself which favours a particular coordinate system where $\eta_{\mu \nu } = diag (-1,+1,+1,+1)$ and constrains the form of our field as we have seen, but also the invariance under permutation $X_{ij}$ of any spatial indices i,j in our metric elements that we might have demanded from the begining.

If our gravitational field is in the isotropic form in the privileged frame defined by a system consisting of many individual sources in a completely anisotropic configuration and does not even depend on the momenta of these sources whatever their motions relative to each other and relative to our frame (as if we were taking photos at high imaging frequency of the system and considering the temporal serie of these photos as the actual source for gravity), we will have for sure a departure from GR predictions testable for instance in the gravitomagnetic sector of the theories. But we consider this possibility to be already a much speculative extension beyond minimal DG. In the next section we will stick to a minimal version of DG following the first way.

\end{itemize}

But of course, both ways are only promising provided there is a simple method to later combine all the elementary gravitational
fields obtained in various privileged frames into a total single one to be injected in the equations of freefall for instance. Fortunately such a combination is natural and easy in DG as we shall later show and this motivates our willing to demonstrate that we need nothing else but the isotropic form of the metric to solve our equations and eventually successfully pass all observational tests of gravity.

If it is derived from fundamental principles, may be our simple isotropic metric form can be exploited to efficiently a priori reduce the number of degrees of freedom of $g_{\mu \nu }$ 
and $\tilde {g}_{\mu \nu }$ thus the number of equations which is much a stronger constraint than in GR. Indeed, in some cases the off diagonal elements will not be varied at all 
(frozen to zero before varying the action) and the other elements will not be varied independently. 
 \footnote{Arguments against the freezing of fields elements before varying the actions have been given by D.Deser and J.Franklin in Ref.~\refcite{Des}  
It is argued that in electromagnetism, isotropy a priori imposes that the magnetic field vanishes, however using this to freeze 
the magnetic field degrees of freedom before varying the action, one looses a fundamental equation which is the electromagnetism equivalent of the 
Birkhoff theorem: the vanishing of fields time derivatives in a spherically symmetric context. However this argument is only tenable provided there is
 no fundamental monopole in nature, since in presence of a monopole, B does not vanish and thus there is no reason to freeze it. But the inexistence 
of the monopole is certainly an a priori dangerous assumption so far. Moreover, our freezing of degrees of freedom does not follow from isometry 
requirements here as we have stressed. 
Ref.~\refcite{Des} also reminds us that freezing the off diagonal terms a priori in GR, one looses a fundamental equation from which the Birkhoff theorem
is derived. But there is a priori no fundamental reason for the Birkhoff theorem to remain valid in DG (no need to protect this theory against radiation of 
scalar modes by an isotropically breathing source for instance).}
\subsection{Space/Time exchange and the metric}
 We are left with the only two degrees of freedom A and B in
\begin{equation}
d\tau ^2=Bdt^2+A {{d\sigma} ^2},\quad
d\tilde {\tau }^2=\frac{1}{B}dt^2+\frac{1}{A} {{d\sigma} ^2} .
\label{eq4}
\end{equation}
where ${d\sigma} ^2=dx^2+dy^2+dz^2$. 
We can now generalize to space-time the kind of arguments we have used to constrain the form of our field in the preferred frame
by imposing a relation between the space-space A and the time-time matrix element B allowing to further reduce the number of independent 
elements to a single one. Then the equations of gravity might reduce to a single one with a single remaining degree of freedom to be varied.
To do so we must identify a symmetry linking time and space coordinates. We know that the meaningful global symmetries on flat space-time
should relate the fundamental representations of the Lorentz group i.e. tachyonic and bradyonic fields as well as positive 
and negative energy fields. The space/time exchange symmetry reversing the signature provides such a natural link between the
 tachyonic Lorentz group representation and the bradyonic one according to the arguments given in Ref.~\refcite{Rec}.
This leads us to suspect that even other conjugate forms should be involved, the ones with the flipped signature such as 
$\hat {g}_{\mu \nu }=-g_{\mu \nu }$ and background $-\eta_{\mu \nu }$. Indeed, following the same method as before i.e. constraining the form of our field to be symmetric under the 
permutation $X_{it}$ of spatial i and temporal t indices, we would find the trivial B=A=C Euclidian form to be the only acceptable one because it 
manifestly involves time and space matrix element indices in a symmetrical way, whatever C(r,t) in 
$d\tau ^2=C\left(dt^2+ {{d\sigma} ^2} \right)$. But if we also have an opposite form such as for instance $\hat {g}_{\mu \nu }=-g_{\mu \nu }$
to exchange the role of A and B and restore in this way a kind of permutation symmetry between space and time we can 
certainly at least admit solutions with B=-A. Let us try to follow this method in an exhaustive way exploiting conjugate forms such as
$\hat {g}_{\mu \nu }$ and $\tilde {g}_{\mu \nu }$ to make our permutation symmetry work. 

The most general expression is
\begin{equation}
d\tau ^2=C\left(\frac{1}{D}dt^2+D {{d\sigma} ^2} \right),
d\tilde {\tau }^2=\frac{1}{C}\left(Ddt^2+\frac{1}{D} {{d\sigma} ^2} \right).
\end{equation}
However D must be purely imaginary if we want a Minkowskian signature so we also need a purely 
imaginary C  term to have a real field.
Thus let us redefine D and C into iD and iC with the new D and C real.
Introducing the opposite form $\hat {g}_{\mu \nu } =-g_{\mu \nu } $, 
we have eventually four conjugate forms
$g_{\mu \nu }$, $\tilde {g}_{\mu \nu }=\eta_{\mu \rho }\eta_{\nu \sigma } \left(g^{-1}\right)^{\rho \sigma}$ 
, $\hat {g}_{\mu \nu }=-g_{\mu \nu }$, 
$\tilde {\hat {g}}_{\mu \nu }=\hat {\tilde {g}}_{\mu \nu }$ to hopefully describe conjugate 
positive/negative energy worlds as well as tachyonic/bradyonic worlds:

\[
\begin{array}{l}
d\tau ^2= C\left(\frac{1}{D}dt^2-D {{d\sigma} ^2} \right),
d\tilde {\tau }^2= \frac{1}{C}\left(Ddt^2-\frac{1}{D} {{d\sigma} ^2} \right)\\ 
 d\hat {\tau }^2= -C\left(\frac{1}{D}dt^2-D {{d\sigma} ^2} \right] , 
 d\hat {\tilde {\tau }}^2= -\frac{1}{C}\left(Ddt^2-\frac{1}{D} {{d\sigma} ^2} \right] \\ 
 \end{array}
\]

Only $C$ and $D$ both different from one breaks down the space / time permutation symmetry so that the symmetry is insured 
provided either C or D equals one (this is because in these particular cases the conjugate fields just exchange the roles of D and -1/D 
or of C and -C) in which cases the field must respectively be of the form $B=-A^{-1}$ or of the form
$B=-A$. 
Thus, our new permutation symmetry principles allow us to accept only two kinds of Minkowskian fields: the B=-1/A and 
B=-A forms.

\section{The Conjugate Fields and their Variations}

In the following sections we require that the field elements be a priori linked by $A=-B$ or $A=-B^{-1}$. 

We postulate that $A=-B$ is such a strong symmetry that a single degree of freedom eventually remains.
This means that A and B elements being related can no longer be varied independently. 
Therefore, given any tensor $R_{\mu \nu }$ and for real elements making use of $\frac{\delta g_{{i}'{i}'} }{\delta g_{ii} 
}=\frac{g_{{i}'{i}'} }{g_{ii} }$ and  $\frac{\delta g_{{t}{t}} }{\delta g_{ii} 
}=\frac{g_{{t}{t}} }{g_{ii} }$ in case B=-A, a typical action variation 
will be proportional to a trace:
\[
\delta (g^{-1})^{rr} R_{rr}+\delta (g^{-1})^{tt} R_{tt}+\delta (g^{-1})^{\theta \theta } 
R_{\theta \theta }+\delta (g^{-1})^{\phi \phi } R_{\phi \phi }=\frac{\delta (g^{-1})^{rr} 
}{(g^{-1})^{rr} }R
\]
 while in case B=-1/A, we will vary independently all field elements to get as many equations. 
We could speak of a weakly broken space-time global symmetry in this case, weakly because 
B=-1/A does not eliminate degrees of freedom when we extremize the action but can be used to
 simplify the equations once we have them (just as we do in GR when isometries allow to work with 
a simplified form of the metric).  

We also need the 
relation between the relative variations of the inverse conjugate forms needed to obtain the gravitational equation
in term of the components of a single form of the field:
$\frac{\delta\tilde{g}_{{x}{x}}}{\tilde{g}_{{x}{x}}}=-\frac{\delta g_{{x}{x}}}{g_{{x}{x}}}$.
Eventually, for B=-A we will therefore obtain the single equation in vacuum :
\[
\sqrt g R-\frac{1}{\sqrt g }R_{g\to 1/g} =0
\]
where R is the familiar Ricci scalar. 

Recall that the opposite $\hat {g}_{\mu \nu }$ and $\tilde {\hat {g}}_{\mu \nu }$ so their respective 
actions should also be involved.
However, because the Einstein-Hilbert action of -g is opposite to the Einstein-Hilbert
 Action of g, the $\hat {g}_{\mu \nu }$ and $\tilde {\hat {g}}_{\mu \nu }$ Einstein-Hilbert 
actions must come into the game with a negative sign i.e. with an opposite gravitational 
constant -G as suggested in Ref.~\refcite{Qui} to avoid a trivial cancellation of the 
total Einstein-Hilbert action. But then the new Einstein-Hilbert 
actions are identical to the ones we already have so we will assume these can be taken into 
account simply by a convenient redefinition of the factor G and it will
 not be necessary to involve explicitely these actions in the following. 

\section{The DG Fundamental Equations in Vacuum}
In this section we start from the rewritten propertime
\[
d\tau ^2=-Bdt^2+Ad\sigma ^2
\]
so that our previous section conditions: B=-1/A and B=-A translate into B=1/A, B=A. We can perform the computation in the more convenient and usual spherical polar coordinates. 
We find for the Ricci tensor elements:
\[
\begin{array}{l}
 R_{rr} =\frac{{A}''}{A}+\frac{{B}''}{2B}-\frac{{B}'^2}{4B^2}-\left( 
{\frac{{A}'}{A}} \right)^2+\frac{{A}'}{Ar}-\frac{1}{4}\frac{{A}'}{A}\left( 
{\frac{{B}'}{B}} \right)+\left( {-\frac{1}{2}\frac{\ddot 
{A}}{B}-\frac{1}{4}\frac{\dot {A}}{B}\left( {\frac{\dot {A}}{A}-\frac{\dot 
{B}}{B}} \right)} \right) \\ 
 R_{\theta \theta } 
=\frac{3}{2}\frac{{A}'r}{A}+\frac{{A}''r^2}{2A}-\frac{{A}'^2r^2}{4A^2}+\frac{1}{2}\frac{{B}'}{B}\left( 
{\frac{{A}'r^2}{2A}+r} \right)+\left( {-\frac{1}{2}\frac{\ddot 
{A}}{B}-\frac{1}{4}\frac{\dot {A}}{B}\left( {\frac{\dot {A}}{A}-\frac{\dot 
{B}}{B}} \right)} \right)r^2 \\ 
 R_{\phi \phi } =\sin ^2\theta R_{\theta \theta } \\ 
 R_{tt} 
=-\frac{1}{2}\frac{{B}''}{A}-\frac{1}{4}\frac{{B}'{A}'}{A^2}+\frac{1}{4}\frac{{B}'^2}{BA}-\frac{1}{A}\frac{{B}'}{r}+\frac{3}{2}\left( 
{\frac{\ddot {A}}{A}-\frac{1}{2}\frac{\dot {A}}{A}\frac{\dot 
{A}}{A}-\frac{1}{2}\frac{\dot {B}}{B}\frac{\dot {A}}{A}} \right) \\ 
 \end{array}
\]
The sum of variations when B=A involves the Ricci scalar:
\[
R_{B=A} =\left( {3\frac{{A}''}{A^2}-\frac{3}{2A}\left( {\frac{{A}'}{A}} 
\right)^2+6\frac{{A}'}{A^2r}} \right)+\frac{3}{A}\left( {-\frac{\ddot 
{A}}{A}+\frac{1}{2}\left( {\frac{\dot {A}}{A}} \right)^2} 
\right)
\]
The equation in vacuum when B=A follows from
\[
\sqrt g R-\frac{1}{\sqrt g }R_{g\to 1/g} =0
\]
giving:
\[
\begin{array}{l}
 3A\left( 
 \frac{{A}''}{A}-\frac{1}{2}\left( \frac{{A}'}{A} 
\right)^2+2\frac{{A}'}{Ar} - \frac{\ddot 
{A}}{A}+\frac{1}{2}\left( \frac{\dot {A}}{A}  \right)^2 
\right) \\ 
 -3\frac{1}{A} \left( -\frac{{A}''}{A}+\frac{3}{2}\left( 
\frac{{A}'}{A} \right)^2-2\frac{{A}'}{Ar} + \frac{\ddot 
{A}}{A}-\frac{3}{2}\left(\frac{\dot {A}}{A} \right)^2 
 \right)=0 \\ 
 \end{array}
\]
For an homogeneous field $A(t)$ satisfies~:
\[
A\left( { {-\frac{\ddot {A}}{A}+\frac{1}{2}\frac{\dot {A}}{A}\left( 
{\frac{\dot {A}}{A}} \right)} } \right)-\frac{1}{A}\left( { 
{\frac{\ddot {A}}{A}-\frac{3}{2}\left( {\frac{\dot 
{A}}{A}} \right)^2} } \right)=0
\]
admitting both time dependent as well as a stationary solutions.

For B=1/A, the time-time equation in vacuum follows from
\[
\sqrt g G_{tt}-\frac{1}{\sqrt g }B^{2}G_{tt, g\to 1/g} =0
\]
giving:
\[
\frac{1}{A}( \left( {2\frac{{A}''}{A}-2\left( {\frac{{A}'}{A}} 
\right)^2+4\frac{{A}'}{Ar}} \right)+\frac{3}{4}\left( { \frac{1}{A^2} -A^2 } 
\right) {\left( \frac{\dot {A}}{A} 
\right)^2}) = 0
\]
The space-time equation reads (B replaced by 1/A)
\[
A \left( \frac{{\dot{A}}'}{A}-\frac{1}{2}\frac{{A}'}{A}\frac{\dot{A}}{A}
\right) +\frac{1}{A}\left( -\frac{{\dot{A}}'}{A}+\frac{3}{2}\frac{{A}'}{A}\frac{\dot{A}}{A}
\right) = 0
\]
and the space-space equations (B replaced by 1/A):
\[
A(A^2 \left( \frac{\ddot{A}}{A}+\frac{1}{4}(\frac{\dot{A}}{A})^2
\right) -\frac{1}{A^2}\left( -\frac{\ddot{A}}{A}+\frac{9}{4}(\frac{\dot{A}}{A})^2
\right)) = 0
\]
Making the assumption that  $\dot{A}$ does not vanish, we redefine $A=e^{a}$. Integrating as much as we can the three equations then 
read:
\[ \dot{a}^2 Sinh(a)=\frac{4}{3} \Delta a 
\]
\[
\dot{a}^2 Sinh(a)=f(t) 
\]
\[
\dot{a}^2 Cosh^{5/4}(2a)= g(r)
\]
but the last two for instance are not compatible unless $\dot{a}=0$. 
 Therefore the solution for the B=1/A field is time independent and in vacuum the only remaining equation reads: 

\[
\Delta a=0 
\]

\section{Summary}

Now that we are in position to solve the DG fundamental equations, let us take sometime to first 
recapitulate the conceptual foundations from which
 these were derived and the main steps in this derivation to show that, though these look very unfamiliar for the GR expert, this derivation 
completely follows from basic principles which themselves are obvious in a flat space-time framework.  
\begin{itemize}
\item Contemporary physics relies on symmetry principles. In a pure Special Relativity framework, time reversal symmetry is fundamental and generates 
negative energy objects solutions of all propagation equations. Quantum theory comes into the game with a binary mathematical choice for time reversal 
symmetry. Either it is Anti-Unitarily performed and may be we are allowed to neglect the negative energy fields solutions or it is Unitarily performed in which case 
the negative energy fields being regenerated by time reversal must be taken serious. Our starting point was the Unitary choice.
\item The Hamiltonian of the most simple field, the scalar field, does not reverse under Unitary time reversal. This is a dead-end unless we extend our thought 
to gravity to investigate its time reversal behaviour.
\item Time reversal is a global discrete symmetry which is only meaningful in a flat space-time and not in GR curved space-time. Thus we need an 
alternative theory of gravity involving al least the flat non dynamical background space-time  $\eta_{\mu\nu}$ and the usual $g_{\mu\nu}$ of GR. For the sake of simplicity we first 
investigate a classical theory for these fields. 
\item But given $\eta_{\mu\nu}$ and $g_{\mu\nu}$, inevitably arises one (and only one) covariant symmetric order two tensor $\tilde{g}_{\mu\nu}$, depending on the latter 
 and having exactly the same status as $g_{\mu\nu}$. So we must respect the symmetry under their permutation in the theory making
 them play equivalent roles in our equations.
\item This is done by adding to the GR action the same one built from $\tilde{g}_{\mu\nu}$. $\tilde{g}_{\mu\nu}$ is eliminated thanks to the relation that links it to
$\eta_{\mu\nu}$ and $g_{\mu\nu}$ and the equations satisfied by $g_{\mu\nu}$ are derived from the least action principle in the coordinate system 
where $\eta_{\mu\nu}=diag(-1,+1,+1,+1)$
\item This frame also imposes global permutation symmetries to the fields which can only take the B=-A and B=-1/A forms that generate two cohabiting sectors of the 
theory.
\item The DG equations are simplified by the compensating  $g_{\mu\nu}$ and $\tilde{g}_{\mu\nu}$ terms as well as by the simple forms 
taken by the fields and a reduced number of degrees of freedom (one in case B=-A). Thus these are easily solved. 
Remains to be investigated the viability of our solutions, interpret them and understand the articulation between the B=-A and B=-1/A sectors
 before confronting the theory predictions with various test results. 
\end{itemize}

\section{B=-1/A, The New Schwarzschild Solution}

\subsection {The non dynamical gravific action}

In a previous section we anticipated that any extended distribution of matter and radiation should not any more be considered as a continuous and dynamical medium when the question is asked how it becomes a source of the gravitational field. We postulated the existence of a network of gravific sites, each one being a center of isotropy and elementary source for each B=-1/A elementary isotropic metric. Of course, it is understood that each site "catches" everything around that could source gravity, probably to be found in the total energy momentum tensor of the standard model fields derived as usual in another independent action where these fields play their dynamics. But the gravific action we want to build now is non dynamical in the sense that all fields indirectly sourcing gravity there have already played their dynamics somewhere else.

All this is motivated by a straightforward method the background metric offers to later combine elementary $[g^{B=-1/A}]_{\mu\nu }$ solutions once each has been obtained from the resolution of the 
equation deriving from its standalone 
action, into a single one. For instance for two metrics $g_1$ and $g_2$ the combination results in 
\begin{equation}
[g^{comb}]_{\mu\nu }=\frac{1}{2}( [g_{1}]_{\mu \rho }[g_{2}]_{\nu \sigma } +
[g_{1}]_{\nu \rho }[g_{2}]_{\mu \sigma } ) \eta^{\rho \sigma} 
\end{equation}
the fields being combined after 
they have been exported to a common coordinate system of course. One can check that the 
procedure works well thanks to the exponential form of the elementary solutions matrix elements found in vacuum in a previous section and that we shall confirm in this section even if the fields to be combined are not anymore in a diagonal form after Lorentz Transformation.

We postulate that in the gravific action, the gravific elementary source is a static perfect fluid element so, adopting the notation conventions of Ref.~\refcite{Will} (page 91), we have for the only non vanishing elements of the corresponding source tensor
$T^{00}= -\rho(1+\Pi)/B_{tot} = - \rho(1+\Pi)/B_{ext}B$ and 
$T^{ij} = p \delta^{ij}/A_{tot} = p \delta^{ij}/A_{ext}A $
where $B_{tot}$ and $A_{tot}$ stand for diagonal matrix elements of the total combination 
of $B =-1/A$ fields and we have separated two contributions 
to these gravitational potential energy terms: an external one and the particular B=-1/A field 
we want to solve the equation for. It is usefull to define the scalars $\bar{p}$ and 
$\bar{\rho}$ 
which in the privileged coordinate system take the value
\[
\bar{\rho} =\rho(1+\Pi)/B_{ext} ;
\bar{p} = p /A_{ext} 
\]
 include internal as well as gravitational potential energy in the external field but
 dont depend on A and B. This allows us to rewrite 
 $T^{\mu\nu}$ in the usual way for a perfect fluid 
$T^{\mu\nu}= \bar{p} g^{-1\mu\nu}+ (\bar{\rho}+ \bar{p}) U^{\mu}U^{\nu}$ with
 $U^{\mu}U^{\nu}g_{\mu\nu} = -1$, 
where
$g_{\mu\nu}$ here only involves the B=-1/A gravitational field we need to compute.
The source action for our side fields then has to be 
\[
\int \sqrt{\eta} T^{\mu\nu}\eta_{\mu\nu} d^4x 
\]
 for the integrand under a variation of $(g^{-1})^{\mu\nu}$ to become
 $\eta_{\nu\sigma}g_{\mu\rho}T^{\rho\sigma}\delta (g^{-1})^{\mu\nu}$.
To recover at least Newtonian gravity, solving 
-2$\triangle a(r) =  n\pi G M\delta (r)$ in case the source is a point mass 
M (total gravific energy as seen from outside the source), must give a(r)=-2U(r) where U(r) is the adimentional Newtonian potential $U(r)=\frac{-MG}{r}$
solution of $\triangle U(r) = 4\pi G M\delta (r)$. Thus n equals 16 
and our total Action $S_{grav}$ for gravity and the gravific sources had to be
\[
S_{grav} = \frac{-1}{16\pi G} \int d^4x(\sqrt g R+\frac{1}{\sqrt g }R_{g\to 1/g})
+ \int d^4x \sqrt{\eta} ( T^{\mu\nu} + \tilde{T}^{\mu\nu} )\eta_{\mu\nu}
\]
provided we have in a single privileged 
frame a perfect fluid at rest at the same place on both sides of the field. If it's not the case, we would have to consider two separate actions of this kind, one with source $T^{\mu\nu}$ and the other with source $\tilde{T}^{\mu\nu}$ in order to independently solve for two elementary B=-1/A gravitational fields, each in the rest frame of its source.
  
Let us write down the complete system of equations for this field.
The time-time equation reads:
\[
 \left( {2\frac{{A}''}{A}-2\left( {\frac{{A}'}{A}} 
\right)^2+4\frac{{A}'}{Ar}} \right)+\frac{3}{4}\left( { \frac{1}{A^2} -A^2 } 
\right) {\left( \frac{\dot {A}}{A} 
\right)^2} = -16\pi G (\bar{\rho}A - \frac{\bar{\tilde{\rho}}}{A})
\]
The space-time equation reads: 
\[
A \left( \frac{{\dot{A}}'}{A}-\frac{1}{2}\frac{{A}'}{A}\frac{\dot{A}}{A}
\right) -\frac{1}{A}\left( -\frac{{\dot{A}}'}{A}+\frac{3}{2}\frac{{A}'}{A}\frac{\dot{A}}{A}
\right) = 0
\]
and the space-space equations:
\[
A^2 \left( \frac{\ddot{A}}{A}+\frac{1}{4}(\frac{\dot{A}}{A})^2
\right) -\frac{1}{A^2}\left( -\frac{\ddot{A}}{A}+\frac{9}{4}(\frac{\dot{A}}{A})^2
\right) = -16\pi G (\frac{ \bar{p}}{A} - \bar{\tilde{p}}A)
\]
This system clearly shows that the gravific energy momentum tensor had 
to be defined as we did but with $\bar{p} = \bar{\tilde{p}} = 0$ whatever p to impose that 
either pressure should not source gravity whatever p (each elementary source "catches" everything around that could source gravity except pressure)
or the perfect fluid at rest that we are considering as a good candidate model
for the gravific elementary source has no pressure, hence it is rather a small spherical Mass or may be a stationary field or standing wave in a spherical volume trapped by a spherical shell. Then, as in vacuum, the space-time and space-space 
equations are trivially satisfied provided our B=-1/A solution does not depend on time.
  
\subsection {The solution for a small spherical mass source M}

Our static solution outside of the small spherical mass source M (understood to include all contributions to the total gravific mass) reads
\begin{equation}
 A=e^{\frac{2MG}{r}}\approx 1+2\frac{MG}{r}+2\frac{M^2G^2}{r^2} \\
\end{equation}
\begin{equation}
B=-\frac{1}{A}=-e^{\frac{-2MG}{r}}\approx 
-1+2\frac{MG}{r}-2\frac{M^2G^2}{r^2}+\frac{4}{3}\frac{M^3G^3}{r^3} \\ .
\end{equation}
different from the GR one though in good agreement up to Post-Newtonian order. 
Notice
 by the way 
that -1/B=A is only satisfied in the standard coordinate system in GR, not in the isotropic one.
No coordinate singularity arises in our frame and it is straightforward to check that 
this Schwarzschild solution involves no horizon. 
The conjugate forms can be transformed into one another through 
 $r \to -r$ or  $M \to -M$. 
Thus, it is tempting to consider that the discrete symmetry reversal involved here reverses mass (gravific rest energy) and 
involves at least space reversal with the new definition: $r \rightarrow -r $ when $x, y, z \rightarrow -x, -y, -z $. Two conjugate fields (at r and -r
) may appear antipodal with 
respect to r=0, the center of our elementary gravific site.

Eventually, it is important to notice that the poisson equation source term involves 
$\bar{\rho}A$ where A should affect importantly how the gravific mass 
varies inside matter but also the total integrated gravific Mass source as seen from outside
 matter because of the exponential behaviour of A in the strong field regime. In case we could apply our metric solution to the gravitational field of a very compact object much more extended 
than the elementary microscopic source we have considered in priority so far, this should contribute to generate a huge gravific mass 
very different from the gravific mass of the same quantity of matter spread over a larger 
volume. This effect is absent in GR because in a static configuration there is always an equilibrium between gravitational potential energy and pressure producing a compensation also in the gravitational field they source. 
The uncompensated huge gravific effect that we get in DG is an interesting feature for an alternative understanding of the origin of the many billion solar masses 
(this is a gravific mass) objects at the center of most galaxies, even in the lightest ones.

\subsection{The PN solution}
\subsubsection{For our elementary gravific source}
Adapting the PN formalism to DG, following the notation conventions of Ref.~\refcite{Will} we
can systematically compare the obtained results with the corresponding ones in GR.
The GR PPN field at PN order reads:
\[
\begin{array}{l}
g_{00}=-1+2U-2U^2 +4\Phi_{1} +4\Phi_{2} +2\Phi_{3} +6\Phi_{4}  \\
g_{0j}= -\frac{7}{2}V_{j}-\frac{1}{2}W_{j}  \\
g_{jk}=(1+2U)\delta_{jk}
\end{array}
\]
where
\[
\begin{array}{l}
U(x,t)= \int\frac{\rho(x',t)}{\mid x-x'\mid } d^3x'
\end{array}
\]
(Warning: this new definition of U makes it opposite to the previous section more usual definition) and $\Phi_{4}$, $V_{j}$, $W_{j}$  terms are respectively 
 pressure and momentum source terms 
 while $\Phi_{1}$, $\Phi_{2}$, $\Phi_{3}$ are kinetic energy, gravitational potential energy
 and internal (non gravitational potentials) energy terms.
The explicit expressions for these terms can be found in Ref.~\refcite{Will} p95.
In comparison, the DG B=-1/A field at PN order in the preferred frame for our elementary 
gravific source is
\[
\begin{array}{l}
g_{00}=-1+2U-2U^2+4\Phi_{1} +2\Phi_{3}+4\Phi_{2}  \\
g_{0j}=0 \\
g_{jk}=(1+2U)\delta_{jk}
\end{array}
\]
We see that in general $\Phi_{4}$, $V_{j}$, $W_{j}$ are present in GR but not in DG. Another difference is that in DG, $\Phi_{1}$ only stands for total microscopic kinetic energy, in other words, heat, because we are in the comoving frame of our gravific system. However, most of these apparent discrepencies are not actually significant because the DG field solution does not apply to any extended source distribution and might only be meaningful for the elementary gravific source we have postulated. If this source exists, then it is at rest in the privileged frame since it is a comoving one so we would also have vanishing $V_{j}$, $W_{j}$ and macroscopic $\Phi_{1}$ terms in this frame for such source (element of perfect fluid at rest) in GR. Then either:
\begin{itemize}
\item
First possibility: The elementary source has no pressure in which case $\Phi_{4}$ also vanishes so DG and GR predictions at PN order are the same for such elementary gravific source. Then the only departure between the two theories that we could imagine would be between the total field predictions for any kind of extended distribution of sources: such field is directly obtained from the resolution of Einstein equations in GR while it is in DG the result of a combination of elementary solutions. But since we took special care to correctly incorporate gravitational potential energies as sources of the DG gravitational field as in GR, we actually expect no measurable difference between the two theories predictions at PN order. Indeed, after Lorentz exporting all elementary solutions to a common frame and combining them, the same pressure, momentum and kinetic energy terms are expected to occur in DG as in GR for the extended source at least at PN order.

Let us stress that even for fields believed to be massless such as light, one could postulate an incredibly tiny mass just to be able to have again a restframe privileged frame to compute as we did the gravitational field sourced by these fields in DG provided elementary gravific sources actually exist as we imagined them.
\item 
Second possibility: The elementary source has pressure but the gravitational field in DG is not sensitive to it, contrary to the GR case. Indeed, the pressure contribution $\Phi_{4}$ can be removed in DG in the preferred frame 
simply by setting $\bar{p} =0$ and $ \bar{\tilde{p}} =0$ in place of the previous section 
definitions.
\end{itemize}
\subsubsection{For an extended gravific source}

The DG static solution is of course the most natural and obvious in the restframe of our elementary gravific source which eventually just appears to be a spherical mass at rest so the reader might wonder 
why we do not specialize from the begining to this most natural and trivial scenario allowing DG and GR to be completely indistinguishable so that all GR observational successes could be readily translated into DG successes.
The answer is that we want to keep open the possibility of making predictions different from GR 
ones in the case the B=-1/A field could be solved for not only for the elementary perfect fluid element we have considered so far but also for an extended distribution of gravific sources in motion relative to each other. Departures are expected in the $\Phi_{4}$, $\Phi_{1}$, $V_{j}$ and $W_{j}$ sectors of the theory. \begin{itemize}
\item $\Phi_{1}$

The relevance of the kinetic term $\Phi_{1}$ even for a microscopic extended and composite object such as a nucleon seems insured for if
all kinetic terms were absent, these could not contribute to the gravific mass of the nucleon.

As a matter of fact, we have just learned that the mass 
of the proton was recently successfully (with good precision) computed and is the total 
rest energy mostly determined by 
internal energies of the partons (QCD and kinetic terms), the quark masses contribution remaining very small. 
Since the equality of gravific and inertial mass is well tested for baryonic matter, it would 
be dangerous to forget the $\Phi_{1}$ contribution to the gravific mass in DG.
However since we were not informed about the exact actual contribution of kinetic terms to 
the proton mass, we cannot exclude that $\Phi_{1}$ be absent in DG.

\item $\Phi_{4}$

Actually, 
$\Phi_{4}$ could never be isolately
measured up to now which according to Ref.~\refcite{Ehl}, is due to an exact compensation between gravific contributions of pressure 
and gravitational potential energy for a star in a static state. Therefore, 
neither DG nor GR are yet constrained in this sector so we could even consider the case where the DG field solution applies to an extended object such as a star. 
 In this case, as we already mentionned at the end of the last subsection, a possible way to test DG against GR would be the evidence for the absence of any gravific pressure to balance the gravific self gravitational potential energy in a 
 highly non equilibrium situation such as that of a Supernova or for very compact objects such as black holes which would be able to acquire an unexpectedly high gravific mass in absence of the above compensation.
Considering again the nucleon as a possible extended gravific source, in GR one would also expect a discrepancy between the gravific mass of the proton and its inertial mass if internal pressure for the partons inside the protons is not completely dominated by the inertial mass because otherwise the gravific pressure contribution would make the gravific and inertial mass very different, contrary to what various precision tests have learned to us. This is only an issue for GR since, as we have just noticed, pressure is not gravific in DG for the extended source of the B=-1/A field.
 
\item $V_{j}$ and $W_{j}$

In the privileged coordinate system where we would compute the B=-1/A solution for an extended source, the elements could read $-1/A=B=e^{2U(r)}$, 
where U(r) is simply the total additive Newtonian gravitational potential 
generated by individual masses (including gravitational potentials in the definition of these gravific masses) belonging to our extended source. 
For a set of masses $M_{i}$ a priori in motion relative to each other, an exact potential such as $U=-{\Sigma_i\frac{M_iG}{r_i}}$ would ignore any possible $V_{j}$ and $W_{j}$ gravitomagnetic effects. The possible consequences for an extended source as big as a rotating planet will be explored in the next section dealing with gravitomagnetism.

\end{itemize}

\subsection{The gravific source Action}

The physical 
mechanism behind the generation of our gravific action in case we have a network of elementary gravific sources,
would be the following: we postulate the existence of a network 
of huge masses with alternate positive and negative values in vacuum at the microscopic scale. 
The strong anti-gravitational 
interaction between each mass and its neighbours stabilizes the network. At the locations of these
mass does the creation and annihilation of spherical waves of QFT take place. In particular, the 
network is responsible for the QM wave collapse. Only at the location of these masses is the 
gravitational field generated when various waves are created and annihilated at these points. 
The huge masses also generate their own gravity but their compensating fields result in 
undetectable gravitational effects at the macroscopic scale.

It is understood that any spherical wave in a wave packet can be trapped (annihilation) in the deep well of one of the network 
masses where it momentarily has the rest total energy M in a comoving frame at this point before being reemitted (creation). In between points of the network a field propagates but never generates any gravitational field. It can do so only when it is trapped in the form 
of a system of standing waves  
 in the deep potential well generated by any mass 
in the vacuum network. This is a possible way to convert the total extended energy distribution of any field into a collection of gravific isotropic (in adequate comoving frames) rest energies, the energies of momentarily 
standing waves trapped in the wells.

This description would be valid even for an almost massless wave propagating at nearly the speed of light in between the nodes of the network and might even apply to light if the photons are not strictly massless. This would allow the radiation 
pressure and energy to contribute to the gravitational total field as in GR after Lorentz exporting 
all elementary fields and combining them in a common frame. 
In this case DG and GR would have the same predictions in all sectors of gravity as we noticed earlier, but another possibility in DG is if the photon field is really massless in which case there is no restframe and may be eventually it's not any more possible to get radiation pressure gravific effect in the theory.

Now the question of course is: why should gravity only originate from the particular sites where our network vacuum masses are sitting and how could these efficiently momentarily trapp the incoming spherical waves in the form of a system of standing waves ? We anticipate that each of these vaccum masses is actually a kind of microscopic Dyson sphere, the mass at the center being surrounded by a spherical field discontinuity, a potential barrier which can indeed trapp a system of standing waves very efficiently inside. Such potential barriers, we dont have to arbitrarily postulate their existence because these are an important prediction of the DG theory that explains the Pioneer anomaly as we shall show later. Such discontinuous potential barriers are also delimitating surfaces at which boundary conditions for the gravitational field are expected to apply. 

Indeed it is well known that the gravitational field solution in general both depends on the actual differential equation it must satisfy but also asymptotic or boundary conditions. Thus it was not mandatory to introduce a gravific source action as we did. We could as well have considered that there is no source action at all and that the equations were the ones we obtained in vacuum for the B=-1/A field, but to avoid the trivial Minkowski solution, we would just have needed an additional condition to the requirement that the field should be asymptotically Minkowskian. This new condition would just be a boundary condition for the field at the frontier defined by the surface where the discontinuity is sitting. Therefore our gravific source action introduced in the previous section was most probably just an heuristic model.

\subsection{Stability}

The system of equations does not admit any wave solution. So the solution is unpropagated and
 instantaneous in our frame and there is no force carrier.
Usually stability is menaced in a theory when there are interacting degrees of freedom 
with both positive and negative energy states opening an infinite phase-space for decay.
Once the DG equations are obtained, the very brutal elimination of all but one degrees 
of freedom, B=-1/A, with direct consequence from our equations that it cannot depend 
on time in vacuum trivially insures stability of this sector of the theory.

One may wonder how the solution could remain static in case the sources were not static. 
First the issue seems real only when we want to describe extended source distributions for an 
elementary source is always static in its restframe. But even for the extended case 
our solution is only static in 
the sense that it has no explicit time dependency. Of course the set of variables $r_{i}(t)$
 entering in our solution can change in time when the sources labelled by i are in motion. 

The phenomenology is simple: masses living in the same form (on the same side of gravity) attract each 
other. Masses living in conjugate forms repel each other, as if the mass living in the conjugate form 
contributes as a negative mass source from our side point of view. This is the same phenomenological stability as in the bimetric theory 
of JP Petit (Ref.~\refcite{petit} and \refcite{petit2} with references therein).
This phenomenological stability is interesting to notice. Indeed, in GR we would naively expect a negative energy object to be attracted by a positive energy object, 
the latter being repeled by the former. They then would accelerate together for ever. Here such kind of 
phenomenological instability is also avoided since masses living in different forms 
just repel each other. Yet, from the point of view 
of each form, this is really the interaction between a local positive mass and an invisible negative mass from the dark side of gravity.  

\subsection{Causality}

Let us follow the point of view of a fictitious observer living in the background spacetime metric which unrenormalized 
rods and clocks would possibly allow him to measure a light (for instance propagating in $g_{\mu\nu}$) speed different from c. 
Indeed, for this observer
a photon or any ultrarelativistic particle propagating in the Schwarzschild $g_{\mu\nu}$ field generated by a nearby spherical Mass
M on our side follows its geodesics hence:
\begin{equation}
 0 = (1-2GM/rc^2) dt^2 - (1+2GM/rc^2) d\sigma^2
 \end{equation}
while another photon on the other side follows the geodesics of $\tilde{g}_{\mu\nu}$
\begin{equation}
0 = (1+2GM/rc^2) dt^2 - (1-2GM/rc^2) d\sigma^2
\end{equation}
where we have retained only some Post
Newtonian approximated metric element terms because we assume that we are in a very weak gravitational field. Hence
from the background point of view the PN approximated speed of light in $g_{\mu\nu}$  is  $\frac{d\sigma}{dt} = 1-2\frac{GM}{rc^2}$ (resp $\frac{d\sigma}{dt} = 1+2\frac{GM}{rc^2}$ 
in $\tilde{g}_{\mu\nu}$ ) so the speed of light in the conjugate metric is $1+4\frac{GM}{rc^2}$
 times faster. This ratio between these two observables
would be the same if we had considered an observer in either of the two conjugate metrics instead of the observer linked to the background. In this case the conjugate light-cone is wider than our side 
lightcone. This remains true from the genuine observer point of view on our side and means 
that the speed 
of light on the conjugate side also appears to him greater than c he measures on his side.
Making appeal to an extra metric was exactly the same kind of idea (except it was ad-hoc) recently advocated by J. W. Moffat (Ref.~\refcite{Moff3}) to try to explain the anomalous
Opera superluminal neutrino velocities (Ref.~\refcite{Ope} and Ref.~\refcite{DGneut}) before the measurement error was discovered. It is also appropriate to recall that even in GR the velocity of light or any ultrarelativistic particle propagating in a gravitational field different from the one our rods and clocks feel on earth, for
instance in the vicinity of a far away compact object, as measured with respect to our local rods and clocks also appears
subluminous or superluminous (if the object has a negative mass for instance) as explained in more details in Ref.~\refcite{Lust} although the locally measured speed of light is
everywhere still of course c. The interesting new phenomenology allowed by $\tilde{g}_{\mu\nu}$  is that the particles need not propagate
in another distant gravitational field (necessarily far away to be very different from our local one) but just here and now
in $\tilde{g}_{\mu\nu}$.

This possibility of faster than c information transfer in a special relativistic 
framework is well known to generate causality issues: some observers might see these 
informations propagating backward in time. 
Of course, since the dominant
contribution to the adimentional potential $GM/rc^2$
is typically determined by the nearby clusters of galaxies potential well which order of magnitude is generally a few $10^{-5}$ it is in practice very difficult to obtain a CTC (closed timelike curve) in this case (though it would become much easier in the strong gravitational field near a very compact object) but having an instantaneous interaction in a theory raises an even more serious causality issue.
 Instantaneity can only be valid in one Lorentz frame.
If the instantaneity frame is the restframe of the emitter, many emitters in 
motion relative to each others define the same number of different 
instantaneity frames (this is the case when we have many privileged frames, one per
 individual source) which makes it possible for a signal exchanged from A to B in 
relative motion, then back from B to A, to arrive in the past of A original 
emission. 

Of course, the issue would disappear in case of a unic frame of instantaneity whatever 
the emitters motion relative to this frame. Indeed A could at most instantaneously send a 
signal to himself via B, never in its proper past, this remaining 
valid in any other frame: $\Delta t_{AB}+\Delta t_{BA}\geq 0$. But a single cosmic 
preferred frame would have produced large and easily detectable LLI violating effects
 in many solar system tests (see our next section devoted to gravitomagnetism).  

Backward in time propagation of information in DG thus appears a real possibility and we shall 
take it serious. First, after all, we know from various tests of quantum non locality 
that there actually exists influences able to propagate faster than light 
(even instantaneously) even if as is usually believed, these cannot be used to transfer
 information. Secondly, propagation of informations from the future to the past is not a concern 
provided the whole story is coherent i.e. the events induced by such transfer of information 
will not affect the future events differently than what is already predetermined : indeed 
the information received from the future means that this future is already written and cannot 
be modified (Novikov self-consistency principle).
Actually, it is only when consciousness and in particular free will come into the game that 
the causality issue arises since it's difficult to see what would forbid this free will to 
act differently than what is already known from the written future. So this kind of issue is 
most probably mainly related to our fundamental ignorance of the deep nature of 
consciousness and may be also of what decides the wave function collapse in QM.
For instance the collapse may not be complete in some cases allowing the separation of two parallel worlds 
if a free will decided to modify the course of the events: instead of such modification 
a distinct parallel future would be generated. Therefore assuming many alternative timelines accessible or many parallel histories in a single time
(taking  serious  the  many-worlds  interpretation  of  quantum  mechanics  for  instance)  would  also  be  a  fascinating
possibility.

\subsection{WEP}
The Weak Equivalence Principle is obviously not menaced 
if once the field solution is established, matter and radiation have to follow its "geodesics"
as in GR since the actions $S_{SM}$ and $S_{\tilde{SM}}$ are the same as in GR except for 
the presence of the B=-A field we shall study later on.

\subsection{Energy of the Gravitational Field }

Most components of the energy momentum tensor $t^{\mu\nu}$ of the B=-1/A gravitational field, a Noether current
computed thanks to the global translational invariance of the action, trivially vanish for the B=-1/A solution since this field is time 
independent. The Lagrangian 
\[
L=\frac{-1}{16\pi G} (-2(\frac{A'}{A})^2 + 
3A^2(-\frac{\ddot{A}}{A}-\frac{1}{2}(\frac{\dot{A}}{A})^2)+ 
3\frac{1}{A^2}(\frac{\ddot{A}}{A}-\frac{3}{2}(\frac{\dot{A}}{A})^2))
\]
has second derivatives
so the expression of the conserved Noether current is a little bit less 
trivial than the usual one: 
\[
t_{\mu\nu}=
\frac{\partial{L}}{\partial(\partial^\mu A)} \partial_\nu A -\eta_{\mu\nu}L 
-\partial^\rho (\frac{\partial{L}}{\partial( \partial^\mu \partial^\rho A)}) \partial_\nu A
+ \frac{\partial{L}}{\partial( \partial^\mu \partial^\rho A) } \partial_\nu \partial^\rho A 
\]
We could obtain this simple expression exploiting the fact that our Lagrangian 
only depends on second derivatives $\partial^\mu \partial^\rho A$ of a single coordinate 
($\mu=\rho$). We find
\[
t_{00}=
-\frac{\partial{L}}{\partial \dot A} \dot A + L 
+\frac{\partial}{\partial t} (\frac{\partial{L}}{\partial \ddot A}) \dot A 
- \frac{\partial{L}}{\partial \ddot A } \ddot A 
\]
but all time derivatives vanish so 
 $t_{00}=L=\frac{1}{16\pi G}(2a'^2)$. 

\subsection{Rotating Frames}

An elementary $g_{\mu\nu}$ (B=-1/A) transforms as usual under a transformation from the privileged frame to any rotating frame.
In the former, we already interpreted the expression of the proper time:
\[
d\tau ^2=A(-dt^2+\frac{1}{A^2}d\sigma ^2)
\]
On a disk uniformly rotating at z=0 about any z-axis with angular velocity $\omega$ the motion of a point P at distance $\rho$ from the axis, allows
to define a rotating polar frame where the P polar coordinates $\rho$, $\theta$, 0, t   can be related to x,y,z,t through
\[
x=\rho cos(\omega t+\theta), y=\rho sin(\omega t+\theta), z=0
\]
In the rotating frame, the proper time now reads:
\[
d\tau ^2=-(A-\frac{\omega^2\rho^2}{A}) dt^2  +\frac{\rho^2d\theta^2}{A} +2\frac{\omega\rho^2}{A}dt d\theta    
\]

All special relativity effects are of course the same in GR and DG but the interpretation 
of these effects from the outside observer (with unrenormalized rods) point of view is 
worth describing in DG: For a clock at rest ($d\theta=0$) in this frame, $d\tau^2=-A (1-\frac{\omega^2\rho^2}{A^2}) dt^2$, and we see that the "speed of time" is modulated
 in a rotating frame by an extra factor $(1-\frac{\omega^2\rho^2}{A^2})^{-1/2}$ relative to its $A^{-1/2}$ behaviour in the frame we started from. 
For a light signal, $d\tau=0$ and the speed of light is given by solving the previous equation for dt to get 
$\frac{\rho d\theta}{dt}=A \frac{1-\frac{\omega^2\rho^2}{A^2}}{1\pm\frac{\omega\rho}{A}}$ with an extra factor 
$\frac{1-\frac{\omega^2\rho^2}{A^2}}{1\pm\frac{\omega\rho}{A}}$ relative to its $A$ behaviour in the
 frame we started from. The speed of light thus appears anisotropic depending on light running clockwise or counterclockwise in the rotating frame.
In GR, there is no such exterior background point of view and the anisotropic 
speed of light interpretation is not meaningful: the Special Relativity Sagnac effect 
is obtained because space and time intervals
 are deformed in the rotating frame. 

\subsection{Speed of Gravity}

Recently, Kopeikin suggested in Ref.~\refcite{Kop} that the time delay of light from a quasar as the light passed by the Jupiter planet 
could be used to measure the finite speed of gravity. However, the analysis of the light propagation in the Jupiter's rest frame
makes it obvious that the speed of gravity is irrelevant in this frame (Ref.~\refcite{Wil2}) so that such measurement represents no more than a test of LLI. 
Up to now there thus exists no evidence for the finite speed of the gravitational interaction.

\subsection{The total B=-1/A field and cosmology}

The total B=-1/A field obtained by combining all elementary such fields cannot drive 
a global evolution because in an homogeneous universe, compensations in such a combination result in Minkowski everywhere. 
Thus our universe would 
be completely static without the contribution of the B(t)=-A(t) cosmological background. 

This is very different from what we would have obtained solving the equation for the B=-1/A field for an extended infinite homogeneous distribution where we would have cutted out a spherical cavity and obtained Minkowski inside as in GR. In this case we would have been led to conclude that in an homogeneous universe the B=-1/A field effect is analogous to what we get in the Newtonian approximation of GR supplemented by the Birkhoff theorem: global expansion! But our investigation led us to interpret the B=-1/A field in a very different way: if we are right it can only be the elementary field we dealed with in this section and again the B=-1/A universe is static. 

In DG a void on our side or a concentration of matter on the conjugate side will both have a repelling gravitational effect on our side and attractive on the conjugate side so that 
such voids will appear to expand relative to their environment and reach a more perfect 
vacuum than in GR thanks to the conjugate structure that soon nonlinearly grows at 
the same time from the initial perturbations. Then, when this globally static picture is complemented by the effect of the field B(t)=-A(t) all
perturbation scales will be submitted to cosmological effects in exactly the same way 
(very different from what happens in GR). 

There is however an alternative that should not be neglected: if our observable universe appears homogeneous on large scales, it remains worth considering the possibility for it being just an island of matter in the universe, in which case this island is not anymore static but should
be contracting or expanding (dont forget antigravitational effects between the two sides of the universe) under the influence of its own gravity. Such effect would need to be combined with the truely global evolution driven by the B(t)=-A(t) field in order to investigate the effects on our cosmological probes!

\section{Russian Dolls Actions}

The background metric provides also the same straightforward procedure to combine the complete $[g^{B=-1/A}]_{\mu\nu }$ field
with $[g^{B=-A}]_{\mu\nu }$ resulting in the single $[g^{tot}]_{\mu\nu }$, the total field to which matter and radiation fields 
will actually couple in another action. 
\begin{equation}
[g^{tot}]_{\mu\nu }=\frac{1}{2}( [g^{B=-A}]_{\mu \rho }[g^{B=-1/A}]_{\nu \sigma } +
[g^{B=-A}]_{\nu \rho }[g^{B=-1/A}]_{\mu \sigma } ) \eta^{\rho \sigma} 
\end{equation}
The dynamics of our side SM fields does not take place in $S_{grav}$  but 
in $S_{SM}$ the familiar SM action 
describing how our dynamical radiation and matter fields 
propagate in the total $g^{tot}_{\mu\nu}$ field (exchanging kinetic 
and gravitational potential energy), interact with each other but also exchange energy 
with the $[g^{B=-A}]_{\mu\nu }$ field. Similarly, the dynamics of the conjugate SM fields 
takes place in another $S_{\tilde{SM}}$ action involving total $\tilde {g}_{\mu\nu}$ 
in place of total $g_{\mu\nu}$. These $S_{\tilde{SM}}$ and $S_{SM}$  are obtained by the usual covariantization procedure 
of the flat space time SM actions except
 that in these actions the $[g^{B=-1/A}]_{\mu\nu }$ field in turn is external 
and non dynamical, all its elementary components having already played their dynamics in 
their own separate standalone $S_{grav}$ actions. 
Therefore the bidirectional dialog between matter and B=-1/A gravitation does not take
 place in a single total action as in GR and as is now the case for the B=-A field
 but each direction is handled in an independent action for the B=-1/A field.  
In that sense, we have what we might call Russian Dolls Actions.

 Since the $S_{\tilde{SM}}$ and $S_{SM}$ actions involve the B=-A field, we expect from the equations derived from these actions
 the background B(t)=-A(t) solution and its B(r,t)=-A(r,t) perturbations 
exchanging energy and momentum with the SM and $\tilde{SM}$ dynamical sources. 
 As we shall show, these perturbative solutions are waves different from GR ones but as
 successful as the latter to describe the quantitative 
decay of the binary pulsar.

\section{B=-1/A, Gravitomagnetism}
\subsection{Theoretical predictions}
We already knew that for a theory involving only elementary microscopic gravific sources, eventually the phenomenology is the same in DG as in GR but if a B=-1/A field can be directly sourced by an extended distribution of matter i.e. without having to combine "more elementary fields" as we did, a new phenomenology is expected particularly in the gravitomagnetic sector of the theory that we want to explore now. Of course, it is still understood that each such B=-1/A field computed in a privileged frame associated with its extended source distribution, will also
eventually need to be combined with other B=-1/A fields from other extended sources as we did for elementary gravific sources.

We shall show later that DG naturally predicts the occurence of discontinuities of the background B=-A 
gravitational field allowing to quantitatively explain the Pioneer anomaly. We will also explain why such discontinuities are probably drifting in the solar system. These discontinuities are delimiting spatial volumes and it might be that each such volume defines the domain of a given extended source distribution hence the domain of validity of its associated preferred frame. In the following we will try to understand what kind of gravitomagnetic effects should be expected even in case we have several such domains and associated privileged frames e.g. one for the sun and one for the earth. 

In a chosen working PPN coordinate system moving at velocity $\vec w = \vec w_{PPN/PF}$ relative to a  
privileged frame, we can get the Lorentz transformed 
$g_{0i} $ metric element to Post-Newtonian order for any point mass source m in the domain of this preferred frame. 
We get 
\[
g_{0i} = -4 w_i \frac{m}{r}
\] 
so the gravitomagnetic metric elements completely ignore the actual motion of the individual sources belonging to the preferred frame domain, as if a serie of static pictures given by shots of what is inside the domain was the actual source. It's only the preferred frame speed relative to our chosen PPN system that matters and this in turn might be (recall that we are investigating such very speculative possibility just to try other predictions than in GR) the consequence of the B=-1/A form of the field being imposed by global space-time symmetry principles rather than isometries determined by the sources as we explained earlier.
The decomposition $-\vec w=\vec w_{PF/PPN}=\vec w_{Source/PPN} +\vec w_{PF/Source}$ will make it easier to compare DG and GR predictions. In GR $\vec w$ would have been the velocity of the PPN frame relative to the point mass source, $\vec w_{PPN/Source}$ hence $-\vec w = \vec w_{Source/PPN}$ while in DG the extra $\vec w_{PF/Source}$ preferred frame term will add to $g_{0i} $ contributions depending on the source velocity relative to the preferred frame where the field was diagonal and satisfied B=-1/A.  

Anyway, in this field and arbitrary chosen PPN frame, the precession of a gyroscope's spin axis $\vec S$ relative 
to distant stars
as the gyroscope orbits the earth at speed $v_{GPB}$, is given as in GR by ( Ref.~\refcite{Will} , p208):
\[
\frac{d\vec S}{d\tau }=\vec \Omega \times \vec S
, \vec \Omega =-\frac{1}{2} \vec \nabla \times \vec g +\frac{3}{2}\vec v_{GPB} \times\vec \nabla \left( {\frac{m}{r}} \right)
\]
the components of the 3-vector $\vec g$ being the $g_{0i}$.
Explicitely, 
\[
\vec \Omega = - 2 \vec w \times \vec \nabla \left( {\frac{m}{r}} \right) +\frac{3}{2}\vec v_{GPB} \times\vec \nabla \left( {\frac{m}{r}} \right)
\]
Replacing $-\vec w=\vec w_{Source/PPN} +\vec w_{PF/Source}$ in DG yields 
\[
\vec \Omega =\vec \Omega _{\mbox{geodetic}} +\vec \Omega _{\mbox{PF}} 
\]
where in addition to the same geodetic precession as expected in GR,
\[
\vec \Omega _{\mbox{geodetic}} = +  2 \vec w_{Source} \times \vec \nabla \left( {\frac{m}{r}} \right) +\frac{3}{2}\vec v_{GPB} \times \vec \nabla \left( {\frac{m}{r}} \right)
\]
\[
=\frac{3}{2}\vec v_{GPB/source} \times \vec \nabla \left( {\frac{m}{r}} \right) +\frac{7}{2} \vec w_{Source} \times \vec \nabla \left( {\frac{m}{r}} \right) 
\]
(which last anomalous term depending on the arbitrary PPN system may be dropped since it does not appear in the truly measurable quantity according Ref.~\refcite{Will}, p211), a preferred frame effect arises in 
\[
\vec \Omega _{\mbox{PF}} =-  2 \vec w_{Source/PF} \times \vec \nabla \left( {\frac{m}{r}} \right)
\]
Now for an extended body of mass M such as a planet fully within the domain of our privileged frame, we were of course free to replace for any elementary point mass m entering in the 
composition of this body, $w_{Source}$ by a single $w_{CM}$ in the decomposition we started from, where $w_{CM}$ stands for the planet center of mass speed and will factorize when summing over all point masses contributions.
Not only are we free to do so but we also must do so to eventually obtain after summing up,
 the well known GR geodetic term expected from such extended body:
 \[
\vec \Omega _{\mbox{geodetic}} = \frac{3}{2}\vec v_{GPB/CM} \times \vec \nabla \left( {\frac{M}{r_{CM}}} \right)  
\]
 But then we also obtain a preferred frame term only depending on the speed of 
 our body center of mass relative to the preferred frame:
 \[
\vec \Omega _{\mbox{PF}} = - 2 \vec w_{CM/PF} \times \vec \nabla \left( {\frac{M}{r_{CM}}} \right)
\]
Eventually, we get no  gravitomagnetic term depending 
on our body angular momentum  
at the contrary to the GR case where terms 
involving the actual speeds of the point sources relative to each others inside the body 
produce an angular momentum contribution to $\Omega$ (Ref.~\refcite{Will}, p104). Hence, we have a Preferred frame Post Newtonian effect that comes in place of the Lense-Thirring precession or ``the dragging of 
inertial frames'' usually interpreted as a genuine 
coupling in GR between the spins of the earth and gyroscope (By the way, the highly non trivial issue of deviation from geodesics for a spinning body in 
general relativity is also avoided if the body angular momentum does not source gravity). Therefore if there is such domain of a privileged frame encompassing our earth, its rotation can have no effect on a Gravity Probe B
 gyroscope axis which means no frame dragging. In place, various preferred frame effects can occur depending on how actually moves our planet Center of Mass relative to the preferred frame and these effects might even change in time, appear or disappear depending on the actual position of all the moving preferred frame domain frontiers defined by the discontinuities if these are indeed drifting in time.
\subsection{Preferred frame effects phenomenology}

It is instructive to compare our DG preferred frame effect to the preferred frame
 effect that arises in the Parameterized Post Newtonian formalism, a general framework making it easier to classify alternative theories of gravity according their predictions:
\[
\vec \Omega _{\mbox{PF}} =\frac{1}{4}\alpha _1 \vec w_{PF} \times \vec \nabla \left( 
{\frac{m}{r}} \right)
\]
Following Ref.~\refcite{Will}, p209, for an earth orbiting satellite, the dominant 
effect comes from the solar term (the source is the sun) leading to a periodic angular precession 
with a one year period, with amplitude:
\[
\delta \theta _{\mbox{PF}} \le 5.10^{-3 \prime \prime } \alpha _1 
\]
thus completely negligible according to the PPN formalism given the 
experimental limit $\alpha _1 <4.10^{-4}$.
Then DG just appears as a huge 
$\alpha _1=-8$ (it is also straightforward to show that $\alpha_{3}=-4$)
 theory so one should wonder why we still take serious the 
possibility of detecting these preferred frame effects with GPB and also in
 other classical tests of GR which up to know were able to constrain very 
strongly the $\alpha_i$ parameters. The answer is simply that we expect 
the preferred frames to depend on local changing conditions such as the position of 
various propagating discontinuities delimitating several domains with different
 B(t)=-A(t) evolutions. So we suspect that various preferred frames effects that did not show up before 
 might suddenly manifest themselves and be huge even in famous classical tests such as the secular change in perihelion 
position and geophysical tests in a very unexpected way for the experts. Some
might even have already been discovered, but considered too unlikely to be taken serious, or too much in contradiction with other precision test results, have been hidden under the carpet or
 classified by scientists working for military black programs.

For GPB we have two possible preferred frame effects:
 \[
\vec \Omega _{\mbox{sun}} = - 2 \vec w_{sun/PF1} \times \vec \nabla \left( {\frac{M_{sun}}{r_{sun}}} \right)
\]
and
\[
\vec \Omega _{\mbox{earth}} = - 2 \vec w_{earth/PF2} \times \vec \nabla \left( {\frac{M_{earth}}{r_{earth}}} \right)
\]
where we have taken into account the fact that the earth and the sun might not be in the same domain in which case there would be two preferred frames, PF1 and PF2 in our problem. So we have to investigate several cases: \begin{itemize}
\item
First the case where PF1=PF2 is attached to the center of our galaxy.
Then the main effect is the solar one
 with an amplitude $\delta \theta _{\mbox{PF}}$ approaching $0.02^{\prime \prime }$
 depending on the velocity of the sun relative to this frame. It would be even greater 
for a preferred frame attached to the CMB, if the CMB actually follows the global motion
of our present universe clusters (which might not be the case according to recent studies
 such as Ref.~\refcite{Kash}).
The prediction for a GPB gyro West Est (counted positive from West to Est) drift is illustrated over one 
year for the case of a preferred frame attached to the center of our galaxy in Figure 1, showing the 
dominant one year periodic drift and the two subdominant 1.6 hour (GPB orbital period) periodic ones. t=0 is the spring 
equinox time. The global quasi linear behaviour on a limited period of up to six months 
around t=0 might well mimic the GR frame-dragging but the ambiguity must disappear once
the full year data are considered. The mathematica plot does not allow to 
resolve the rapid 1.6 hour periodic oscillations and one should not pay 
attention to the seemingly chaotic behaviour of the curve but only to the 
smoothly varying upper and lower limits. On a short period of time, the rapid oscillation would even 
give a contribution on both North-South and Est-West directions slightly but significantly shifting the expected geodetic effect, and mimicking a frame-dragging effect but would disappear integrated over a complete year. 
\item

In case PF1=PF2 is the sun restframe, the previously dominant sun term obviously vanishes.
The earth term then produces a small periodic drift of the gyros with
a GPB orbit period modulated by a term depending on the velocity of the earth relative to the 
sun. 
\item
In case PF1=PF2 is the earth restframe, the sun yields a small effect comparable to the sun geodetic effect and the earth term vanishes.

\item
In case PF1 is the earth restframe and PF2 the sun restframe, there are no preferred frame effects. Since DG also predicts the same geodetic effect as GR, the only difference that remains between the two theories in this case is the absence of frame-dragging, implying that we should see no drift at all in the West Est direction.
\end{itemize}
Eventually, the preferred frame effects are only significative for frames 
associated with very large spatial domains such as our galaxy, large
scale structures or the universe as a whole.
But even in those cases we have only periodic oscillations, which effects cancel upon integration over an integer number of periods. 

\subsection{Experiments}

Lunar Laser Ranging did not highlight frame-dragging so far.
Although there were recent claims by Ciufolini that frame-dragging effects due to the rotating earth have been already evidenced at the 10 percent
 level of accuracy in an analysis exploiting the LAGEOS satellites data and indirectly in the binary pulsar PSR 1913+16, these are complicated 
analysis with undedicated data which sources of systematical effects may not be under total control so that it was more
 commonly believed that only the dedicated Gravity Probe B experiment measurement could hopefully give us the final answer. The frame dragging but also most possible preferred frame effects were a priori well reachable given the experimental accuracy 
($5.10^{-4 \prime \prime } /year)$ of the GPB experiment designed 
to measure for the first time gravitomagnetism isolated from other Post-Newtonian effects. 
However, GPB encountered unexpected systematical effects orders of magnitude larger than the GR predictions the experiment was designed to probe. So even after the final announcement by the GPB collaboration that the GR frame dragging was eventually discovered at five standard deviations from the null result, there is still place for doubt all the more since:\begin{itemize}
 \item The preferred
 frame effect of Figure 1 described above might also mimick the GR frame-dragging (it has almost the same amplitude) for a limited period of less than six monthes.

\item What was the probability for the GPB team to get, even on a single gyroscope, and even at an early stage of the analysis where not all systematical effects had been fully understood and corrected, the West Est drift presented in a poster conference in April 2007 (Ref.~\refcite{GPB1}) totally compatible with zero (at most 0.1 mArcsec/year) and 160 times less than the one GPB eventually published in Ref.~\refcite{GPB2} (that this zero drift was not putted by hand in the plot and could indeed be interpreted as zero frame dragging at this time was confirmed to me by email by F Everitt, physicist leader of the GPB team)? What kind of conspiracy could have produced such extremely unlikely compensation between huge systematical effects?
\item If at a given time a local discontinuity was encompassing the earth but not the sun, the local earth preferred frame did 
not allow any significant preferred frame effect to be detected as we explained above. If more recently, the discontinuity has drifted and reached an equipotential surrounding 
a much larger domain including the sun and the earth, we might witness the sudden appearance of associated preferred frame effects described above.  If the drift took place during 
the GPB data taking then GPB gyros could have registered the succession of the two regimes, and this would have left the experimentalists rather perplexed and awfully confused. 
\item The existence of discontinuities even allows to further speculate along the possibility that the gravitational field of an extended object such as a planet be obtained in its vicinity by combining elementary microscopic gravific masses fields resulting in exactly the same predictions as in GR : frame dragging and geodetic effect, while sufficiently far away from the planet to cross its surrounding discontinuity, the B=-1/A field considered in this section would kick in implying no more frame dragging outside. Then the GPB probe crossing the frontier defined by such drifting discontinuity also would have resulted in a succession of two regimes. 
\end{itemize}
\section{B=-A, the Cosmological Field}

A B=-A isotropic and spatially homogeneous gravitational field can only depend on
 time. But even if we did not impose the B=-A condition, the requirement that
both forms of the field be spatially homogeneous and isotropic enforces spatial flatness!
In the global preferred coordinate system where $\eta_{\mu\nu}=diag(-1,+1,+1,+1)$, we already derived the equation satisfied by the B=-A field in vacuum. Now, including the source terms that follow from the SM and $\tilde{SM}$ source actions extremization, for the single remaining degree of freedom A(t), we get :
\begin{equation}
3A\left( {-\frac{\ddot {A}}{A}+\frac{1}{2}\left( {\frac{\dot {A}}{A}} 
\right)^2} \right)-\frac{3}{A}\left( {\frac{\ddot {A}}{A}-\frac{3}{2}\left( 
{\frac{\dot {A}}{A}} \right)^2} \right)=n\pi G (A^2(\rho-3p)-\frac{1}{A^2}(\tilde{\rho}-3\tilde{p}))
\label{eqcosmo}
\end{equation}
The right hand side perfect fluid source term is just 
$\sqrt{g}T-\frac{1}{\sqrt{g}}T_{g \Rightarrow  1/g}$ obtained by the extremization of 
the usual SM and $\tilde{SM}$ actions where the fields are dynamical except the B=-1/A 
fields obtained in a previous section. 
The covariant 
energy momentum conservation equations deriving from these actions can be written:
$T^{\mu\nu}_{;\nu}=0$ and $\tilde{T}^{\mu\nu}_{;\nu}=0$.
They result in the evolution laws $\tilde{\rho} =3\tilde{p} \propto 1/\tilde{A}^2$ 
and $\rho = 3p \propto 1/A^2$ in the 
radiative eras while $\tilde{\rho}$ and $\tilde{p}$ vary
 as $\propto 1/\tilde{A}^{3/2}$ and $\rho$ and p as $1/A^{3/2}$ in the 
cold eras (p=$\tilde{p}$=0). 
Therefore, in the radiative eras (if the two sides could be hot at the same time which is not necessarily the case) the source terms vanish and the evolution of the universe 
is simply driven by its self interaction. 
 In the cold eras (if the two sides could be cold at the same time which is not necessarily the case) the source term varies as 
$(A^{1/2}-A^{-1/2}) \rho_{0}$ where $\rho_{0}$ is presumably the common positive 
density the two conjugate sides started from.  

The trivial a(t)=1 stationary solution describes a self 
conjugate world. A perturbation to this stationary solution,  
probably occured for the birth of times to take place and see a couple of 
conjugate universes start evolving from a(0)=1.
We are now looking for these less trivial evolution laws and solutions in some particular ranges for the scale factor a(t) (A(t)=$a^2(t)$). 
When $a(t)\approx 1$, the source vanishes to first order even in case we are in a cold universes scenario and vanishes exactly for hot universes. The solution in both cases is (from now on, non dimensional time units are used).
 \begin{equation}
\mbox{ }A\approx 1\Rightarrow \ddot {A} = \frac{\dot 
{A}^2}{A} \Rightarrow A\propto e^t \Rightarrow a\propto e^{t/2}  
\end{equation}
We notice that $a(t)\approx 1$ implies $t \approx 0$.

As long as both remain hot, the conjugate worlds have simple evolution laws in
the particular ranges $a(t)<<1,a(t)>>1$. Indeed, the scale factor evolution is then driven
by the following differential equations: 
\begin{equation}
 \mbox{ }a<<1 \Rightarrow \ddot {(1/a)}  = 0  \Rightarrow a \propto 1/t\mbox{ where }t<0 ,
\end{equation}
\begin{equation}
 \mbox{ }a>>1\Rightarrow \ddot {a} = 0  \Rightarrow a\propto t  \mbox{ where $t$ }>\mbox{ 0}  \\ 
\end{equation}
If one or both are evolving in a cold era, there is a dominant source term determined by the content of the side with greater scale factor. The differential equations read:
\begin{equation}
 \mbox{ }a<<1 \Rightarrow \ddot {(1/a)}  = \frac{ -n\pi G \rho_{0} }{6} 
\Rightarrow a \propto \frac{-12}{n\pi G \rho_{0}}\frac{1}{t^2} \mbox{ where }t<0 ,
\end{equation}
 \begin{equation}
 \mbox{ }a>>1 \Rightarrow \ddot {a} = \frac{ -n\pi G \rho_{0} }{6} 
\Rightarrow a \propto  \frac{-n\pi G \rho_{0}}{12} t^2 \mbox{ where $t$ }>\mbox{ 0} \\ 
\end{equation}

Of course we can check that $t \to -t$ implies 
$1/t^2 \to  t^2$, $1/t \to  t$ but also $e^t\to e^{-t}$ thus $A\to 1/A, B\to 1/B$ when t reverses for all 
our solutions i.e. remarkably, we have $A^{-1}(-t)=A(t)$. 
The conjugate background is therefore obtained by time reversal. Two opposite times joined each over at t=0. The existence of a time reversal conjugate universe 
was also suggested a long time ago in Ref.~\refcite{Sak}.
But here time reversal does not mean going backward in time anymore.  As shown in 
Figure 2,
 reversing time twice can never make you reappear in the past in a given side of the
 universe.
Not only our universe can be accelerated thanks to the $1/t^2$ 
solution (which translates into a $t^2$ evolution 
in standard comoving coordinates) without any need for a cosmological
constant or dark energy component, not only can it also decelerate in a standard model (with no cosmological constant) way thanks to the $t^2$ 
solution (which translates into a $t^{2/3}$ regime 
in standard comoving coordinates) but it is flat without inflation and gets rid of the big-bang singularity at t=0.
The $1/t$ and $t$ radiative eras solutions respectively translate into
 $e^{t}$ and $t^{1/2}$ evolutions in standard comoving coordinates while the exponential solution near t=0 translates into a "freely coasting" one, $a(t)\propto t$. 
 Strangely enough, cosmic data seem to favour a global transition as shown in Figure 3 (red square) to the new solid line evolution with reversed time arrow while the Pioneer anomaly suggests
 a recent local transition to the dashed line evolution (Figure 3).
 Indeed, if the two sides exchanged their scale factor regimes at the standard time $t_0$ of the so called "turn around redshift", so that on our side the cosmological field switched from 
 $a^2(t)$ to $\frac{a^4(t_0)}{a^2(t)}$
then, provided the arrow of time also reversed (that it started to flow backward is mandatory otherwise we would have measured contraction instead of expansion!!), we have in standard cosmological time coordinate the succession of two expanding regimes: $t^{2/3}$ then $t^2$ and this would perfectly mimick the LCDM story. This surprising scenario, even in the DG framework, requires very strong evidences for us to admit it, but since many cosmologists believe that the evidence for LCDM is already compelling, let's keep open minded... By the way, figure 3 also opens the way to conceive a cyclic evolution of our universe. 

We can also notice that the coordinate system where $\eta_{\mu\nu}=diag(-1,+1,+1,+1)$ defines two fundamental scales or constants of nature. The first one is the 
speed of light c=1 in this system. On the second one, $a(t=0)=1$, depends the speed of the absolute time marked by clocks of the background metric. Considered as fundamental constants both are for the time being arbitrary but measurable.

We expect no suppression of the background effects here as in GR in the vicinity of a  massive object. 
But we shall show in a forthcoming section why atoms and planetary systems typical sizes and periods are affected in the same 
way so that it is not possible to locally detect any effect 
of the cosmological background on planet trajectories from our reference rods and clocks point of view.
 Only the comparison of atomic periods with free photons periods, i.e. the redshift,  can allow us to probe the cosmological expansion anywhere. 

A striking and very uncommon feature is that for two sides of the universe evolving simultaneously in a radiative era, the evolution of the scale factor is driven
 by the gravitational 
interaction between the conjugate sides of the universe corresponding to the two forms independently
of their matter and radiation content.
 Also in the cold era, probing the flatness or the variations of the expansion rate of our universe 
should no longer allow us to draw any conclusion about its content except in case 
among the three possible solutions (expanding, contracting and static in the true time), one side does not choose a single one everywhere, but instead
is spatially divided into many domains delimited by discontinuities, with different evolution regimes of A(t) depending
 on local density perturbations. This discussion will
be reopened in a next section.  

\section{B=-A Perturbations: Gravitational Waves}
\subsection{The Perturbative Solutions}
The proper times read:
\[
d\tau^2={A(r,t)}\left[ {{d\sigma}^2-dt^2} \right]
\]
\[
d\tilde {\tau }^2= {A^{-1}(r,t)} \left[ {{d\sigma} ^2-dt^2} \right]
\] 

The expressions will be simpler if we adopt the dynamical variable h(r,t) defined by
 $A=e^{h(r,t)}$. In vaccum we get 
\[
\begin{array}{l}
\Rightarrow \Box h(r,t)=\frac{Tanh(h)}{2}(h'^2-\dot h^2) \\ 
\end{array}
\]
having massless plane wave solutions:
\[
h(r,t)=\sin (Et-\vec {p}\vec {r});\left| {\vec {p}} \right|=E
\]
The nonlinear term on the right hand side of the h(r,t) differential equation vanishes 
for any propagating h(r-t) or h(r+t) but not for a superposition of such functions. 
For instance, the superposition of an outgoing (retarded) and ingoing (advanced) spherical waves with the same frequencies is a standing wave producing
 a nonzero nonlinear term which can act as a source term. 
Indeed, though on large scales the averaged ${h}'^2-\dot {h}^2$ vanishes, on small space-time scales relative to the wavelength, 
the Zitterbewegund of ${h}'^2-\dot {h}^2  $ may have been the perturbation needed to start a non stationary evolution of the background, 
i.e. for the birth of a couple of time reversal conjugate universes which scale factors evolved as shown in the previous section.
Note that such nonlinear effects are also efficiently reduced by the Tanh(h) factor for small 
h. 
The basic small perturbation solution is thus a plane wave: 
\[
A(r,t)=e^{sin(Et-\vec p\vec r  )}
;
\mid \vec p\mid =E 
\]

\subsection{The Binary Pulsar Decay}
 We now demonstrate that these waves can be responsible for the decay 
of the binary pulsar. 
We follow Weinberg's computation of the 
power emitted per unit solid angle and adopt the same notations to obtain 
the same energy lost through our gravitational 
waves radiation by the binary pulsar as in GR in good agreement with the 
observed decay of the orbital period. 
The Lagrangian satisfies:
\[
-n\pi G L=\sqrt g R+\frac{1}{\sqrt g }R_{g\to 1/g} = 3Cosh(h)(h'^2-\dot h^2)+6(Sinh(h))(\triangle h-\ddot h) 
\]
The conserved Noether current being
\[
t_{\mu\nu}=
\frac{\partial{L}}{\partial(\partial^\mu h)} \partial_\nu h -\eta_{\mu\nu}L 
-\partial^\rho (\frac{\partial{L}}{\partial( \partial^\mu \partial^\rho h)}) \partial_\nu h 
+ \frac{\partial{L}}{\partial( \partial^\mu \partial^\rho h) } \partial_\nu \partial^\rho h 
\]
we obtain
\[
t_{00}= \frac{1}{-n\pi G}( 3Cosh(h)(h'^2-\dot h^2)+6Sinh(h)\triangle h)
\]
Thus for any propagating solution h(r-t) or h(r+t):
\[
t_{00} = \frac{1}{-n\pi G}( 6Sinh(h) \triangle h \approx 6h \triangle h =6h \ddot h )
\]
 the approximation being for weak h(r,t) fields.

The radial momentum component of our gravitational wave energy momentum 
tensor reads:
\[
t_{r0} = \frac{6}{-n\pi G}Cosh(h) {h}'\dot {h} \approx \frac{6}{-n\pi G}{h}'\dot {h}\]

We have used the genuine energy-momentum tensor derived from the Noether theorem which
 is not the pseudo-tensor of GR. It exploits the global Lorentz invariance of our flat spacetime theory.
  
So we have all in hands to compute the energy radiated by the small h(r,t) field from the binary pulsar.
The field is solution of 
6$\left(\triangle -\partial_{0}^{2}\right) h(r,t)=n\pi G \delta (r)\delta (t)$ when 
excited by a dirac impulse $\delta (r)\delta (t)$. 
Again, as for the B=-1/A case we want to determine n that insures the compatibility 
condition in the static case:  h(r)=2U(r) 
where U(r) would again be solution of 
$\triangle U(r) = 4\pi G \delta (r)$.  
This requirement is of course mandatory to insure that in both B=-A and B=-1/A sectors, the coupling between gravity and mass is the same! Thus now n equals 48, the energy carried by the field h(r,t) is positive  
and our Lagrangian for gravity in the B=-A sector was
\[
L= \frac{-1}{48\pi G} (\sqrt g R+\frac{1}{\sqrt g }R_{g\to 1/g}) 
\]

For any extended non relativistic source $\delta T^{\mu}_{\mu}$ (remember that our single 
equation only involved tensor traces) the solution is the "retarded potential":
\[
h(x,t)=2G \int {d^3{x}'\frac{ \delta T^{\mu}_{\mu} 
({x}',t-\left| {x-{x}'} \right|)}{\left| {x-{x}'} \right|}} 
\]
Replacing by the expression of our wave solution,
\[
 h=\sum\limits_{\omega ,k} {h\left( {\omega 
,k} \right)e^{i\left( {\omega t-kr} \right)}+h^\ast \left( {\omega ,k} 
\right)e^{-i\left( {\omega t-kr} \right)}} 
\]

\[
\left\langle {t_{r0} \left( {\omega ,k} \right)} \right\rangle 
=\frac{-1}{8\pi G}\left\langle {{h}'\dot {h}} \right\rangle _{\omega ,k} 
=\frac{1}{4\pi G}\omega ^2\left| {h(\omega ,k)} \right|^2
\]
allowing to get the total power emitted per unit solid angle
 in the direction of $k$ :
\[
\frac{dP}{d\Omega }(\omega ,k)=r^2\left\langle {t_{r0} (\omega ,k)} 
\right\rangle =r^2\frac{1}{4\pi G}\omega ^2\left| {h(\omega ,k)} \right|^2=
\frac{G}{\pi }\omega ^2\left|\delta T^{\mu}_{\mu}\right|  ^2 (\omega ,k) 
\]
Then following Weinberg, we use $T_{0i} (k,\omega) =-\hat{k}^jT_{ji}(k,\omega)$ and
$T_{00} (k,\omega) =\hat{k}^i\hat{k}^jT_{ji}(k,\omega)$ (where $\hat{k}$ is the unit vector 
in direction of the vector k), 
 assume that the source radius is much smaller than 
the wavelength 1/$\omega$ to
write $T_{ij}(k,\omega)\approx -\frac{\omega^2}{2} D_{ij}(\omega)$ to obtain
 in terms of the moment of inertia Fourier transforms $D_{ij}(\omega)$ :
\[
\begin{array}{l}
 \frac{dP}{d\Omega }(\omega ,k) =
\frac{G\omega ^6}{4\pi }( \hat{k}_i \hat{k}_j \hat{k}_l \hat{k}_m +\delta_{ij}
 \delta_{lm} -2\hat{k}_i \hat{k}_j \delta_{lm}) D_{ij}^* (\omega )D_{lm} (\omega ) 
\\ 
 \end{array}
\]
 The next step is to do the integral over solid angle
\[
 P=( \frac{1}{15}(\delta_{ij}
 \delta_{lm} +\delta_{il}
 \delta_{jm} + \delta_{im}
 \delta_{jl})+ \frac{1}{3}\delta_{lm}\delta_{ij}
)G\omega ^6D_{ij}^* (\omega )D_{lm} (\omega ) 
\]
$\delta_{ij}D_{ij}$ terms do not contribute for a rotating body and we get
\[
P=\frac{1}{15}G\omega ^6\left[ {2D_{ij}^* (\omega )D_{ij} (\omega )} \right]  = 
\frac{1}{15} G\omega ^6 8 D_{11}^2 (\omega ) \\ 
\]
For a rotating body with angular speed $\Omega$, equatorial ellipticity e, 
moment of inertia I in the rotating coordinates, $\omega =2\Omega$, 
$D_{11}(\omega )=\frac{eI}{4}$ and the radiated power reads:
\[
P=\frac{8}{15}64 G\Omega ^6e^2\frac{I^2}{16}=\frac{32}{15}G\Omega^6e^2I^2
\]
We find that an extra factor three eventually is needed to get the same lost total power 
3P as in General Relativity. We interpret this as a strong indication that the h fields 
exist in 3 kinds. Though it is not obvious at first sight which kind of  
symmetry is involved here, the analogy with the three mesons $\pi^+$,
$\pi^-$, $\pi^0$, is attractive. These are Lorentz scalars just as the field 
$\Phi$ we may have defined by $g_{\mu\nu}=\Phi \eta_{\mu\nu}$ and even though 
these are massive while DG fields are massless, we shall investigate later 
how our fields might acquire mass. 

For the B=-A wave perturbation we just computed, we of course assumed that the background metric is $\eta$ as we always did. One might wonder why now the B=-A field does not include the homogeneous cosmological solution B(t)=-A(t) we got in the previous section i.e why we didnt even try to find solutions for the total field about $\eta$ with conjugate elements:
\[
A_{tot}(r,t)= A(t)e^{h(r,t)} ; \tilde{A}_{tot}(r,t)= A^{-1}(t)e^{-h(r,t)} 
\]
The reason is that the source term is of course still the same that we wrote for the cosmological differential equation which reduces to the dominant term $A^{1/2}( \rho_{0}+ \delta\rho)$ in the cold eras in presence of a perturbation $\delta\rho$, since $A>>1$ in the present universe.
Thus keeping the dominant terms, the B=-A differential equation in this case is:
\[
\begin{array}{l}
 3A\left( 
 \frac{{A}''}{A}-\frac{1}{2}\left( \frac{{A}'}{A} 
\right)^2+2\frac{{A}'}{Ar} - \frac{\ddot 
{A}}{A}+\frac{1}{2}\left( \frac{\dot {A}}{A}  \right)^2 
\right) = A^{1/2}( \rho_{0}+ \delta\rho) \\ 
 \end{array}
\]
hence our perturbation $\delta\rho$ is damped by a huge $A^{1/2}$ cosmological factor as a source for gravitational waves.
Therefore the gravitational waves we dealed with when computing the decay of the binary pulsar are not at all perturbations of the cosmological field B(t)=-A(t). They are actually perturbations of the $\eta$ field which was also a trivial solution of our cosmological equations, a static self conjugate one ! 

But the needed factor three strongly suggests the existence of two additional independent simultaneously propagating solutions: the ones we could have computed in exactly the same way, assuming that not only $\eta$ can play the role of the background but also our conjugate cosmological solutions themselves for a couple of new fields (thus new genuine degrees of freedom) of the B=-A kind on top of them. For the time being lets only investigate a  perturbative solution for such new fields $h_A(r,t)$ and $h_{\tilde{A}}(r,t)$.
What we mean is that on the geometric side of the action, we have three independent fields : about non dynamical homogeneous $A(t)\eta$ we have the field and its conjugate:
\[
g_A=A(t)e^{h_A(r,t)}\eta ;  \tilde{g}_{A}=A(t)e^{-h_A(r,t)}\eta
\]
and we will solve for the dynamical $h_A(r,t)$ (we omit tensor indices for the sake of readability).
About non dynamical $\tilde{A}(t)\eta$ the conjugate fields are:
\[
g_{\tilde{A}}=\tilde{A}(t)e^{h_{\tilde{A}}(r,t)}\eta ; \tilde{g}_{\tilde{A}}=\tilde{A}(t)e^{-h_{\tilde{A}}(r,t)}\eta
\]
and we will solve for dynamical $h_{\tilde{A}}(r,t)$ and about  $\eta$ the conjugate fields are:
\[
g_1=e^{h(r,t)}\eta ;  \tilde{g_1}=e^{-h(r,t)}\eta
\]
Eventually, our understanding is that any distribution of matter can source the B=-1/A field
as we explained in previous sections but can also source independent $h(r,t)$, $h_A(r,t)$ and $h_{\tilde{A}}(r,t)$ perturbation waves. The total action for undamped gravitational waves would read:
\begin{equation}
I_{g} - \int  (\sqrt{g} T^{\mu\nu}_{SM}g_{\mu\nu} + \sqrt{\tilde{g}} \tilde{T}^{\mu\nu}_{\tilde{SM}}\tilde{g}_{\mu\nu}) d^4x  
\label{waveaction}
\end{equation}
where $g_{\mu\nu}$ here involves the three independently dynamical B=-A fields we need to compute: $g_1$, $g_A$ and $g_{\tilde{A}}$. The action must be extremized with respect to independent variations of these three fields to get their differential equations.  
Since $g_1$, $g_A$ and $g_{\tilde{A}}$ are combined together as prescribed in section 10.1 resulting in a quasi Minkowskian field, they will obviously satisfy the same equations with no damping factor hence will contribute to the same amount of the decay thus the needed factor 3.

\subsection{Phenomenological Implications}

Conjugate waves are in dotted line in Figure 4. For instance the red dotted line involves
\[
e^{-h_A(r,t)} = e^{h_A(-r,-t)} = e^{sin(-Et+\vec p\vec r  )} 
\]
which has a negative frequency but is still 
propagating on the same side of the B=-A cosmological field as $e^{h_A(r,t)}$ in undamped solid red line in Figure 4.
It also has a reversed momentum, so that the discrete symmetry involved 
here is space and time reversal (r,t to -r,-t).

It's important to understand how the DG framework allows us to overcome instability 
issues. There is no real instability issue for the damped waves because the negative energy side of the damped wave in blue dotted line is propagating in the conjugate side of the universe where it can only meet negative energy fields from our side point of view. 
The issue is only a priori serious for the negative energy side of fields, as the red dotted line wave, propagating on our side of the universe.
We have two possibilities to interpret the negative frequency conjugate waves for the undamped waves: the first possibility to consider is to deny any
 physical role to be played by this conjugate 
form because it is not another degree of freedom but just the other side, the negative energy side, of a regular object, the two sided (or Janus) red wave. From the point of view of the source mass which carries positive energy, it is only coupled to a positive energy $h$ or $h_A$ or $h_{\tilde{A}}$ respectively in black, red and blue plain lines for the undamped waves, the only degrees of freedom that remain after eliminating all the conjugate fields during the extremization procedure.
Of course, for the negative energy sides (dotted lines) of all those field any interaction with (coupling to) a positive energy side (plain line)
of another field or any other positive energy field should be forbidden i.e. interaction would only be allowed between for instance red waves which are both in plain lines or both in dotted lines.
However, may be any negative energy wave in dotted line can be associated with an annihilation operator of a positive energy state in second quantization and then be given the right to propagate and interact with any kind of positive energy waves. In this case, it may be interpreted as a negative energy going backward 
intime equivalent to a positive energy antiwave going forward intime hence an antiparticle following the Feynmann interpretation.

 We have considered here the interaction between  
matter fields on our side, here the matter of our free falling pulsar, and the h, $h_A$, $h_{\tilde{A}}$ fields which also carry the same sign of the energy.
Our understanding is that although these matter fields are confined to our side of the universe they were able to source gravitational waves also in the background $\eta$ and conjugate universe. This confirms that the coupling is indirect. This is not so disturbing  because any point mass sources a B=-1/A field which can be feeled on both sides of the universe so that when this mass is moving, for instance in rotation as the pulsar about its companion star, this B=-1/A field follows this motion of the mass on both sides (and probably so does a field discontinuity) 
so even though we have matter (the star) on one side only, we have the corresponding gravific mass on both sides (and also in $\eta$ in between) and it is actually these gravific masses that source the three B=-A field waves!
Hence we have the gravific shadow of the pulsar at the same time on the other universe side interacting with
the conjugate side $h_{\tilde{A}}$ field waves, both carrying again the same sign of the energy. 
Notice that this new understanding implies that the Standard Models matter and radiation fields contributing to $T^{\mu\nu}$ and $\tilde{T}^{\mu\nu}$ cant actually play the whole of their dynamics in this new action (\ref{waveaction}) for the B=-A wave fields because the source action in it does not specify the geodesics of which order two tensor fields among the three possible ones these are following. This means that in order to describe the free fall of Standard Model fields (and interactions with each others) in the total external and non dynamical metric fields they actually propagate in, the needed action was the one we used in the last section to get the equation (\ref{eqcosmo}) satisfied by the cosmological B(t)=-A(t) background (and damped gravitational waves). We could have written it explicitely:
\begin{equation}
I_{g} + \int  (\sqrt{g} L_{SM} + \sqrt{\tilde{g}} L_{\tilde{SM}}) d^4x  
\label{cosmoaction}
\end{equation}
where g stands for the dynamical cosmological field B(t)=-A(t) and its damped perturbations about $\eta$ and all SM and $\tilde{SM}$ fields are also dynamical.
This also means that the physics behind the action (\ref{waveaction}) we have eventually built for the B=-A undamped wave fields is similar to the physics we investigated behind the B=-1/A fields in that in both cases we only describe the generation of gravity from the sources and not how matter in turn falls in the gravitational field. There is a difference however: as we have seen there is a real exchange of energy and momentum between B=-A field waves and their sources which was not the case for the B=-1/A fields. 

The main difference as opposed to the GR case is that our gravitational wave 
is found to propagate pure scalar modes: $-g_{00} =g_{11} =g_{22} 
=g_{33} =e^h$.
The wave clearly affects all spatial directions in the same way so that no interferometer 
will be able to detect it. To see this we can for instance adopt the experimentalist 
proper time as our new time 
coordinate t' and in this new coordinate system the only non vanishing field perturbation elements
are $h_{11} = h_{22} =h_{33} $ depending on t' and the direction of propagation z.
The Riemann tensor elements we need to characterize the observational effects of our wave 
are $R_{0i0j}$ with only non vanishing elements in the small h approximation appropriate for 
a wave, $R_{0i0i}=\frac{1}{2}\ddot h_{ii}$. Hence we have the same effect on any arm of an 
interferometer whatever the propagation direction of the wave. 

The h field once quantized is expected to generate a new gravitational propagated interaction in addition to the Schwarzschild 
non propagated solution we obtained in the previous section. We cannot add its potential to the Schwarzschild one since this would severely 
conflict with observations except, may be, if the B=-A solution is quantized with another Planck constant $h_G$ much larger than the 
one used to quantize electromagnetic interactions. Indeed, it was pointed out by J.M Souriau that the planetary periods in the solar system
are Fibonacci multiples of an approximate 30 days period (see Ref. \refcite{sou}), which we interpret in the framework of our 
quantum gravity with huge gravity quanta exchanges (see last sections and Ref. \refcite{Ira} with references 
therein). 
Then postulating a suitable energy cutoff for the virtual gravitons (DG itself is a suitable 
framework to generate such a cutoff as we explain in a next section) at distances smaller than the corresponding threshold the force can be made negligible. Following this way of thinking an extra contribution to 
gravity might only arise beyond at least interstellar distances (this was also postulated 
in Ref.~\refcite{Dru}) since no deviation from 
Newtonian gravity was observed at smaller distances. At smaller distances the extra 
interaction would thus exhibit a kind of asymptotic freedom. Notice that the cutoff would only 
apply to virtual gravitons, not the real ones participating in the binary pulsar gravitational
 radiation. 

An even simpler alternative proposal becomes possible in case the fields are understood to acquire 
mass and we may even identify our field to be that of pion like particles. Indeed, in this 
case the Yukawa potential $e^{-mr}/r$ makes the interaction a very short distance one. Then, because 
 its coupling constant is G, its potential 
is completely negligible as compared to the nuclear interaction ones taking place at the same
scales. On the other hand, adding a mass term is possible without introducing
 any new dynamical degree of freedom as is unavoidable when one adds massive gravitons 
in GR so our result for the power emitted through gravitational waves radiation should not 
change.
Following this way of thinking it is also tempting to speculate about the validity of our wave 
equations at the nuclear scale with another coupling constant much higher than G to account for the Yukawa 
interaction though nowadays the latter is rather understood to be an effective description,
 QCD being the accepted fundamental theory for the strong interactions.

\section{Gravity and the Quantum: Strong Theoretical Motivations for DG Field Discontinuities}

    Genuine discontinuities are completely banned from modern physics. Indeed the derivation of all our fundamental interactions differential equations and conservation laws can only follow from the postulated actions invariance under various fundamental symmetries provided there are no field discontinuities. For example, the absence of discontinuities belong to the set of mandatory conditions for the Noether Theorem to be valid so even the local conservation of energy and momentum is not in principle granted anywhere we would encounter a field discontinuity. However, there are strong clues that at a fundamental level field discontinuities should be taken very serious. We know that an extremely enigmatic process, discontinuous and non local, the collapse of the wave-function, is one of the fundamental postulates of Quantum Mechanics and all modern physics of course must respect the rules of QM just because Nature was found to behave according to these rules (even the non-local essence of the collapse is now very firmly established by many beautiful Quantum Optics experiments). It is important to realize that QM describes physical phenomena by two very distinct sets of rules. Let us stress this.
    
	The first set of rules drives the continuous space-time evolution of various field solutions of the fundamental differential propagation equations of the fields (Klein-Gordon, Dirac ...) which can also be understood as local conservation equations and can also describe the interactions between all the fundamental fields once various local Gauge Symmetries are demanded. 
    
	The second set of rules was completely unexpected and very disturbing because it seemed to incredibly ignore all the beloved principles underlying the first set: these are the strange projection that mathematically describes the QM collapse and the Planck-Einstein relations, E = h.f, both completely unfamiliar to all the rest of physics. Both are  discontinuous and non local in essence!  Many physicists were indeed soon very dissatisfied with QM and Einstein himself believed that a more fundamental theory were to be found, also because QM is fundamentally undeterministic. This hope seems to be now completely given up just because we now know for sure that any such more fundamental theory underlying QM, a so called hidden variable theory, would have to be explicitely non local and most physicists prefer QM as it is (a set of rules that should not too much be taken serious thus a positivist interpretation rather than a realistic one) rather than trying to build a new framework with a set of explicitely non-local and discontinuous rules and principles drastically different from everything else we were used to think about seriously when constructing classical theories. Yet my conviction is that discontinuous fields and non local interactions are absolutely mandatory if one would really want to elucidate the origin of the as well discontinuous and non local rules of QM (and hopefully compute the value of the Planck constant h from more fundamental space-time parameters). So, from this point of view, it should actually be considered as an incredible advantage to have a new theoretical framework in which discontinuities are natural and necessary and not a drawback as most theoretical physicist would think nowadays. 
    
	The initial motivation for a theory such as Dark Gravity was not at all to stage a priori shocking new rules such as non local interactions and field discontinuities but the very constraining principles of the theory led to it and eventually this is unhoped-for. At a more fundamental level the two sets of rules we found in QM, one continuous and local and the other discontinuous and non local will hopefully emerge from the structure of the DG theory which admits both a sector of usual propagated interactions but also field discontinuities and a non local sector for gravity. This makes it a «dream» theory not only to unify QM and gravity but more importantly to really explain where the QM strange discrete and non local rules come from and derive them.
    
	Why are field discontinuities naturally expected in DG ? Just because the theory follows from a new treatment and understanding of space-time discrete symmetries, and therefore its fundamental equations admit two time reversal conjugate solutions to describe the background (cosmological type solutions) for instance.  Now because time reversal can occur a priori anywhere, there is no reason why a single of these two solutions should be valid everywhere on our side of the universe i.e there is no reason why the solution on our side should be $a(t)$ (or  $a^{-1}(t)$) everywhere and should be $a^{-1}(t)$ (resp $a(t)$) everywhere on the conjugate side. Instead, we naturally expect the universe to be divided in spatial zones where different solutions were chosen. For instance there might be an expanding solution on our side in the solar system (and the conjugate contracting solution on the conjugate side) replaced by a contracting solution outside the solar system (and the conjugate expanding solution on the conjugate side). Of course this implies a genuine discontinuity of the background field at the frontier between the two zones at which a genuine time reversal occurs and the conjugate solutions are exchanged. 

\section{ The Pioneer Effect}
  \subsection{The Pioneer Effect: Strong Observational Motivation for DG Field Discontinuities}
The so called Pioneer anomaly arose as an anomalous drift in time of the radiowave frequency received from both 
Pioneer 10 and 11 as compared to reference clocks on earth (Ref. \refcite{And}). The anomalous drift was found constant over more than 10 years and 
from 40 to 60 AU 
\[
\begin{array}{l}
 \frac{\dot{f}}{f}=(5.6\pm 0.9) 10^{-18}/s\\ 
 \end{array}
\]
including in the error all systematical sources of uncertainties.
Since all possible systematical origins investigated by Anderson and collaborators (Ref. \refcite{And}) were excluded or found very unlikely to account for 
the Pioneer anomaly, there have been increasing evidence that it was of fundamental physics origin until recent years when we have seen the efforts of several research teams attempting through detailed simulations and publications to convince us that the Pioneer anomaly was merely due to an asymmetrical radiation of the spacecrafts. I will adress these allegations at the end of this section to show why these appear groundless to me. 

The Pioneer anomaly is often accompanied by irritated disbelief of gravity experts which is not surprising given that for GR the Pioneer anomaly is a fatal one. Indeed, not only the sign of the shift is opposite to what we would have expected from 
a background effect but moreover the best effort in GR to melt together a background metric with a Schwarzschild metric is the McVittie metric 
(see Ref. \refcite{Giu} and references therein):

\begin{equation}
 d\tau^2=-\frac{1-\frac{Gm}{2ra(t)}}{1+\frac{Gm}{2ra(t)}}dt^2+({1+\frac{Gm}{2ra(t)}})^4a^2(t)d\sigma^2
\end{equation}
where a(t) can be absorbed by a coordinate transformation $r a(t)\rightarrow r $ so that 
the background is completely suppressed except in the space-time $g_{ti}$ elements where its effect
is at a much lower level than in the reported anomaly magnitude. In the combined
DG solution, there is superposition without suppression:
\begin{equation}
 d\tau^2=-e^{-\frac{2Gm}{r}} a^2(t)dt^2 + e^{\frac{2Gm}{r}}a^{2}(t)d\sigma^2                     
\end{equation}
While in such A(t)=-B(t) background, photons keep unaffected, the reference atomic periods contract or expand 
 resulting
 in an apparent cosmological red or blue shift.
Therefore the Pioneer photons periods should also appear shifted with respect to atomic clock references due to the B=-A background but 
due to their very small time of flight, the effect is very tiny. 
At best it would result in the good sign uniformly changing shift but with a rate too low 
by a factor $\frac{v}{c}$ to account for the reported anomaly showing 
up from the Pioneer spacecrafts data. However we are well motivated to keep searching for a Background effect solution because  
of the strange coincidence
$ \frac{\dot {f}}{f}=2\frac{1}{11.10^9\mbox{years}} \sim 2\frac{\dot {a}}{a}\;\equiv 2H_0$
at a roughly 15 percent confidence level coming from both the uncertainty in the directly measured Hubble parameter $H_{0}$ and in the effect.

Several authors have argued that, because of its weakness, the Pioneer 
anomaly seen as such a time dependent metric effect (due to the lack of any fundamental theory to account for it
 the issue is always examined from a phenomenological point of view), a clock acceleration rather than a mere
 acceleration, would actually entail no contradiction with the four classical tests of General Relativity 
since the trajectories are unperturbed (Ref. \refcite{Pio1}, \refcite{Pio2} and references therein). 
 At the contrary, all theories appealing to extra accelerations to produce a Doppler effect in order to explain the Pioneer anomaly 
are in serious conflict with recent tests of the outest planets trajectories in the solar system. An exception may be a modification of 
the r dependence of space-space metric elements having a very little influence on almost circular planet trajectories but accelerating 
significantly radial Pioneer like trajectories as proposed in Ref. \refcite{Rey} (see also Ref. \refcite{Gef}). But this is a very ad-hoc 
and fine-tuned proposal.     

For significant effects to appear in DG we are led to consider the case where the emitter and receiver are not in the same background which
happens if somewhere between the earth and the spacecraft there is a discontinuity of the background. 
Of course, light is trivially not affected by conformal metrics ($a(t)$ or $a^{-1}(t)$ have no effect when $d\tau$ =  0) therefore light wave lengthes and its propagation are a fortiori not sensitive in any possible ways to the presence of background discontinuities.
Then if we imagine two identical clocks exchanging light signals from both sides of the border line where we have a discontinuity, they can compare the speed of time in the two zones, one zone where the background metric field element is $a(t)$ and the other where it is $a^{-1}(t)$. The frequency shift one clock will see comparing the frequency of the other clock with its own  can be computed easily because on one side:
\begin{equation}
d\tau^2= \frac{1}{a^2(t)}(dt^2-d\sigma^2) 
\label{earth}
\end{equation}
which yields for a clock at rest ($d\sigma^2 = 0$) there (suppose on earth): 
\begin{equation}
dt_{Earthclock} = a(t)d\tau
\end{equation}
while for a clock at rest on the other side (suppose Pioneer is there):    
\begin{equation}
d\tau^2= a^{2}(t)(dt^2-d\sigma^2)  
\label{pioneer}
\end{equation}
yielding
\begin{equation}
dt_{Pioneerclock} = \frac{d\tau}{a(t)}
\end{equation}
This obviously implies that the Pioneer clock frequency will drift in time as compared to our earth clock. But then shouldn't the period of the Pioneer clock have been suddenly rescaled by a huge $a^2(t)$ factor when the spacecraft crossed the discontinuity. Not necessarily if time reversal and exchange of the conjugate fields only occured at a recent time $t_0$ so that in the Pioneer zone the background field started an evolution of the kind 
\begin{equation}
d\tau^2= \frac{a^{2}(t)}{a^{4}(t_0)}(dt^2-d\sigma^2) 
\label{pioneerone}
\end{equation}
rather than (\ref{pioneer}) while on earth (\ref{earth}) remained valid.
Then for our Pioneer and Earth clocks in two zones exchanging the roles of metric conjugate solutions (\ref{earth}) and (\ref{pioneerone}), from (\ref{earth})      $dt_{Earthclock} = a(t) d\tau $	
and from (\ref{pioneerone}), $dt_{Pioneerclock} = \frac{a^2(t_0)}{a(t)}d\tau$.

Fortunately, the Pioneer effect is instructive: it tells us that clocks periods are not instantaneously rescaled by a huge $a^2(t)$ scale factor that would have followed from (\ref{earth}) and (\ref{pioneer})  when crossing a discontinuity but rather the much smaller rescaling $dt_{Earthclock} = dt_{Pioneerclock}  \frac{a^2(t)}{a^2(t_0)}$ following from (\ref{earth}) and (\ref{pioneerone}) which absolute effect, $f_{Pioneer} = f_{earth} \frac{a^2(t)}{ a^2(t_0)}$ may be just implied a very small correction to the probe speed hardly distinguishable and identifiable to an anomaly given those probably much larger implied by the thrusters during manoeuvers, at the contrary to the continuous frequency drift in time, genuine acceleration of Pioneer clocks relative to our earth clocks, the anomalous: 
\begin{equation} 
\frac{\dot {f} }{f } = 2H_0 = 4.8. 10^{-18} s^{-1}
\end{equation} 
where we have used the expression $H_0=\frac{\dot{a}}{a}$ for the Hubble parameter in conformal coordinates which is easy to check. This result is remarkably compatible with the one that was measured,$\frac{\dot{f}}{f}=(5.6\pm 0.9) 10^{-18}/s$, when analyzing the Pioneer spacecraft radiowaves. Therefore, we are tempted to conclude that there must have been a discrete jump from $\frac{a(t)}{a(t_0)}$ to $\frac{a(t_0)}{a(t)}$ in the background field between us and Pioneer so that the effect could only start to be seen after the crossing of this frontier by the spacecraft. Within the error bars the jump (see the steep rise up of the effect in \refcite{And}) could not have been better evidenced than it was around 15 AU in 1983 by Pioneer 11. This extraordinary evidence, the perfect expected signature for a background discontinuity is the fact that convinced me that such discontinuities, a priori naturally expected in DG are actually real and have observational consequences. Indeed, up to now nothing else appart from this kind of very particular shift in time can account for the Pioneer anomaly without conflicting with many other precision tests of gravity in the solar system and the main possible systematical effect, an anisotropic radiation from the spacecraft, can only account for a small fraction of the effect as i will argue later. 

 \subsection{Drifting clocks? Really? }
 
The data acquisition system of the Pioneer probes is detailed in Ref \refcite{And} page 8. We learn that: "To ensure that the reception signal does not interfere with the transmission, the spacecraft has a turnaround transponder with a ratio of 240/221. The spacecraft transmitter local oscillator is phase locked (Phase Lock Loop) to the up-link carrier. It multiplies the received carrier frequency by the above ratio and then retransmits the signal to earth."
The PLL data acquisition mode clearly rules out our interpretation of the Pioneer anomaly and explains why the drifting of the Pioneer spacecrafts clocks (at the contrary to ground clocks) was never considered nor investigated anywhere else. Indeed, the PLL method insures that the downlink signal could not be sensitive to any drifting of on board clocks because these are slave clocks fully synchronized to the up-link frequency: typically such a PLL has an oscillator working at 240 times the up-link frequency and this oscillator signal frequency is digitally divided by 240 in order to be continually compared to the uplink one, any deviation being counter-reacted to keep the oscillator output locked on 240 times the up link signal frequency. This frequency is afterwards divided by 221, which again must be done digitally with a digital counter able to select only one cycle every 221 to generate the downlink carrier. 

The analog frequency multiplier is may be an interesting alternative solution to consider because an analog V=240/221 might have hopefully introduced the required background drifting factor $a(t)$ according to the necessary ElectroMagnetic extension of DG developped in the following main section. But we unfortunately must give up this possibility because an analog frequency multiplication by V would have made the result sensitive to any possible drift in time of this analog signal V and 
therefore would not have been tolerated given the required precisions for the probes to complete their program i.e. explore the gravitational environment in the solar system up to a limit never before possible (we are reminded in Ref \refcite{And} that the Pioneers were mainly designed to be high accuracy celestial mechanics experiments). 

Eventually, if the probes had been at rest with respect to the earth resulting in no Doppler effect, the comparison of the received downlink carrier frequency at the DSN (on earth) to the current uplink transmitter frequency which was performed there by the Doppler extractor, would merely have been, in the closed loop mode, a comparison of the emitters frequencies at a round trip delay interval which was about 3 hours and a half when the anomaly started to manifest itself (when the probes were at 15 AU from the earth) and has reached beyond 20 hours in the latest years of the mission when the anomaly was still observed.    
This also ruins any hope to explain the anomaly as the result of a drift in time of the clock on earth at the time of the reception relative to the clock on earth at the time of the emission. Indeed, the anomaly was also there for instance at the time when the round trip delay was about 8 hours, and during this period, even though the reception and emission of a given radiowave most of the time could not be performed by the same station on earth 
(there are three ground DSN complexes respectively near Madrid/Spain, Goldstone/California, Canberra/Australia separated by approximately 8 time zones), 
we could not have the Australia station clock accelerated by an $a(t)$ factor with respect to the Spain one, itself accelerated by the same factor 
with respect to the California one, itself accelerated by the same factor 
with respect to the Australia one (again but at one complete day interval)...
This is obviously because our theoretical understanding of the drifting of clocks in the background does not allow arbitrary integer powers $a^n(t)$ factors to apply between clocks!
\subsection{The case for a fraud}
Eventually our theoretical proposal to ideally explain the manifested anomaly could only work in case the anomaly was detected in data acquired in the open loop mode (such data was also recorded on tape as we learn from page 9 in Ref \refcite{And}) and not in the closed-loop mode tracked by phase lock loop hardware. Indeed, in the open mode, the local on board Pioneer oscillator frequency is used on the down-link and it is free to drift as implied by the background a(t) driving factor. However, still according to Ref \refcite{And}, the analysis that revealed the Pioneer anomaly was performed on closed-loop mode data exclusively so is this a total dead end ?
At this level of our investigation several serious questions remain to be answered:

\begin{itemize}
\item From our discussion above we realize that the open loop mode is the only one to be sensitive to effects such as those that unavoidably arise when not all clocks are submitted to the same background or not in the same way. Such question is one of the most important ones in our quest for a satisfying understanding of gravity and in particular it is a well known open question even in GR. So is it really conceivable that the theorists and gravity experts who designed the Pioneer experiment totally missed the interest of being able to perform tests in the open loop mode? Difficult to believe!
Anyway, the experimentalists would not have missed such effect even if it was there in the open loop mode data only.
\item If theorists were aware about the incredible fundamental interest of tests in the open mode, were these actually feasable in practice given the poor intrinsic stability of free quartz oscillators eligible on board of the Pioneers? Indeed, we are reminded many times in Ref \refcite{And} (in pages 30 and 31 for instance) that what made the program (accuracy tests of gravity in the solar system) possible was among other features (such as the spin stabilization of the spacecrafts), the impressive performances of the hydrogen Masers at each station to which all clocks were synchronized, insuring Allan variances of the order of $10^{-12}$ for a 1000s Doppler integration time for the S-band frequencies of the Pioneers. Given that within a year one can have thousands independent single measurements with duration 1000 seconds, the above stability of the Pioneer S band frequencies allowed to estimate in Ref \refcite{And} page 31 the corresponding contribution to the systematical error, a contribution 3000 times less than the reported total systematical error on the anomaly. This means that the Pioneer anomaly could have been evidenced as well with frequencies 3000 times less stable than the ones discussed in Ref \refcite{And} i.e. a free oscillator frequency multiplied to 2.29 GHz in the spacecraft with stability better than a 9 Hertz deviation (4 parts per $10^9$) for a 1000 seconds integration time would have successfully done the job. This represents a stability of better than $10^{-4}$ per year which for a quartz oscillator was not at all unrealistic: the quartz aging amounts to 0.5 ppm per year and the required stability over 1000 s implies temperature variations controlled to less than one degree Celcius according to Ref \refcite{quartz}. It's essentially the extremely stable environment of deep space (chemically neutral, negligible pressure, humidity, vibrations and temperature variations as well as free fall) that would have insured the as well extreme stability of the quartz at relatively low efforts. We also noticed that some of the most important discoveries that improved Crystal Quartz oscillators performances were published in 1974 (SC cut) and 1976 (BVA) so just a few years after the Pioneers launch while such or equivalent technologies were certainly developped many years in advance in military US labs since the military issue of the race to high performance clocks is obvious.

Thus eventually not only the open loop test was extremely attractive from a theoretical point of view and well within the technical possibilities in the early seventies but it could also have been used in alternance of short runs with the closed loop mode.
\item
An important fundamental discovery potentially means a technological military supremacy in the short term. No doubt that the Pioneer anomaly as we understand it could have been such a discovery as it was requiring a complete refunding of the theory of gravity. What was the probability during the cold war for such a discovery to be shared with other countries? I think vanishingly small! 
\item
The case for a scientific fraud is almost perfect, necessarily a fraud because to convert open loop data for which the Doppler effect is one way into closed loop data for which the Doppler effect is two ways one had to introduce by hand in the ODFILE shared with independent scientific teams, half of the Doppler effect carefully taking into account motions of the DSN at emission, which apparantly was not performed perfectly given two additional annual and diurnal anomalies also found in the data and discussed in Ref \refcite{And} page 41 (most probably due to errors in the navigation programs' determinations of the direction of the spacecraft's orbital inclination to the ecliptic according to the authors)!
\end{itemize}
But let's keep open minded to a less compromising possibility: if some malfunction of the closed loop mode was soon discovered after Pioneer 10' launch (yet we are being told that the harware on board of Pioneer 11 was the same at launch one year later), the engineers in fault (a serious fault since it was potentially jeopardizing all the scientific program) might have discreetly converted the open loop mode into closed loop mode data, just not to be charged for that...

 \subsection{Can an asymmetrical radiation explain the Pioneer anomaly? }

The very simple reasons why an asymmetrical radiation from the spacecrafts cant alone explain the Pioneer anomaly were repeatedly detailed in almost all Anderson's publications. Even in the latest one (Ref \refcite{And2}), he does not adhere to the view published almost at the same time by his colleague Turyshev from the same lab who claims that according to recent thermal simulations an asymmetrical radiation can account for the anomaly. Not only such simulations are intrinsically obscure, but the results published by independent teams are conflicting : for instance in Ref. \refcite{Bert} it is found that the asymmetrical radiation contribution to the anomaly can only be significant provided there were an aperture anomaly of the louvers while nothing such is clearly stated in the recent Turyshev (Ref. \refcite{Turyshev}) publication. Also the fraction of the anomaly explained by 
asymmetrical reflexion off the spacecraft's body and antenna of the IR light radiated by the RTGs is much less in Ref. \refcite{Bert} (as expected) than according to Ref. \refcite{Turyshev}.

At last, it's not difficult to exploit the temperature maps published in Ref \refcite{Turyshev} to compute the radiated powers from all faces knowing their emissivities and to check in this way and as i did in Ref \refcite{fhcasym}, that their final results dont add-up. Indeed, one easily finds that the alleged contribution to the anomaly due to the spacecraft main body asymmetrical radiation is only tenable provided it's the totality of the power radiated from the back face projected onto the z direction, that was reflected from the antenna (no absorption nor radiation in free space) and back did not intercept again the main body!
 
\section{Effects of discontinuities}
\subsection{Effects on solar system outer planets orbital periods}

The Pioneer anomaly in our framework is obviously a Local Position Invariance ("position in time") 
violation so the simple superposition of the background and perturbation readable in DG's 
solution might be in conflict with the usual tests of LPI in time constraining variations of 
the gravitational constant G at a much lower level than $H_0$.
These tests were performed in the inner part of the solar system. We shall reconsider this issue later after a careful study of the 
behaviour of an atom in the DG background field to prove that 
if planet trajectories and our local reference atomic rods and clocks are in the same background there is no detectable effect 
at all, and therefore no conflict with constraints on the time variation of G.
We are mainly concerned here with the effect on the outer solar system trajectories with respect to our local clocks if these  
are not in the same background.
We find that the expected apparant effect of reduction of the outer planet period is too small to be detected since the outer planet 
trajectories were not monitored by accurate Doppler 
or Laser Ranging methods but simply by following their visual position in the sky. By far, the best precision on the pericenter
 precession according to Ref. \refcite{Io2} were a 0.036 Arcsec/century for Jupiter to be compared with the 0.0024 Arcsec/(10years) 
expected in DG from 1997 to 2007 if, as we can suspect, there was a single discontinuity of the background between the 
earth and Jupiter during that period implying that the Jupiter orbit and our Earth clocks were not in the same background.

\subsection{Light and discontinuities}
Remind that photons are not sensitive to an homogeneous background B=-A field wether continuous or not, except the possibility that might be given to them to jump 
to the conjugate side of the universe thanks to a discontinuity understood as a genuine switch. Then their wave lengthes should be shifted when jumping from one side of the universe to the other by the potential difference $2GM/rc^2$ implied between the emitter side where the feeled gravitational potential is $-GM/rc^2$ and the receiver side where the feeled gravitational potential is the opposite $GM/rc^2$. Hence in this case it is rather a discontinuity of the feeled B=-1/A field that would lead the game. We suspect that such gravitational wavelength shift of CMB photons transferred from one side to the conjugate one through discontinuities might have contributed to or might even have been responsible for most CMB fluctuations since the order of magnitude expected for such shifts corresponds to typical temperature fluctuations in the CMB : a few $10^{-5}$. Indeed, the dominant potentials are typically those from galaxy clusters and the gravitational redshifting effect of such potentials was
even recently measured to be in good agreement with GR expectations for nearby clusters (Ref. \refcite{clustershift}). The order of magnitude of the adimentional potential 
$GM/rc^2$ is generally a few $10^{-5}$ which is not a surprise since we expect 
$GM/rc^2 \approx (v/c)^2$ and we know the typical galaxy velocities relative to the CMB to be 1000 km/s.  
\subsection{Matter and discontinuities}
For matter, a discontinuity represents a potential barrier
 that might produce caustic like effects when matter is trapped along the discontinuity forming matter rings or shells. 
The seemingly time dependent flux of Ultra High Energy cosmic rays as a consequence 
of a discontinuity drifting in time in the vicinity of the Earth is also an interesting possibility if some of these particles originate somehow
 at the discontinuity. 

\subsection{Cosmological implications}

The Pioneer effect also tells us that because all clocks in the universe did not have exactly the same history, this could eventually lead to huge redshift anomalies: frequency deviations even between identical clocks relatively closed to each other i.e. not at cosmological distances from each other. Because such anomalies were not observed we must conclude that on the long term, particularly on cosmological times, all clocks have experienced almost the same relative elapsed time in regime $a(t)$ and $a^{-1}(t)$ on average. This in turn is only possible provided discontinuities such as the one responsible for the Pioneer effect are themselves moving, drifting everywhere probably cyclically. 

According to this new understanding, the effect of the background on local rods might have reversed many times and even periodically so that integrated along the total duration of a cosmological path, its impact is not trivial. In other words, photons emitted from distant SNs might have propagated with their periods relative to local clocks alternating between a $t^2$ 
regime and $1/t^2$ regime. The amount of time spent in the two different regimes should then determine
 the long term behaviour and the expected Hubble diagram parameters.
 \subsection{Discontinuities and LENR Phenomena}
 We expect instantaneous boosts of massive particles crossing the potential barrier implied by $a(t)/a(t_0)$ and searched for such impressive signatures in the so-called Low Energy Nuclear Reactions. This fascinating phenomenology is explored at the end of our article Ref. \refcite{DGneut}.
\subsection{The Pseudo Black Hole Horizon}

Our exponential solution tells us that there is no more BH horizon in DG.
However a test mass approaching the Schwarzschild radius of a massive body is propagating in the background 
and perturbation superposition:
\begin{equation}
d\tau ^2=\frac{1}{A(r)}A(t) dt^2-A(r)A(t) {{d\sigma} ^2}
\label{eq5}
\end{equation}
But for a gravitational field such that the A(r) term "crosses" the background term we locally get $g_{00}=1=\tilde{g}_{00}$, the only matrix element of the gravitational field that a test mass at rest ($d\sigma=0$) reaching this radius could feel: then we see no reason why this local crossing of the total "feeled" conjugate metrics would not be a necessary and sufficient condition to allow the switch to operate the transfer of the test mass to the conjugate universe.  As well as a genuine black hole horizon this mechanism
 would account for the absence of thermonuclear bursts when matter falls down into BH candidates.
\subsection{Drifting discontinuities?}
However, depending on the actual values taken by $A(r)$ and $A(t)$, it may happen that no particular $r_{0}$ eventually 
allows the crossing $A^{-1}(r_{0})A(t)=1$. Instead, we could have $A^{-1}(r_{0})A^{-1}(t)=1$ but this crossing can only 
occur provided the background jumps from $A(t)$ to $1/A(t)$ at $r_{0}$, a discontinuity believed to be 
at the origin of the Pioneer anomaly as explained in the previous section. To avoid further speculations at this point we may stick to the minimal view that discontinuities just occur to maintain connections (crossings) between conjugate metrics or, in other words, surfaces where the total feeled gravitational "potential" vanishes (or reaches a minimum in absolute value). Then, interestingly, because of the temporal evolution of $a(t)$, $r_{0}$ should propagate to maintain:
\begin{equation}
e^{Gm/r_{0}(t)}=a(t)
\end{equation}
Over a negligible timescale compared to the universe age and in the weak field approximation, 
the propagation time $t_{2}-t_{1}$ from $r_{1}$ to $r_{2}$ is given by: 
\begin{equation}
H_{0}(t_{2}-t_{1})= \phi (r_{2})-\phi (r_{1})
\end{equation}
where $\phi(r)$ stands for the gravitational potential at r. 
The immediate consequence of this is that the discontinuities are expected to scan very rapidly regions of weak gravitational fields and spend much more time where the gravitational fields are the strongest i.e. in the vicinity of matter where we should find them most of the time.

We find that a 
discontinuity in the solar system would take 26000 and 9 years to travel through the solar and earth potentials respectively i.e. from 
vanishing potential at infinity to the potential reached at the surface of the objects. 
The Pioneer effect seems to tell us that the discontinuity was at around 12.5 A.U from the sun in 1983, the time 
when Pioneer 11 detected the very steep rise up of the effect. It follows that it must have reached 
the Jovian potential well and started to fall inside it in 1997 while it will reach the earth potential in 
2104. It is thus now already falling in the potential wells of Jupiter and the outer planets Saturn, Uranus and Neptune, which 
might be correlated with the unexplained activity of Saturn's satellite Encelade and perturbed magnetic field of Uranus.

Recently a Bubbling behaviour accompanied by ripples visible in the hot gas of the Perseus cluster 
has been reported by Chandra Ref.~\refcite{Chan}. We interpret this as the first evidence for a periodic
 path of a gravitational discontinuity which in turn produces our scalar (under rotation in a plane perpendicular to 
the direction of propagation)
 longitudinal gravitational wave ripples rather than sound waves as interpreted in Ref.~\refcite{Chan}. The Bubbling frequency 
is also the gravitational 
waves frequency which can be computed accurately
knowing the measured wavelength (approximately 11kpc) and the GWs speed which is the speed of light. One obtains that the 
frequency is just $2^{49}$ times lower than 475 cycles per second, the frequency corresponding to B-flat above middle C and the 
corresponding period is roughly 30000 years which compares very well with the time needed by the discontinuity to travel accross 
the gravitational potential (from infinity to the surface of the star) of a typical star as is our sun. Therefore 30000 years is 
a good estimation of the time needed for a gravitational discontinuity travelling through the 
almost flat potential of a galaxy to scan the gravitational potential of the majority of stars in the galaxy
 (only a minority of white dwarf stars have much larger surface potentials). This seems to indicate that when most of the stars surfaces have 
been reached by the discontinuity, an instability makes the discontinuity bounce somewhere in the deepest potential zone near a galaxy 
center to initiate a 30000 years return.

\subsection{Discontinuities and boundary conditions}
When the discontinuity reaches the vicinity of a planet or star, the isopotential is roughly a spherical surface surrounding the object.
An isopotential where we find a discontinuity or a crossing of the two forms of the metric has an other interesting
effect: because it acts as a boundary for a spatial domain, the asymptotic decrease of local field perturbations 
at infinity required by the Gauss theorem is a condition which may not be possible to
 fulfill anymore, so that the finite field value at the boundary
 might behave as a central effective source mass. This mechanism might 
create the illusion of a dark several billion solar masses object with huge effects on stellar dynamics in the central arcsecond 
of our galaxy as reported in Ref.~\refcite{bho}.
\section{General equations of motion: toward the unification of gravity and electromagnetism}
With the metric in our usual isotropic form, the free fall equations of 
motion read
\[
\begin{array}{l}
 \frac{d^2r}{dp^2}+\frac{1}{2}\frac{{A}'}{A}\left( {\frac{dr}{dp}} 
\right)^2-\left( {\frac{d\theta }{dp}} \right)^2\left( 
{\frac{{A}'}{2A}r^2+r} \right)-\left( {\frac{d\phi }{dp}} \right)^2\sin 
^2\theta \left( {\frac{{A}'}{2A}r^2+r} 
\right)+\frac{1}{2}\frac{{B}'}{A}\left( {\frac{dt}{dp}} \right)^2+\frac{\dot 
{A}}{A}\frac{dr}{dp}\frac{dt}{dp}=0 \\ 
 \frac{d^2\phi }{dp^2}+\frac{\dot {A}}{A}\frac{d\phi 
}{dp}\frac{dt}{dp}+2\left( {\frac{{A}'}{2A}+\frac{1}{r}} 
\right)\frac{dr}{dp}\frac{d\phi }{dp}+2\cot \theta \frac{d\theta 
}{dp}\frac{d\phi }{dp}=0 \\ 
 \frac{d^2\theta }{dp^2}+2\left( {\frac{{A}'}{2A}+\frac{1}{r}} 
\right)\frac{dr}{dp}\frac{d\theta }{dp}-\sin \theta \cos \theta \left( 
{\frac{d\phi }{dp}} \right)^2+\frac{\dot {A}}{A}\frac{d\theta 
}{dp}\frac{dt}{dp}=0 \\ 
 \frac{d^2t}{dp^2}+\frac{{B}'}{B}\left( {\frac{dr}{dp}} \right)\left( 
{\frac{dt}{dp}} \right)+\frac{1}{2}\frac{\dot {B}}{B}\left( {\frac{dt}{dp}} 
\right)^2+\frac{1}{2}\left( {\frac{d\phi }{dp}} \right)^2\frac{\dot 
{A}}{B}r^2\sin ^2\theta +\frac{1}{2}\left( {\frac{dr}{dp}} 
\right)^2\frac{\dot {A}}{B}+\frac{1}{2}\left( {\frac{d\theta }{dp}} 
\right)^2\frac{\dot {A}}{B}r^2=0 \\ 
 \end{array}
\]
With $d\theta =0,\theta =\pi /2$, $\theta $ is not anymore a dynamical 
variable, expliciting the spatial and temporal parts~: A(r,t)= 
A(r)$a^{2}$(t), B(r,t)=B(r)$a^{2}$(t)~, where B(r)=1/A(r) stands for our 
time independent Schwarzschild solution. 
\[
\begin{array}{l}
 \frac{d^2r}{dp^2}+\frac{1}{2}\frac{{A}'}{A}\left( {\frac{dr}{dp}} 
\right)^2-\left( {\frac{d\phi }{dp}} \right)^2\left( {\frac{{A}'}{2A}r^2+r} 
\right)+\frac{1}{2}\frac{{B}'}{A}\left( 
{\frac{dt}{dp}} \right)^2+2\frac{\dot {a}}{a}\frac{dr}{dp}\frac{dt}{dp}=0 \\ 
 \frac{d^2\phi }{dp^2}+2\frac{\dot {a}}{a}\frac{d\phi 
}{dp}\frac{dt}{dp}+2\left( {\frac{{A}'}{2A}+\frac{1}{r}} 
\right)\frac{dr}{dp}\frac{d\phi }{dp}=0 \\ 
 \frac{d^2t}{dp^2}+\frac{{B}'}{B}\left( {\frac{dr}{dp}} \right)\left( 
{\frac{dt}{dp}} \right)+ {\frac{\dot {a}}{a}}\left( 
{\frac{dt}{dp}} \right)^2+\left( {\frac{d\phi }{dp}} 
\right)^2\frac{A}{B}\frac{a\dot {a}}{{a^2}}r^2+\left( 
{\frac{dr}{dp}} \right)^2\frac{A}{B}\frac{a\dot {a}}{{a^2}}=0 
\\ 
 \end{array}
\]
In our privileged frame, light propagation is obviously not 
sensitive to the scale factor evolution.
In the following we are mainly interested in the $a(t)$ leading terms influencing typical 
sizes and times of non relativistic bound systems. So we work with 
slow velocity approximated equations.
We integrate the second equation after dividing it by $d\phi /dp$
\[
\frac{d}{dp}\left[ {\ln (\frac{d\phi }{dp})+\ln A+\ln a^2+2\ln r} 
\right]=0\Rightarrow \frac{d\phi }{dp}\propto \frac{1}{Aa^2r^2}
\]
In the small speed approximation we can neglect the $(a(t) dr/dt)^{2}$ terms 
and after dividing by dt/dp the third equation integration leads to
\[
\frac{d}{dp}\left( {\ln \frac{dt}{dp}+\ln B+\ln a } 
\right)=0\Rightarrow \frac{dt}{dp} \propto \frac{1}{a B} \]
The first equation in the small speed limit and using 
$\frac{d}{dt}( \frac{dt}{dp} )=-\frac{\dot{a}}{a}\frac{dt}{dp}$ gives 
\[
\begin{array}{l}
 \frac{d}{dt}\left[ {\frac{dt}{dp}\frac{dr}{dt}} 
\right]\frac{dt}{dp}+\frac{1}{2} \frac{{B}'}{A} (\frac{dt}{dp})^2
+2\frac{\dot {a}}{
{a}}\dot{r}(\frac{dt}{dp})^2=0 \\ \Rightarrow
 \left[ {\ddot {r}- {\frac{\dot {a}}{a}\dot {r}} } 
\right]+\frac{1}{2} \frac{{B}'}{A} +2\frac{\dot {a}}{a}\dot {r}=0 \\ 
 \end{array}
\]
Therefore
\[
\ddot {r}=-\frac{1}{2} \frac{{B}'}{A} -\frac{\dot {a}}{a}\dot 
{r}\approx -\frac{Gm}{r^2}-\frac{\dot {a}}{a}\dot {r}
\]
Using $\frac{dt}{dp}=\frac{1}{aB}$ to eliminate dp also gives 
$\dot{\phi}=\frac{B}{Aar^2}$

The effect of the expansion on light or gravitationally bound system or any hyperbolic trajectory 
can of course only be probed with respect to the effect of expansion on an electromagnetically 
bound system:the reference atomic sizes and periods.  

The equations of motion of any body of charge q and mass m in a gravitational field and electric
potential field $V(r,t)$ are the one written above which already include in the gravitational field the effect
 of a local static gravitational potential and the scale factor evolution and must now be corrected for an extra electric acceleration 
$\left[d^2x^\mu/dp^2\right]_{EM}=\frac{e}{m}g^{\mu\rho}( \partial_\rho A_\nu -\partial_{\nu} A_\rho )\frac{dx^\nu}{dp}$ where 
$A_\mu=(V(r,t),0,0,0)$. Explicitly
\[
\begin{array}{l}
 \left[\frac{d^2r}{dp^2}\right]_{EM}= \frac{q}{m} \frac{{V(r,t)}'}{A a^2(t)}{\frac{dt}{dp}}  \\ 

 \left[\frac{d^2\phi }{dp^2}\right]_{EM} =0 \\ 

 \left[\frac{d^2t}{dp^2}\right]_{EM}=  \frac{q}{m} \frac{{V(r,t)}'}{B a^2(t)}{\frac{dr}{dp}}   
\\ 
 \end{array}
\]

The $d\phi/dp$ equation is thus integrated as in the purely gravitational case to give the same result.
We now take it for granted that V(r,t) on its own side also includes the effect of expansion and shall 
show later that $V(r,t)=a(t)V(r)$ where V(r) stands for the usual electrostatic potential.

We first perfom the integration of
\begin{equation}
\frac{dt}{dp}\frac{d}{dp}\left( {\ln \frac{dt}{dp}+\ln B+\ln a } 
\right) = \frac{q}{m} {\frac{dV}{B(r)a(t)dp}} 
\end{equation}
or defining $u = B(r)a(t)\frac{dt}{dp}$:
\begin{equation}
u \frac{d}{dp}\left( {\ln u} 
\right) = \frac{q}{m} {\frac{dV}{dp}} 
\end{equation}
hence now $dt/dp =\frac{1}{a(t)B(r)} (\frac{q}{m} V(r) + 1) $ and in the case of 
atomic energy levels $dt/dp \approx \frac{1}{a(t)B(r)} $.  
Replacing as before the first equation simplifies and eventually we have the two 
equations of motion including all effects :
\[
\ddot {r} = \frac{q}{m} {V'(r)}\frac{B}{A}(\frac{q}{m} V(r) + 1) -\frac{1}{2} \frac{B'}{A} -
\frac{\dot {a}}{a}\dot {r} \approx  \frac{q}{m} {V'(r)} -\frac{1}{2} \frac{B'}{A} -
\frac{\dot {a}}{a}\dot {r}  \]
\[
\dot{\phi} = \frac{B}{Aar^2}(\frac{q}{m} V(r)+1) \approx \frac{B}{Aar^2}
\]
where the approximative results are given in the case of atomic energy levels. 

We can now interpret these results. 
\begin{itemize}
\item  The time evolution of $\phi$ for an atom tells us that the 
characteristic period of a clock behaves as 
$a(t)$. Let us write this $dt_{clock} \propto a(t) dt$. Since
 $d\tau \propto a(t) dt$, the time given by physical clocks is just 
the propertime. 

Analyzing the period of a distant Supernova radiation with respect 
to such a clock, it is straightforward to check that in this time coordinate, it appears redshifted by an accelerated
$t_{clock}^{2}$ (resp decelerated $t_{clock}^{2/3}$ ) scale factor in case $a(t)=1/t^2$ (resp $a(t)=t^2$). 
This result is most easily obtained by a change of variable from the preferred 
time t to $t_{clock}$. Under this transformation the expression of the
 background field metric which was 
\begin{equation}
d\tau^2= (1/t^2)^2(dt^2-d\sigma^2) 
\end{equation}
becomes
\begin{equation}
d\tau^2= dt_{clock}^2-t_{clock}^{4} d\sigma^2 
\end{equation}
so the cosmological phenomenology is just given by the scale factor
$a( t_{clock} )\propto t_{clock} ^{2}$ while $a(t_{clock} )\propto t_{clock} $ 
corresponds to the exponential regime of a(t).
\item
We find that the $a(t)$ dependency of r and $\phi$ for a bound system do not actually depend on the
 nature of the bounding force but rather on the radial speed $\dot{r}$ only. 
This means that for circular orbits there is no way to see the effect of expansion by studying
circular or small excentricity planet trajectories (with respect to our reference atomic standards).
Our assumption that $V(r,t)=a(t)V(r)$ was a crucial one to obtain this result. Had V(r,t) been static, we would
have obtained an $a(t)$ dependency in the final electric acceleration but not the gavitational one, a dead end since this would 
have led to effects usually referred to as variation of the constant G effects well constrained at the level 
of a percent of the Hubble factor (notice however that $\dot{\phi}$ behaves as in the 
standard picture detailed in Ref.~\refcite{Giu}, hence there is nothing really new as for 
the characteristic time periods and energy levels).
In the preferred coordinate system the atomic and gravitationally bound systems
are only subjected to the tiny effect of $-\frac{\dot {a}}{a}\dot 
{r}$ relative to the unaffected light wavelengths. In the physical observer 
clocks coordinate system all are submitted to the same expansion effects except for 
the tiny friction effect.  

\item The $H_0\dot{r}$ acceleration must be compared with $H_{0}c$ needed to account for the 
Pioneer anomaly if it could be interpreted as an acceleration. It is roughly $3.10^4$ smaller so that obviously 
it could not have been detected by solar system spacecrafts up to now. Because $\dot{r}$ vanishes on a full period
no advance of perihely effect can be seen. But on the long term a variation of the excentricity might 
be searched for with foreseenable long term effects on our earth climate. On the short term, because planets and satellites 
radial speeds are so small in the solar system $H_0\dot{r}$ is certainly too small to be detected since it doesn't lead to
 an integrative effect. 
\end{itemize} 

We can now justify the crucial $V(r,t)=a(t)V(r)$ assumption. This result is obtained when one realises that 
in DG the equations of the h(r,t) field are not only the gravitational waves ones but could also 
describe the generation of the electromagnetic or other interaction field with the appropriate coupling constant 
of course. Indeed h(r,t) 
satisfies quasi linear wave equations e.g. for instance in the static case, the equation satisfied 
by h(r) is just the Poisson equation so h(r) might be identified to V(r) multiplied by an approriate factor $\alpha$ to insure that h(r) remains adimentional.  In other words, we suggest the existence of another B=-A field which diagonal components B = $e^{h(r,t)}a^2(t)$ = -A would involve the EM four-vector potential.
It's easy to identify a four-vector of interest because one can always write $g^{\mu\nu}=A^\mu_\alpha A^\nu_\beta \eta^{\alpha \beta}$ in terms of a  vielbein field $A^\mu_\alpha$ from which we can define the four-vector $A^\mu$ taking, in the static electrostatic case the value :
$A^\mu=A^\mu_0=(\frac{1}{\alpha} e^{\alpha V(r)/2}a(t),0,0,0)$ in the preferred frame where an extra factor $\alpha$ comes in again to insure that our potential has the required dimension for an electromagnetic potential four-vector. Therefore we could replace V(r,t) by $a(t)\frac{1}{\alpha} e^{\alpha V(r)/2}$ hence V(r) by $\frac{1}{\alpha} e^{\alpha V(r)/2}$ and V'(r) by $\frac{V'(r)}{2} e^{\alpha V(r)/2}$ in our above equations to get:
 
\[
\ddot {r} = \frac{q}{m} {\frac{V'(r)}{2} e^{\alpha V(r)/2}}\frac{B}{A}(\frac{q}{m} \frac{1}{\alpha} e^{\alpha V(r)/2} + 1) -\frac{1}{2} \frac{B'}{A} -
\frac{\dot {a}}{a}\dot {r} \approx  \frac{q}{2m} {V'(r)} -\frac{1}{2} \frac{B'}{A} -
\frac{\dot {a}}{a}\dot {r}  \]
\[
\dot{\phi} = \frac{B}{Aar^2}(\frac{q}{m} \frac{1}{\alpha} e^{\alpha V(r)/2}+1) \approx \frac{B}{Aar^2}
\]
 Where we have used the strong $\alpha$ where typically $\frac{q}{m\alpha}<<1$ and weak field approximation $\alpha V(r)<<1$. Notice that the unexpected factor 1/2 to V(r) is here not a real issue since it can be reabsorbed in the redefinition of the EM coupling constant.
Thus our crucial assumption that the potential entering in our equations brought the needed $a(t)$ factor is now justified and we by the way 
realise that this simple road toward the unification of gravity and electromagnetism made possible by the B=-A DG sector is necessary
 to avoid being in conflict with accurate tests constraining the variation of G ! However for strong fields and/or strong $q/m$ a new kind of electromagnetic nonlinear effects are expected
!

\section{Mass, Cosmological Constant Terms, Divergences}

Since one of our main motivations for DG was to avoid an epicyclic construction, 
cosmological constant terms are a priori forbidden in our framework unless
these are unavoidable from a fundamental particle physics point of view.  
Thus, the question arises wether we should expect this kind of contribution 
to our gravitational sources from either symmetry breaking or vacuum quantum fluctuations.

It was shown in Ref.~\refcite{fhc1} that a left-handed kinetic and interaction Lagrangian 
can satisfactorily describe all known physics except mass terms which anyway remain
 very problematic in modern physics. This strongly supports the idea that the right handed
 chiral fields are not on our side : the most straightforward  way, as explained
 in Ref.~\refcite{fhc1} to explain maximal parity violation observed in the weak
 interaction. According to this idea, there would be no fundamental massive fields and 
mass would always be effective as is the mass of a photon propagating in a transparent medium.
In particular we would like to give up the idea that mass may be generated by a symmetry 
breaking mechanism based on the interaction with an extra field as is the Higgs field.  
Therefore we are well motivated for trying to investigate the possibility of a vacuum 
ether which would be responsible for generating the effective masses of the fundamental
 massless fields propagating through and interacting with this medium. Of course work 
is needed to build the postulated ether and new mass generating mechanisms but we will 
start to show in a forthcoming section why
it is clearly easier to follow this line of thinking in our framework. Of course one 
then avoids vacuum energy contributions from spontaneous symmetry breaking.

Now let us investigate quantum fluctuations. One usually assumes that there is a 
cut off at the unification scale or at the Planck
 scale. According to Ref.~\refcite{Akh}, if care is taken to regularize the UV divergent integrals 
in a Lorentz invariant way, only massive fields (mass m) vacuum quantum fluctuations contribute
 to a source term in the Einstein equation and this is a cosmological constant term. 
Thus our main hope is to completely avoid such contributions if our fundamental fields 
are only massless as we explained earlier.

However may be could we live more easily with huge vacuum quantum fluctuation cosmological 
constant terms than we can in standard model cosmology. Indeed, one important result 
we got is that whatever its value the source term enters our
 cosmological scale factor solutions $a(t)$ in such a way that it only defines 
a particular time scale $T=\sqrt{\frac{-12}{n\pi G(\rho_{vac}-3p_{vac})}}$, the threshold 
(t must be $>>$ T) for the validity of our $a(t)<<1$ and $a(t)>>1$ solutions.
Thus there remains only one serious issue : is the vacuum fluctuations contribution 
to $\rho_{vac}-3p_{vac}$ finite ? 
We could also rely on Planck scale physics to get finite vacuum fluctuations 
contribution but our framework also seems to come with its own interesting possibilities 
to be explored.

For instance, if the barrier between the two conjugate worlds can be jumped over by particles above a 
given energy threshold, then QED loop divergences hopefully might cancel thanks to the 
positive and negative energy virtual propagators compensation. Such reconnection might take place through 
 allowing particles to jump from one form to the conjugate one presumably through discontinuities where
 the conjugate forms meet each other. 

Alternatively, we already saw that the time reversed  wave taken at -t, $\tilde{g}_{\mu\nu}(-t)$ may well describe an
 anti wave. But if we follow our first possible interpretation the wave has a negative energy and is on our side.
 We didn't allow it to couple to positive energy waves because it is a conjugate form but if for some reason (let us keep open minded),
 above a given energy threshold, the direct and conjugate forms flip and exchange their roles and if this can occur for any kind of 
wave, since propagators for any virtual interaction (electromagnetic, weak or strong) boson exchange are opposite for opposite energy 
virtual carriers, their UV contributions above threshold hence their UV divergences should cancel out leading to a finite theory.

\section{Phenomenological Outlooks: Cosmology}

Of course, if we expect discontinuities of the background
to be everywhere present in the universe, the first task before trying to do cosmology should be a detailed study of the phenomenology associated with these discontinuities 
 to be able to predict where and when do they occur in the universe, if they affect the element abundances and so on. In this section we will just briefly consider  
the main observables of cosmology trying to anticipate wether the new components involved in our formalism might help us to understand the universe and provide 
a promising alternative to the concordant model which being effective anyway should not be considered as more than a good parametrisation of our ignorance.
If we are distrustful of the standard model construction, this attitude
 is even more supported by recent works (Ref. \refcite{Ali}) calling into question the cosmological principle and showing that the background inhomogeneous 
evolution is completely unpredictible due to GR non linearities so that the associated back-reaction effects might mimic any of our effective parametric 
models (any mixture of Dark energy and Dark matter)! 

A few remarks will prove useful for the coming analysis. In the privileged coordinate system, we have seen that there is nothing spatially expanding at all: neither the gravitationally bound systems nor the electromagnetic ones such as atoms constituting our reference rods, nor even free light wavelengths or any distances 
travelled by light rays. If we now choose to work in the standard coordinate system using our clocks (on our side) time, everything is expanding in the same way so that there is of course still no way to probe such spatial expansion locally. All we can see is a contraction or expansion of light time periods relative to atomic time periods and in the opposite way for frequencies. But the spatial expansion effect can still be probed of course by comparing the same scale at two different redshifts which is the principle of the BAO test and this could also be seen for smaller scales such as galaxy sizes which, neglecting possible evolution effects, are also subjected to expansion according DG (at the contrary to the GR prediction). Indeed, the oldest galaxies discovered so far appear to have been 1000 times denser than nowadays galaxies.

Eventually, we can write down a few coordinate independent relations betwen observables such as the redshift z, the energy densities of baryonic matter $\rho_{DarkBaryons}$ and $\rho_{Baryons}$ respectively on the conjugate and our side and the same for radiation, $\rho_{DarkRadiation}$ and $\rho_{Radiation}$. In a straightforward way we can indeed establish that:
\begin{equation}
\frac{\rho_{DarkBaryons}}{\rho_{DarkRadiation}} \propto 1+z
\end{equation}
Then we shall call this a "contracting" universe because the matter dominated era tend to be in the past (high z).
\begin{equation}
\frac{\rho_{Baryons}}{\rho_{Radiation}} \propto \frac{1}{1+z}
\end{equation}
Then we shall call this an "expanding" universe because it is now the radiation dominated era that tend to be in the past.
We also have the important relations:
\begin{equation}
\frac{\rho_{DarkBaryons}}{\rho_{Baryons}} \propto (1+z)^2
\end{equation}
which means that Dark matter fluctuations effects might have been relatively higher in the past as we shall see. Some claim to have evidenced this Dark Matter disappearing effect (Ref \refcite{DMdis}). At last, we also have:
\begin{equation}
\frac{\rho_{DarkRadiation}}{\rho_{Radiation}} \propto 1
\end{equation}
Let's close the parenthesis.

We already noticed earlier
 that we can have a succession of expansion
laws that could perfectly mimick the LCDM cosmology:  starting from
$\sqrt{t_{clock}}$ in the radiative era and followed by $t_{clock}^{2/3}$ in the cold era, then, at the turnaround redshift when the Comological constant started to be dominant in LCDM we would just need to switch to our accelerated $t_{clock}^2$ expansion regime.
However, to insure the $\sqrt{t_{clock}}$ regime in the radiative era, necessary to get the same nucleosynthesis results as predicted within the Standard LCDM, both sides of the universe had to be hot at the time of nucleosynthesis so the temperature at this time had to be at least 4000 K in the conjugate side. Since, from this time, the latter contracted by the same factor as our side expanded i.e. more than $10^9$, the conjugate side remained hot during all the history that we can probe in this case and has now reached a state of extremely high temperature. In such plasma a galaxy on our side might be able to create by its repelling effect a small relative fluctuation of density $\delta\rho/\rho << 1$ on the conjugate side 
which gravitational effect mainly determined by the negative $\delta\rho$ might be important (because $\rho$ might be huge on the conjugate side) and equivalent to a positive fluctuation on our side, i.e a halo of hot darkmatter. At the same time we know that in the very hot state of the conjugate side in this scenario, the Jean's Mass and Length can become small so that it could have its own growing overdensities (the more since it is in contraction) resulting in hopefully the large voids we see in the distribution of matter in our side on the largest scales. 

However, a first possible drawback of this attractive picture (attractive in the sense that we could extrapolate many results from LCDM so that many successes of LCDM in cosmology would translate into DG successes for free) is that we might still have one serious coincidence problem as in LCDM : why do the dark side
gravitational effects needed to understand all missing mass effects have magnitudes not many orders greater than the baryonic matter gravitational effects given the huge mean density on the conjugate side as compared to our side mean density in this scenario. One could argue that the order of magnitude of the negative $\delta\rho$ on the conjugate side (interpreted as DM in LCDM) is mainly determined by the baryonic overdensity on our side  (BTW such link between DM and baryonic matter is confirmed by observations for most type of galaxies and is a challenge for LCDM) rather than by the mean density on the conjugate side, and indeed this might be the case if the negative $\delta\rho$ could not grow by itself to reach the nonlinear regime where $\delta\rho/\rho \approx 1$. But the physics that would prevent the growing of such under-dense fluctuation toward a void in an extremely high (and unknown) temperature conjugate universe which Jean mass might be very small, is far from being obvious. The mechanism would need to prevent as well the nonlinear growing of positive fluctuations i.e over-densities toward unholy values on the conjugate side that would have produced huge initial fluctuations in the CMB.

The second coincidence problem of LCDM is more serious in this scenario: why did a single global switch from deceleration to acceleration occured so recently (at z=0.6)? A first possible solution would be to deny that such global switch occured as everybody thinks it did, but recent measurements using QSO to probe H(z) at high z seem to confirm it (Ref \refcite{Lyalpha}). The alternative solution is to argue that such switches are common i.e. took place many times and may be even periodically in the history of the universe which is not unreasonable in our framework but necessitates to give up the simple scenario which had all the advantages of an expansion history that perfectly mimicked that of LCDM.

The other price to pay to avoid the coincidence problems is to give up the $\sqrt{t_{clock}}$ regime during our side radiative era because if the conjugate side is cold in the contemporary Universe, this necessarily implies that it was even colder in the past since it has contracted ever since while our side has been expanding.
Then there is no epoch of the universe where both sides are simultaneously in the radiative era which was required for the $\sqrt{t_{clock}}$ regime to take place. In the following subsections we thus follow the scenario of a succession of alternating $t_{clock}^{2}$ and $t_{clock}^{2/3}$ regimes (implying time resets as in Figure 3) in the cold era but also in our 
radiative era since the conjugate side remained cold during all this part of the history of our Universe.
This will help to get a much smaller expansion rate in the radiative era making also plausible an alternative BBN and implying an older universe. 

Eventually, it seems that there is an attractive alternative to the cosmological scenario that would have mimicked almost perfectly LCDM, included the possibility of a kind of inflationary phase, the exponential expansion also allowed in DG when both sides are radiation dominated, an exponential phase which would have preceded 
the $t_{clock}^{1/2}$ expansion phase itself with the appropriate duration to get an acoustic horizon peak exactly where we see it in the CMB spectrum.

In the DG cosmology that we want to investigate now, we cannot rely on acoustic oscillations to explain the position of the first peak in the CMB and BBN must also be reinvestigated. But DG then opens the way to its own original solutions and remains a very promising alternative to Dark Matter Models as we shall show. At the present time, we indeed expect typical effects of SM Dark matter because any galaxy should create a void in the conjugate side matter distribution, a void which is equivalent to an overdensity on our side: a cold dark matter halo. Where effects of our repelling "Dark matter" (of course it's just normal matter but on the conjugate side of our universe) manifest themselves, these will hopefully produce a distinct phenomenology that might rule out LCDM: a more rarefied vacuum in the large scale voids, better understanding of galaxy rotation curves, etc ... 

The ratio DM density over baryonic density (LCDM interpretation) in the Universe, which is now around 6 according estimates of the Deuterium abundance in high redshidt baryonic clouds using QSO absorption lines, is instructive because it tells us that the mean density on the conjugate side is at least 6 times (12 times if we prefer to trust the baryonic estimations in the present Universe where there seems to be 50 percent less baryons) that on our side in the present universe. Then, because our side is expanding while the conjugate side is contracting, the crossing might occur soon but this interpretation is not attractive because it would imply another coincidence problem so that the picture of a conjugate universe with a much larger density than our but small relative fluctuations is favoured.

\subsection{The Hubble Test of the Accelerating Universe}

The final answer from the Hubble test will obviously depend on our ability to separate possibly 
new evolution effects driven by the 
physics of discontinuities in Supernovae explosions. 
We obtain the correct Pioneer blue-drifting shift by assuming that the earth was a few years ago in the $t^2$ regime while Pioneer spacecrafts was 
experiencing the $1/t^2$ regime.
But the $t^2$ regime on earth 
 means a decelerating solution in standard coordinates which could not take place during most of the  
universe recent history according to the SN data that show acceleration at least 
from z=0.6 to the present time.  
This implies that either the discontinuities are stationary and delimiting zones in which we can have a succession of the two solutions (power 2 and 2/3) or discontinuities are moving, drifting in such a way that at a given place one will periodically  cross the frontier between two zones delimited by the discontinuity which again lead to the succession of the two solutions (power 2 and 2/3) 
manifesting itself as a mean $t_{clock}^{\alpha}$,
$\alpha$ depending on the relative time spent in the two alternating regimes on the mean. 
   
We can imagine that discontinuities are scanning all gravitational potentials in the universe switching all the objects they encounter to another background until reaching the bottom of the deepest potential well. To reach the very deep surface potential of a BH candidate would require a time exceeding the age of our 
universe while if the discontinuities stop and bounce at a threshold corresponding
to the typical potential level at the 
surface of most stars before returning (reverse drifting) to their starting point and disappear, the duration of such probably periodic journey would be about 2x30000 years. 
Alternatively discontinuities might propagate in pairs separated by a small potential difference, as suggested by a recent New Horizon's observation at 0.4 A.U from Jupiter (Ref . \refcite{NHobserv}), in which case there would be no need for a bounce and the period would be about 30000 years.
The typical maximum potentials encountered on the largest
 scales might also determine the threshold level at which a discontinuity completes its course but these 
potentials are typically 10 times greater, meaning much longer cycles.

Anyway, if we imagine for instance that a local discontinuity runs through our solar system 
with a ~26000 years period, an atom on earth appears to be at the 
edge of the potential well of the sun, in the sense that
 the time needed by a discontinuity to scan the potential difference between that potential at 
the earth position and the one found at the sun surface is much larger (~26000 years) than the time needed to scan the potential difference between the 
asymptotic value of the potential and the sun potential near earth (~ 150 years). 
Therefore, over cosmological time scales 
this discontinuity is more than 99 percent of the time between the earth and the sun.
This seems to imply that our reference earth clocks drifts
 are mainly determined by one of the two possible B(t)=-A(t) field regimes 
while an atom in the vicinity of the sun surface would mainly be submitted to 
the other regime. Such scenario would
produce huge shifts on cosmological time scales between frequencies of atoms 
far away from the stars and near the stars and very large and common redshift anomalies.
Though there are claims of such redshift anomalies, these are very uncommon (for instance between quasars 
and galaxies that seem to be interacting though at incompatible redshifts).
But actually the problem is not real because when a discontinuity scans a potential it brings all the atoms it crosses during its course to the same new background so that no atom frequency will eventually keep track of the fact that this particular atom was the first or the lattest to be reached by the discontinuity.

However, there should remain anomalous fluctuations most easily detectable in regions 
of deep potentials. These expected fluctuations due to
all stars not being at the same stage of their cycles evolution might produce a small 
dispersion in the redshifts, a new systematical but very small effect in the Hubble diagram 
easier to detect with very low redshift samples of SNIA (Ref. \refcite{snif}).
Though possible non-monotonicities in the Hubble diagram were already investigated in 
Ref. \refcite{schu1}, no significant wiggle was detected. These might lead to greater 
effects than the small scalar perturbations in 
the Robertson-Walker metric of Ref. \refcite{schu2}.
  
In case the discontinuity bounces and comes back to its starting point where it disappears or in case of a pair of discontinuities, eventually at the end of the cycle the background is everywhere the same as if nothing had happened so that the Hubble diagram would only be sensitive to a single expansion regime and Pioneer like effects would have no effects on it except the tiny dispersions and systematical effects we considered above. But if the discontinuity is single and does not bounce back it might succeed to switch the whole universe to another regime. 

Thus we would be as well justified to retain the
 $t_{clock}^{\alpha}$ mean regime resulting from the succession of periodically alternating 
$t_{clock}^{2}$ and $t_{clock}^{2/3}$ for cosmology, as well
as we would be justified to consider the succession of only two 
regimes : $t_{clock}^{2/3}$ at least from last scattering to the turnaround redshift, and $t_{clock}^{2}$ since then to have the same expansion history as in LCDM in the cold era.

The power law $t^{\alpha}$ with $\alpha$ between $1.52 \pm 0.11$ (SCP 2008) well fits  
the Hubble diagram up to redshift 1.8. 
Any $\alpha$ could a priori have been  obtained depending on the relative time elapsed in the two regimes. 
This power law suggests that our side has been mainly in the 
constantly accelerated regime appart from small excursions in the
 $t_{clock}^{2/3}$ regime during the last billion years. 

However we should rather wait for the nearby SN factory results to get 
confidence 
in the Hubble diagram method, since the physics of discontinuities might be involved in 
the processes leading to the Supernovae explosions and this in turn directly depends on
 the scale factor behaviour. 
Such an unexpected evolution mechanism might complicate the extraction of the 
cosmological information from the SN Hubble diagram. 

\subsection{Nucleosynthesis}

In this section we assume that the alternating expansion regimes $t_{clock}^{2/3}$ and $t_{clock}^{2}$ result in a mean $t_{clock}^{4/3}$
(4/3 is the arythmetic mean of the two power laws) from BBN to the present time.

BBN theory predicts the universal abundances of the light elements D, $He^3$, $He^4$ and $Li^7$ which are essentially 
determined by $t \approx 180 s$ (Ref. \refcite{bbn}). The best observable to probe these predictions is the abundance of $He^4$ relative to H. 
The mass fraction $Yp \approx 1/4$ is tested at a $\approx 5 $ percent level of accuracy and is very sensitive to the expansion rate at the 
time of primordial nucleosynthesis. A two orders of magnitude smaller expansion rate would result in a negligible abundance of 
Helium since there would be much more time for the disintegration of neutrons between the time when the neutron-proton 
inter-conversion rate were frozen by the expansion and the later time when nuclear reactions could proceed to incorporate 
protons and the remaining neutrons into light nuclei. In DG, because the expansion rate extrapolated at $10^{9}K$ is
orders of magnitude lower than in the standard model because of the accelerated ($\alpha \approx 4/3$) expansion rate, it thus seems 
hopeless to get a non negligible Yp. However, as we could check with in simulation, a kind of miracle happens: with an expansion rate many 
orders of magnitudes below the SM one, Yp rises and reaches $Yp \approx 1/4$ again. 
Authors have investigated the possibility of having the good proportion of Helium in universes expanding much faster than in 
the SM and found that for an exponent in the power law greater than 1, it is indeed possible to obtain the correct $Yp$ with a baryon 
to photon ratio significantly different from the Standard Model one (Ref. \refcite{ind} and references therein). 

For instance,
Figure 2 in Ref. \refcite{kap} shows that $\alpha=4/3$ gives the correct $Yp$ for a density of baryons 
at the level of 2 percent of the critical density! Since this is half of the LCDM BBN prediction but in better agreement with the density 
of visible (actually detected) baryons in the present Universe, we would not even need to explain an important missing baryonic mass and search for it.

Let us notice that $\alpha=1$ gives the correct $Yp$ for a density of baryons an order of magnitude greater 
than that 
of the Standard Model. However, whatever $\alpha $, the baryon fraction just after BBN 
could be considered an upper bound for the present 
one since an unknown fraction of them might have annihilated with antibaryons from
the conjugate side between BBN (most probably just after BBN) and now.

If DG is successful in this confrontation, we must conclude that the SM ability to predict 
the good abundances was a pure coincidence. Let us list the reasons why this coincidence is not as improbable as it might 
appear at first sight. Indeed, the only two other observables on which BBN rests up to now are the $D/H$ and $Li^7/H$
fractions. The systematical effects are at the level of the measured value for $Li^7/H$ so it is not clear at all whether 
we are here sensitive to a primordial abundance. Unfortunately, the same conclusion applies to $D/H$ since its systematics (dispersion 
between measures) are not understood, in particular the $D/H$ large inhomogeneities, a very serious anomaly in 
the absence of any known astrophysical mechanism to explain either the 
creation or the depletion of D (Ref. \refcite{Kirk}). The physics of discontinuities would likely be at the origin of such anomalies 
since nuclear transmutations are commonly observed in discharge experiments in relation with the presence of micro lighting balls that 
represent the most perfect signatures of a spherical discontinuity. See for instance Ref. \refcite{Elis}, Ref. \refcite{Shou}
 and Ref. \refcite{Urut}.

\subsection{Structure Formation and the Early Universe}
We here again assume the case of an homogeneous and $t^{4/3}$  accelerated universe.
The qualitative conclusions would remain the same for any $\alpha$ between 1 an 2. 
Taking the dominating mass density 
contribution to be the baryonic matter density ($\rho _0 = 0.02 \rho_c $ given by the nucleosynthesis constraint from the previous section and by most estimations of the baryonic mass in the present Universe) we get at the time $t_R $ of hydrogen 
recombination the density through:
\[
\begin{array}{l}
 \rho (t_R )=\rho (t_0 )\left( {\frac{a(t_0 )}{a(t_R )}} \right)^3 \\ 
 =\rho (t_0 )\left( {1+z_R } \right)^3=\rho (t_0 )\left( {1500} \right)^3 \\ 
 \end{array}
\]
Also, we can link the Hubble parameter at $t_R $ and nowadays value at $t_0$ through:
\[
\frac{H(t_0 )}{H(t_R )}=\left( {\frac{a(t_R 
)}{a(t_0 )}} \right)^{3/4}
\]
thus:
\[
H(t_R )\approx H(t_0 )\left( {1500} \right)^{3/4}
\]
Neglecting the effect of the universe expansion in the evolution equation of 
density fluctuations $\delta (t)$ on our side and making use of $p\ll \rho$ after 
recombination leads us to the following differential equation:
\[
\ddot {\delta }-4\pi G\rho \delta =0
\]
and the exponentially growing fluctuations:
\[
\delta _+ =e^{\sqrt {4\pi G\rho } t}
\]
Then we can check that:
\[
\frac{H(t_R) }{\sqrt {4\pi G\rho(t_R) } }=\sqrt {\frac{2}{3}} \frac{1}{1500^{3/4}}\sqrt 
{\frac{\rho _c }{\rho _0 }} \approx 2.5 10^{-2}
\]
insuring that the universe expanding rate was indeed negligible compared to 
the density fluctuations growing rate thereby justifying our previous 
approximation. 
This accelerating universe from the CMB last scattering up to now is 
 older than the standard model universe: $4/3H_{0}\approx 18$ billion years. 
 
But our universe is 
also much older at the time of Hydrogen recombination than it was usually believed to be in the Standard Model
 framework: 
\[
t_R=\frac{4}{3 H_R} = t_0 \frac{H(t_0 )}{H(t_R )}=18. 10^{9}  \left( {\frac{1}{1500}} \right)^{3/4}\approx 75 .10^{6} years. 
\]
provided we can neglect the earlier phases duration.
At $t_R $, the 
universe expanding rate was already so small that it did not affect at all the growing of 
primordial fluctuations so that we could soon reach the non-linear regime starting from $10^{-5}$ density 
fluctuations in the CMB assuming such fluctuations was originally present in the CMB and are not due to unexpected foregrounds having left their imprint through the physics of discontinuities. 
We notice also that the smallest mass fluctuation able to grow after recombination for the present 
universe density $\rho _0 =3.10^{-31}g/cm^3$ is the Jean mass $\approx 
10^8M_\odot$  (see in Ref.~\refcite{Wein} the plotted Jean Mass as a function of temperature for different values of the density)
 while the typical dwarf galaxy baryonic visible mass is $\approx 10^9 M_\odot$.

The larger dimensions of voids interpreted as structures in the conjugate form indicate an  asymmetry between the two conjugate 
universes evolutions 
as first suggested by JP Petit (Ref.~\refcite{petit}) whose simulations have confirmed the appearance of large voids with 
galaxies concentrated at the periphery of these repelling conjugate structures. The matter/anti-matter symmetry was satisfied 
at the origin of time when the two sides of the universe merged. After separation, the annihilation of antimatter and matter on both sides must 
have produced as much matter remaining on our side than antimatter on the conjugate side and an a priori different density of radiation but stastistically 
negligible relative difference $\propto \frac{1}{\sqrt{N}}$ considering the large number N of annihilating particles involved.
Therefore the densities of (anti)baryons and photons are the same on both side. 
The origin of the asymmetrical evolution is obvious if one side is contracting while the other side is expanding.
 This is probably the main reason for the nowadays very large asymmetry between the typical sizes and shapes of 
structures on the conjugate sides.

Therefore, we found that our model is not only successful in explaining the growing of the smallest
 initial CMB fluctuations in the linear regime without any need for dark matter nor dark energy but also could lead to an efficient formation of galaxies and 
universe voids interpreted as over-densities in the conjugate form. In this derivation the very small expanding rate in the early universe played a crucial role.

\subsection{The CMB and its fluctuations}
We of course still could interpret the CMB fluctuations as gravitational redshifting effects due to the network of masses or any quasi periodic structure of the hidden side at the time of decoupling and this would be the simplest option making it even possible to relate the scale of the fluctuations to known scales at low z such as the BAO scale provided the expansion history of the universe was exactly that of LCDM from last scattering up to now. But we are now going to investigate more revolutionary hence fascinating other options suggested by some CMB well known anomalies.

\subsubsection{Fluctuations imprinted by foregrounds}
The CMB signal is most probably the signature of a hot and dense primordial state of the whole observable universe.
Its most remarkable properties are its black body spectrum and its homegeneity far beyond the horizon predicted by the 
Standard Model at decoupling. For this reason the 
inflationary scenario has been proposed to save the model. It is therefore important to notice that the DG universe evolution
started from an almost stationary regime and has been accelerated ever since. Consequently it has maintained a much smaller 
expansion rate than in the SM in its earlier ages and was already much older than the SM universe at the time of decoupling 
so there were enough time to let the fluid become homogeneous over a much 
greater scale than that defined by the Standard Model Horizon scale at decoupling.  
In the absence of any horizon mechanism to explain the scale associated to the peaks of the CMB spectrum, our context is  
very different from the Standard Model one (Ref.~\refcite{Hu}). We would be in trouble to obtain 
the right position of acoustic peaks relying on the Standard Model mechanism (the attempt for a freely coasting 
cosmology is in Ref.~\refcite{Kum}). However, recent studies show that the cosmological constant assumption is essential in the
 framework of the LCDM model to reach the percent level of accuracy on the curvature. Giving up this hypothesis, 
 a more reasonable 15 percent error on the curvature parameter (Ref \refcite{Vir}) is obtained.
 This along with various claimed anomalies of the CMB especially at the largest scales but also
 the too many parameters entering in the CMB spectrum fit allow us to 
seriously doubt that the standard model and inflation actually predicted the position of the first peak 
or the correct shape of the CMB spectrum. 

In our framework, the fluctuations we now see in the CMB spectrum are most probably not related to properties of the CMB at
 decoupling since we have no obvious mechanism to predict the observed scales of the peaks in this case. Rather the explanation has to be
 searched for in the inhomogeneous universe in between last scattering and now. By the way, let us repeat that DG does not 
even need detectable primordial inhomogeneities to insure the subsequent formation of structures thanks to the almost 
exponential growing of any inhomogeneity after decoupling in a very slowly expanding universe.

Looking for a mechanism that could affect our side CMB photons wavelengths at a relative level
 of $10^{-5}$ at the one degree angular scale during their path from the surface of last scattering 
to our detectors, we cannot rely on the usual ISW effects 
because they are absent in DG. As for the SZ effects, their contribution is only significant on much 
smaller scales than the degree. 
DG discontinuities and Gavitational waves also cannot inhomogeneously affect our side CMB photons (in a 
B=-A field). The solution comes from the fact that the CMB radiation is also present on the conjugate side 
and that before the conjugate sides separated the photons fluid were common to both 
sides. Therefore the CMB photons from both sides should appear redshifted by cosmological expansion by exactly the same amount 
at a given time from the point of view of an observer on
 our side though from the point of view of an observer on the conjugate side both must appear much hotter. The important point is
 that at any time from our side point of view the CMB is present on both sides of the universe at the same temperature appart from 
fluctuations which origin we are going to explain and suggest now in relation to local
 B=-1/A field inhomogeneities.

Our suggestion is that inhomogeneities of the CMB are produced when the CMB radiation 
from the conjugate side appear on our side via discontinuities that we expect to encounter 
more frequently in the vicinity of various foregrounds : indeed structures at any scale 
generate potential wells and the deeper the potential wells the lower the
 drifting speed of the discontinuities in their vicinity hence the longer the mean time
 such discontinuities will remain in these regions as we explained earlier.

When CMB photons from the conjugate side cross the gap and appear on our side
for instance at the bottom of a deep potential
 well, their frequency is shifted relative to the frequency they had on the corresponding
 hill of the conjugate side. This shift does not seem obvious because photon frequencies 
 are never affected in a stationary gravitational field wether it is discontinuous or not. Indeed for instance the usual gravitational 
redshift is just a comparison of two clock rates at different gravitational potential levels, comparison 
made possible precisely because photons retain the frequency at which they were emitted during 
their path (in a stationary field) from one clock to the other. But when they reach a discontinuity and cross the gap, 
photons are not understood to be in a stationary gravitational potential there (at the discontinuity) but in a field 
which time $\tilde{t}$ reverses into t=- $\tilde{t}$. This reversal might be thought of as a continuous complex rotation by an angle $\theta$
(thus involving an extra dimension) from $\tilde{t}$ to $t= e^{i \theta} \tilde{t}$ where 
$\theta = \pi $. 
During this time reversal the field $ \tilde{g} $ is transformed into 
g=1/$\tilde{g} $ thus the photon frequency is shifted by twice the local gravitational potential.  
 Therefore, contrary to the case of photons staying on the same side 
which are never affected by local potentials they travel through (in part because there
 are no ISW effects in DG) photons crossing the gap from the conjugate to our side are shifted.  
The peculiar galaxy velocities (1000 km/s) tell us about the typical level of the potential wells at work in the 
universe $Gm/rc^2 \approx v^2/c^2 \approx 10^{-5}$ : exactly what we need 
to account for the amplitude of the CMB temperature fluctuations. It is also encouraging 
that a recent analysis (Ref.~\refcite{Str}) has shown that cosmic strings may account for roughly 10 percent of the CMB map of 
fluctuations. These may rather be the first signature of the conjugate side CMB shifted 
photons emerging on our side through cosmic string like discontinuities. 

When the two sides background 
fields cross each other the wave function of a photon can split at the discontinuity and a QM 
computation should allow to compute the probability for the photon to remain (quantum collapse of the wave function)
 on the same side or alternatively be transferred to the conjugate side.
When the transfer occurs, light rays are deviated by an angle which is of the order of several $10^{-5}$ the incident angle to the normal of the discontinuity surface. Therefore the double Doppler effect is almost cancelling as a result of a compensation at the $10^{-5}$ level between Doppler effect at reception and Doppler effect at reemission.
This is fortunate because the discontinuities are in motion in the universe and without such simple compensation their 
speeds relative to the observer would produce much larger temperature fluctuations due to the Doppler 
effects than what is needed to account for the CMB fluctuations amplitudes. 

If the CMB foregrounds are various structures 
with the needed discontinuities in their vicinity responsible for shifting the frequencies of 
conjugate side photons 
and in this way producing all apparent CMB fluctuations, let us examine if it is possible to
obtain the correct typical 1 degree angular scales from known structures. While the large and small magellanic clouds in the vicinity of our galaxy and Andromeda correspond to 10, 3 and 2 degrees
 apparent angles 
respectively, the number of galactic objects really explodes at angles smaller than 0.1. Neighbour clusters of galaxies with their 
typical apparant 1 degree or subdegree scales which is also the scale of the typical 
distances between them are thus interesting candidates to account for the CMB power spectrum at this scale.
The repartition in the sky of these more distant objects is also naturally 
much more homogeneous but how could we explain a succession of peaks in the spectrum. 
Very interestingly there are evidences, even in the more recent works, of cluster concentrations at particular redshifts, 
roughly multiples of 0.03: z=0.03, 0.06, 0.08, 1.1 corresponding to apparent sub-multiples of an angular scale about 1 degree
corresponding to 2Mpcs diameter clusters at z=0.03. Roughly the same scale appears in the 
typical distances between those clusters. However we would expect a clear correlation between 
the distribution of these clusters and CMB maps if these were the main contributors. Instead 
recent estimation of the ISW effects (Ref.~\refcite{Isw})have shown that the correlations are much lower
(even lower than the SM predictions for the ISW effect) than needed
to establish a link with the known amplitude of the CMB main peak.

But there is may be an even more interesting candidate: the scale of the superclusters and supervoids 
of the universe (Ref.\refcite{SupV}) because at this scale a quasi-periodic feature in the repartion of these superstructures 
would hopefully result in a succession of harmonic peaks in the CMB power spectrum. These structures 
are seen under a four degrees apparant angle at z=0.35 (this is our alternative interpretation of the 
so-called BAO peak) and we compute that in DG the corresponding angular scale was that of the first CMB peak at 
 $z \approx 8.4$ (the age of reionization) rather than $z \approx 1100$ in the Standard Model.
It might be that the discontinuities left their imprint on the CMB at this epoch. 
 
At last, a recent study has shown what might be considered as a kind of honeycomb (Ref.\refcite{Honey}) structure in our own 
galaxy , interstellar clouds delimitating empty bubbles (rarefied gas) with a typical diameter of 
150 pcs, like our own local bubble centered on the solar system a new map has revealed. Such periodical 
pattern would arise in the CMB spectrum at the one degree scale if its contribution to the fluctuations 
is the greater at a distance of nearly 9000pcs (comparable to the distance between the sun and the 
galactic center) from us in all directions. The "cold fingers" joining the galactic pole in WMAP 
maps also suggest a local contribution.

Whatever the actual scale of a periodic structure that would have imprinted the CMB, the ressemblance of the WMAP power spectrum with vanishing lower multipoles to the spectrum 
of an AMI bipolar (or order 2 bipolar) random signal is striking. In potential 
wells (resp hills) where the discontinuities are more numerous we would get the most 
important fluctuations (bit 1 of the AMI signal) from alternating zones of redshifted 
(resp blueshifted) CMB photons with AMI polarity=+1 (resp AMI polarity=-1) and 
unberturbed zones in between (bits 0 of the AMI signal) far away from the structures on
 both sides. 

Now what about the contributions on the largest scales of the lower multipoles?
The energy of the quadrupole was found to be concentrated in a subtracted foreground 
(Ref. ~\refcite{fgd}) of WMAP, a restricted area 
corresponding to the direction toward the galactic disk which is almost the perfect
evidence that at least some fluctuations scales in the CMB power spectrum that was believed to be of cosmological origin might instead be completely produced by foregrounds as the largest concentration of stars (i.e. many potential wells thus discontinuities) in our vicinity: that of the galactic disk. 
Thus the lack of energy at the highest CMB scales would simply be the consequence of our galaxy stars being concentrated in a plane, resulting in an easily removable foreground just 
by applying a suitable mask.

A local, anisotropic with respect to the earth observer, discontinuity in the solar system
would ideally account for the correlations with the equinoxes directions and 
the ecliptic plane seen in the CMB low multipoles by WMAP (see Ref.~\refcite{wmap})
but a systematical effect or software error is not excluded.
Indeed the Wmap data have recently been re-analyzed and the authors claim that they succeeded 
to have their quadrupole almost 
completely removed by better correcting Doppler effects depending on the WMAP spacecraft
 attitude parameters (see Ref.~\refcite{liu}, Ref.~\refcite{liu2} and references therein). May be this is the better confirmation we could 
hope for that there is no cosmological signal in the lowest multipoles corresponding to 
scales concerned by the foreground which is the most easily removable by 
a simple mask. This reinforces our conviction that the cosmological fluctuations signal would 
disappear at all scales if we could remove foreground effects at all scales. 
Even the Wmap Ref.~\refcite{Wmap5}
collaboration admits that there also exists a kind of quadrupolar modulation of
 all the multipoles (detected at a 9 sigma significance level) that most probably is related
 to a systematical effect since it is strongly correlated with the direction defined by the
 ecliptic plane.
 
Many other possible consequences of discontinuities, in particular in our solar system, 
were explored in Ref.\refcite{dga}. 

\subsubsection{An almost LCDM scenario}

At last, we of course still could interpret the CMB fluctuations as gravitational redshifting effects due to the network of masses or any quasi periodic structure of the hidden side at the time of decoupling and this would be the simplest option making it even possible to relate the scale of the CMB fluctuations to known scales after expansion at low z such as the BAO scale provided the expansion history of the universe was exactly that of LCDM from last scattering up to now. However, this would left unexplained the intringuing anomaly, among many others, of a CMB power vanishing exactly at the scale corresponding to large angles in the nowadays sky, thus neighbour foregrounds such as our galaxy: the extraordinary coincidence confirmed by Ti pei Li and Hao Liu (see Ref.~\refcite{liu3}) in their most recent analysis. Read also Ref \refcite{QSOfil} about the recent discovery of a QSO filament extending over 4 billion light years.

A recent publication (see Ref.~\refcite{tor}) using the WMAP 3 years data gives a very satisfactory description of the 
CMB spectrum on the largest scales by postulating the topology of a flat torus for our universe. An L=3.86 Gigaparsecs torus 
gives our universe a finite volume of roughly 5.$10^{3} Gpc^{3}$ but can explain the spectrum at scales larger than the Horizon 
scales without the need for inflation simply because by generating the optical illusion of the infinite repetition of this pattern, 
it produces higher wavelength contributions. The scenario is very interesting for us because it proves that a repetitive pattern on 
the CMB can lead to a succession of peaks 
which is the idea we rely on to get the main CMB oscillations starting from the one degree
 peak. These regularities might be the signature of a universe structured according a periodical network and one should be aware 
that such regular pattern may be apparantly lost beyond some distance because of the particular angle at which the network is rotated
with respect to the observer line of sight. A periodical structure would be an important confirmation of another prediction of DG:
the structure of vacuum according a periodical network which would have left very small fluctuations (invisible) in the primordial 
plasma, at the origin of the nowadays, after expansion, 100Mpcs typical cells.
Of course, any network also manifests preferred directions
and this is why some of the most intriguing CMB properties are very promising for us 
(See the recent Ref.~\refcite{riaz} that seems to confirm the existence of preferred directions).

\subsection{Galaxy Rotation Curves}

The subsequent structures nonlinear evolution is also facilitated given that an older universe
$(4/3)H_{0}\approx 18$ billion years from the CMB last scattering up to now, to be compared with the oldest (z=5) galaxies ages
approaching 13.5 billion years, in quite a good agreement with the oldest stars ages, provides more time for galaxy formation.
Notice by the way that our model is not in trouble with the oldest stars and recently observed old galaxies 
 as is the Standard Model with the new very accurate estimation of the standard universe age by WMAP.
Though this should be checked in details with numerical simulations, it appears already almost certain that the interaction between conjugate
 density fluctuations is all we need to solve missing mass issues at any scale. The problem was already adressed in some details in the work of 
JP Petit, particularly in Ref.~\refcite{petit} where he could show that conjugate matter is a powerful alternative idea to dark matter or dark energy.
In DG, the repelling effect of 
a galaxy in our metric must generate a large void in the conjugate universe. This in turn is completely equivalent to a huge halo of perfectly homogeneous
matter source density on our side. This halo is of course completely dark and only interacts gravitationally with our galaxy.
Because similar halos of weakly interactive dark matter can help to obtain satisfactory galaxy flat rotation curves, our model is very promising.
The main difference with the Halo of weakly interactive dark matter is the perfect homogeneity of a void which prevents any small scale gravitational 
collapse to enter into the game. Very remarquably, it has been shown in Ref.~\refcite{mar}, that a model which suppresses gravitational interactions
 (GraS) between dark matter and baryons on small, subgalactic scales (1 kpc) allows to get the correct inner rotation curves of small galaxies and avoid 
the formation of too many substructures in galactic halos. The issue is that the CDM Standard Model, though it is very successful on the large scales
 predicts very compact density cusps in the Dark matter halo and significantly more object of dwarf galaxy mass in a typical galaxy halo than what is seen. 
Both predictions are related to the gravitational effect of DM aggregates on small scales but conflict with observations.
In our framework, the gravitational artificial suppression is not needed since the smooth central potential profile is a natural consequence 
of the void nature of the conjugate fluctuation (nothing concentrated at the galaxy center!).
Moreover, as in the GraS model, baryons here do not fill dark matter substructures simply because the latter do not exist in the void of the conjugate metric.
Many came to believe that the missing satellites of dwarf galaxy mass are there but very faint so that these should progressively show up thanks to the increasing sensitivity of our detectors. However, the recent discovery that most such objects already detected are aligned (both in the Milky way and Andromeda) in a new vast Disk structure at a large angle from the galactic disk implies that these originated from tidal forces at work in galaxy collisions rather than from primordial Dark matter fluctuations. So all these candidates are lost for LCDM and their large fractions of missing mass is more enigmatic than ever (in the standard model but not in DG) if it is tidal forces that produced them.

The recently heralded discovery of Dark Matter by NASA in Ref.~\refcite{chan} is a bad new for MOND and TeVeS since these theories rely on a modified gravity to account for
the rotation of galaxies. This is not the case in DG which in the weak field approximation is as Newtonian as GR. In a sense, DG also has its "Dark Matter" but it is just normal matter from the Dark side which anti-gravitationally 
repels our side matter and is definitely dark, so the Chandra X ray data will have to be re-examined in this perspective as any other data sources.

A shell of Dark repelling matter may be also very effective helping obtaining a bar spiral galaxy stable structure as shown in the 2d simulations by JP.Petit and F.Landsheat. 
Such shells were predicted in the work of Ref.~\refcite{petit}. 
One of the most impressive hints that there actually is a fundamental scale of mass surface density at which the gravitational accelerations suffer 
a drastic change in regime is the ability of MOND theories (\refcite{Bek}) to accurately fit the measured galaxy rotation curves by switching to a 
modified gravity when accelerations fall below a fundamental  threshold. In our framework it is very tempting to identify this density as being the 
 one that compensates perfectly the conjugate side one (negative from our point of view) at some particular distances from the center of a galaxy. 
A recent simulation result trying to mimick the good performances of MOND fits through an ad hoc distribution of DM 
can, according its authors, both reproduce our galactic rotation curve and observed 
diffuse galactic gamma rays excess by EGRET (see Ref.~\refcite{Kaza} and references therein). Along with the usual distribution of 
Dark matter, this study assumes the existence of two rings of Dark matter in the galactic plane at 4kpc and 14kpc where Susy Wimps annihilation 
would account for the observed gamma ray excess. Though Ref.~\refcite{Sal} shows that this model is excluded by a wide margin from the measured
 flux of antiprotons the attempt is of interest for us.
DG would quite differently try to fit the data. In our framework, we would need shells of the conjugate form anti-matter to produce both 
the gamma rays and some modulations in the galactic rotation curves provided, as in the vicinity of our galactic center where positrons 
excess were observed, 
antimatter from the conjugate form can cross the gap toward our form and annihilate with normal matter near these rings. 
Obtaining rings with antimatter would also be much more natural since it is not collisionless as is Dark matter.
The stability of  antimatter shells would be insured because these would be surrounded by normal 
matter which is repelling from their point of view but since these shells repell in turn our matter these should
be located at somewhat different places than the DM rings of Ref.~\refcite{Kaza} to get the correct deformation of 
the galactic rotation curves.

\subsection{Large Scale Dynamics}

At last, we know that the standard LCDM model is able to correctly predict clusters abundances and galaxy peculiar velocities.
Since our framework allows us to take into account the effective large mass induced by the dark side voids, our galaxies are much heavier 
and extend very far away from the visible part which in some case might be to faint to be detectable. Because again the equivalence between the
 CDM dynamics and our dark gravity with voids dynamics seems natural, successes of the 
CDM model should hopefully translate straightforwardly  
into our model success on the largest scales provided there is no bad surprise from more 
detailed simulation studies. 

\subsection{A cosmological scenario to satisfy the Baryonic Oscillation constraint ?}

The bayonic peak was discovered (Ref.~\refcite{eisen}) recently in the spectrum of galaxy clusters at a redshift about 0.35. 
The general opinion is that it corresponds exactly to the CMB main peak at 1 degree in the sense that we are dealing with 
exactly the same scale which were only submitted to the effects of cosmological expansion between 
z=1100 and z=0.35. 
The ratio between the 
angular distances to the BAO (z=0.35) peak and to the first CMB (z=1100) peak was measured to be $R=0.0979\pm 0.0036$ and 
this is believed to be a strong constraint. This constraint is obviously not satisfied by a uniformly accelerating solution from last scattering up to know which would prefer a
much smaller R. However, it is not obvious that the two observables are directly linked to each 
other. 
Indeed the 143 Mparsec "baryonic oscillation" peak scale is very close to the typical scale of the supervoids in the universe 
(133Mpcs for the diameter of the included sphere) and the strange scale of more intense galaxy clustering at 
  distances multiples of 180 Mpcs (z=0.03) which both might have nothing to do with the CMB first peak. Our privileged interpretation 
is that the 3 length scales are related to each other since they correspond to distances in a network structure.
It might be that this structure left its imprint on the CMB at z=8.4 which would produce the 
first CMB peak at the observed angular scale or even more simply that this structure is the same that left its imprint at the z of decoupling if the expansion history has been that of LCDM ever since.

\section{Precession of Equinoxes and GPB Gyroscopes}

In the context of a theory involving discontinuities, it is worth reconsidering the old question that opposed Newton, Mach and Einstein of what 
should determine the inertial frames. In the spatial volume enclosed within a discontinuity,
let us call this a bubble, we suspect that 
there are associated globally inertial frames, where globally here means valid everywhere inside 
such bubble : a globally inertial frame is a frame where inertial pseudo-forces 
(but in general not gravitational forces) vanish 
everywhere in the bubble. Usually such frames are understood to be not 
rotating but not rotating with respect to what? 
Newton would have answered with respect to absolute space, Mach would have answered with 
respect to the fixed stars,  Einstein would have 
 answered with respect to the total gravitational field generated by local sources plus 
global background. 
The presence of a discontinuity that isolates some part of the universe from 
all the rest of it, leads us to another likely answer. In DG, a globally inertial frame 
of a given bubble is one which is not rotating with respect to this bubble
but in general possibly rotating with respect to the fixed stars.
If such frames exist different from the universe restframe, the expected 
phenomenological consequences should be simple: any gyroscope axis (earth, earth-moon 
system, earth-sun system, Gravity-probeB gyroscopes ) 
within the same bubble should follow the rotation of the globally inertial frame 
associated to this bubble.

It is well understood that the differential gravity forces from the sun and moon on the 
earth oblate spheroid are responsible for the precession
 of the earth spin axis with a well established 25580 years periodicity, which translates into the annual precession of equinoxes of 
$360.3600/25580=50.3''$ per year. If the precession frequency of the earth natural gyroscope 
was somehow determined by the rotation 
of a local bubble with respect to the universe restframe rather than by these tidal 
earth-moon effects, then the relative position of the earth moon and sun would also 
have been stabilized just in such a way as to have the 
frequency of the tidal induced precession rate matching $50.3''$ per year
 in a resonnant way. The motivation for seriously considering this scenario in DG is 
obvious, since for the solar system we have a discontinuity scanning the whole solar system precisely with the same period (at the 15 percent level of
accuracy) determined earlier from the Hubble constant $H0$ and solar surface potential. Hence, we are just postulating that the discontinuity path period 
is also that of a rotating frame associated to a bubble encompassing the whole solar system.
The consequences are obvious for the four Gravity-ProbeB gyroscopes which should exactly mimic the deviation of the earth spin axis: thus
 a $50.3''$/year equinox like precession which is hopefully the effect 
(it has the same magnitude) the collaboration described as a classical 
electromagnetic effect.
By the way, the same globally inertial effect on WMAP and COBE satellites might be 
responsible for a drift in the CMB Dipole apex which would also
take the same value of $\approx 50.3''$ per year 
between COBE and WMAP1. If Ref.~\refcite{for} is correct the effect, though not yet 
statistically significant between COBE and WMAP1 might already have been observed
 between WMAP1 and WMAP7. 

\section{Highly speculative further developpments}

\subsection{ The elementary source and field}

As we already noticed earlier, we have on the source part of our 
DG B=-1/A field equation $\rho A$ rather than $\rho$ and the resulting effect is 
not exactly balanced by gravific pressure effects as it is in GR. $A=e^a$ in the source term 
clearly stands for the self gravitational potential energy of the source and its gravific
 effect. This can only be completely understood by solving the equation satisfied 
by a(r) inside matter, which happens to be the Liouville equation (still in units
such that c=1):
\[
\Delta a = -8 \pi \bar{G} \rho e^a
\]
where we anticipate that $\bar{G}$ is not necessarily G but must be unambiguously related 
to the familiar G constant.
Let us assume that nothing but a spherical source with constant density $\rho$, radius 
R hence energy $M=\rho\frac{4}{3} \pi R^3$ is the elementary source for a(r).

The solution which appears to be unique inside the source is 
 $a_{in}(r)= - Ln (4 \pi \bar{G} \rho r^2)$ or $a_{in}(r)= -Ln(3\bar{G}M(r)/r)$ where M(r) is the 
total energy within radius r. Of course the solution outside the source of $\Delta a = 0$ 
must be $a_{out}(r)=2GM/r$ if we want to recover the familiar Newtonian potential in the weak
 field approximation as well as the asymptotic Minkowskian behaviour of the field elements 
B and A. But at r equal R, $a_{in}(r)$ and its derivative must match $a_{out}(r)$ 
and derivative. $a'_{in}(R)=a'_{out}(R)$ can only be satisfied provided $GM/R=1$ i.e. the elementary source 
radius must also be its Schwarzschild radius. Then $a_{in}(R)=a_{out}(R)$ can only be 
satisfied provided  $2=-Ln(3\bar{G}/G)$ or $\bar{G}=G/(3e^2)$.

The condition $GM/R=1$ is certainly one that 
cannot be satistified by the vast majority of known extended sources as we understand 
them usually and yet our solution seems to be the only valid one we can propose in DG in 
 case of spherical symmetry and constant density inside our source.
However remember that our B=-1/A field is only the field generated by what we have called an
 elementary source and it is tempting to identify our elementary source with an
 elementary particle which indeed might be conceived as such a micro object with 
Schwarzschild radius just equal to its radius (this might even be applied to a photon 
if it has even an extremely small rest energy). With this understanding of elementary
 sources, as we already explained earlier we recover all predictions of GR including 
gravitomagnetic 
and pressure effects after exporting all elementary solutions and combining them 
in any common coordinate system, except of course the departure from GR at the 
PostPostNewtonian order of our Schwarzschild solution. 

However there is an even much more attractive possibility which we are going to 
present and explore now, the possibility that the mechanism that relates 
$T^{\mu\nu}$ to $S^{\mu\nu}$ might involve a discrete mode 
of vacuum, a physical ether made of a genuine network of masses in gravitational interaction 
( worth reading are Ref. \refcite{cha} and Ref. \refcite{cha2} though i'm not sure the author's idea is the 
same as that explained here). We postulate that what we usually call vacuum is actually a physical entity, a genuine 
physical ether which mass is not continuously 
distributed over space but is concentrated in a discrete network of the kind of masses our
previous investigation of the B=-1/A equation led us to: each mass $m_0$ has a Radius 
$R_0$ equal to its Schwarzschild radius G$m_0$. Of course the gravitational stability of such a network is only granted provided
positive and negative masses alternate in such a way that any mass has only repelling 
closest neighbours in its vicinity. Even the observational effects expected from this background
of huge alternating masses may not have been evidenced so far because of the globally 
compensating positive and negative gravitational effects particularly if these are separated by 
small distances d in the network and if we happen to be ourselves in motion at high speed 
with respect to the restframe of this network. Now because the needed objects satisfying $R_0=Gm_0$ are already there everywhere in the vacuum and 
because we also have in DG discontinuities which if located at radius R could explain why some rest 
energy can indeed be trapped within the sphere of radius R we can come back on and strongly support 
 our earlier intuition that gravific energy i.e. mass is only generated by such objects 
and gives rise to $S^{\mu\nu}$ which only non vanishing component is $S^{00}$: the total energy trapped
by the spherical discontinuity at r=R.  
Whenever a wave packet of any field propagates in the network, some of its energy E (a quantum 
as we shall argue later) can be trapped (absorbed) then reemited 
by such objects which radius R vary accordingly to maintain the condition $R=G(m_0+E)$. 
Only in this way can the energy of the field source gravity while the energy it carries in between 
the Masses $m_0$ of the network does not source gravity : of course it cant because the 
conditions required by the B=-1/A field equation are not satisfied for this to be possible.
 
Eventually, we should be aware that by solving the elementary field equation we already 
took into account self gravitational potential energy effect of the B=-1/A field itself
 and simply recovered our beloved 
solution $g_{00} = e^{-2 \frac{G m_0}{rc^2}}$ outside the elementary source where $m_0$ is
only $\rho \frac{4}{3}\pi R_0^3$. However, remember that $\rho$ itself does include also the 
the gravitational potential energy of our source in the total external field of all other 
elementary sources. If in a compact star, there is a single unique privileged frame, for instance 
the restframe of the star where we are allowed to compute a myriad of elementary B=-1/A fields then 
combine them, this will result in a total field that will really take into account the gravific effect 
of the self gravitational energy of the star in its own field but not the pressure effects 
and as we stressed earlier this can have very important observational consequences such as 
the generation of a corrected gravific mass $m_0 A_{ext}$ which can be huge if $A_{ext}$ is strong.   
   
With our expression for $g_{00} = e^{-2 \frac{G A_{ext}m_0}{rc^2}}$ in the freefall equation, 
one can derive the following exact general expression for the gravitational
 potential $ \frac{c^2}{4}(e^{-4\frac{G m_0 A_{ext} }{dc^2}}-1)$ created by $m_0$ which
 in the weak field approximation gives back the Newtonian potential $ -G m_0 A_{ext} /r $.

\subsection{ The network and its vanishing energy conditions}

The B=-1/A field is only generated by density perturbations relative
to a mean homogeneous density of the universe. Thus if the vacuuum mass was homogeneous,
there would be no associated gravitational potential energy of a single newly created mass at rest in 
the B=-1/A Minkowskian field generated by the universe vacuum. The energy 
needed to create this mass $m_0$ would simply be 
the sum of its rest energy and the total integrated 
energy over space of the B=-1/A gravitostatic field it generates which $t_{00}$ we 
computed earlier is $\frac{2 a'^2}{16 \pi \bar{G}}$. 
Using $m(r > R_0) = m_0$  and 
$m(r < R_0) = m_0 (r/R_0)^3$. The total integrated $t_{00}$ then yields 
$12e^2m_0$. As a consequence, the creation of such a single mass 
at rest on one side of the Janus universe 
cannot be at free cost because $m_0+12e^2m_0$ does not vanish.

The only possibility to explain the generation of energy and radiation starting from 
an empty universe is a theory allowing rest mass to be created at free cost. Then the 
subsequent decay of such primordial rest mass into less massive 
particles could transfer part of this rest mass energy into kinetic energies and therefore
 all forms of the energy could be generated 
in the universe starting from nothing (at free cost). 
Let's try to see if a vanishing energy condition can be satisfied 
by each mass $m_0$ in our vacuum, if we consider the vacuum mass not continuously 
distributed over space but concentrated in a discrete network of such $m_0$ masses.
The vanishing total energy condition for each mass must also include its
 gravitational potential energy in the total field generated by all the other masses.
For instance in case we would just want to create a pair of opposite masses $m_0$ 
at a distance d from each other, one on each side of an otherwise empty universe, 
the vanishing energy condition is satisfied provided the distance d 
between them is constrained by (avoiding double counting of the potential energies
and neglecting $A_{ext}$ effects):
\[
m_0 \frac{1}{4}(e^{4 \frac{G m_0} {d} }-1)\frac{1}{2} + m_0(1+12e^2) = 0 
\] 
a condition impossible to satisfy because all terms on the left hand side contribute positively.
Playing the same game but taking into account 
all other masses in an infinite network, the vanishing energy condition would involve
\[
\left[ { \sum\limits_i {\frac{\pm 1}{d_i }} } \right] 
\]
in place of $1/d$. With alternating signs from a mass point to its neighbours 
in a cubic network, the serie converges to $ 1.698/d$ in a cube of side L when 
L tends to infinity. But the vanishing energy condition still cannot be satisfied.

A possible way to get it all right is if we could add a negative contribution from our B=-A 
field which is problematic since this field interacts with the positive energy fields of matter 
and radiation, and we know that the interacting fields also have to carry the same sign of the energy 
to avoid instability issues.
There is however a solution if the negative energy density contribution 
is provided by a cosmological constant, because it is non dynamical and the vanishing energy 
condition could be considered a global one and not a local one.
Then the
 important result if confirmed would be that eventually the global network of mass points can 
be created at free cost with 
masses $m_0$ in spheres of radius R and corresponding step 
distances $d$ between them satisfying a simple relation linking them to cosmological constant term 
or any other kind of global non dynamical term.

The network topology must be well constrained. Indeed, the simplest polyhedron to pave a flat space is a prism with 
two equilateral triangle faces and three square faces. However it does not allow the alternate mass signs from one 
one point to any neighbour which is necessary to insure the gravitational stability of the network. A Cubic network 
or a network paved with hexagonal based polyhedrons or even Penrose networks are 
particularly interesting among other cristallographic possibilities.

We saw that the DG B=-A sector has massless longitudinal wave solutions. Extending 
this framework to include 
extra dimensions, the massless waves can acquire mass from the point of view of our 
3+1 space-time provided the 
extra-dimensional path is quantized. This is just the result of the usual assumption 
of a compact extra-dimension 
which length scale $l$ determines the lower mass of the Kaluza-Klein tower of states. We can
 alternatively consider 
a non compact extra dimension but the fraction of our wave momentum in the extra dimension 
would still be quantized if the wave is periodically 
created and annihilated at the network points in such a way that its trajectory is zigzagging 
in the extra-dimension,
its momentum component in this extra dimension being transverse to the mean observed momentum. 
This "missing" transverse momentum might generate a mass and a complete spectrum would follow 
from the geometrical properties of the network (characteristic lengths and angles) in the 3+extra spatial dimensions. 

The idea is attractive but unavoidably amounts to admit fundamental massive fields in our 3+1 dimensions while 
we earlier explained why our framework instead favours fundamental massless fields with the left and 
right fields being respectively restricted to our and the conjugate side of our universe.

Without appealing to extra-dimensions, the interaction of fundamental massless fields with 
our ether, the vacuum network of mass 
points, might be another possibility to generate an effective mass (why not an entire spectrum?) for these fields
just in the same way as photons acquire an effective mass propagating in a transparant medium.
But then also Lorentz violating effects associated to the privileged frame of this ether are expected 
and it would hardly be explained why such Lorentz violating effects remained unobserved so far. 
This however remains an option to generate very tiny thus still unobserved effective mass 
terms but hardly the masses of the standard model fundamental fermions. 

Other options to get mass? We earlier already expressed our frustration as regards spontaneous symmetry 
breaking or other theoretical ideas that just 
help to understand the generation of mass in our universe but up to now are not able to
 predict the actual mass spectrum starting from a reduced number of fundamental constants.

Eventually remains the possibility that fundamental fermions be massive composite objects made up of 
bound states of fundamental massless fields. This is nearly the case for the proton which mass is 
essentially determined by the gluons interactions rather than the masses of its quark 
components. 

By the way, let us mention that this is an intriguing result and interesting from our framework 
point of view. Indeed if the  
treatment of a bound system as a source for gravity was the same at this small scale of the nucleon 
as it is at the scale of a star, one woud expect the gravific mass of the nucleon 
to be determined by the mass of the quarks, the other contributions, i.e. potential energies and 
internal pressure terms (pressure of the partons inside the nucleon) cancelling each other just as 
pressure and gravitational energy potential terms do in a static configuration of a star (see again Ref \refcite{Ehl}).
In other words the equation of state for cold baryonic matter could not be $p \approx  0$ as usually assumed!
Such an observation is clearly favouring a framework where pressure terms do not contribute 
to the gravific mass: just what we have in DG! 

\subsection {Quantizing Gravity}

We have introduced a network of vacuum masses. The next step is to realize that even if gravity 
is only sourced by the energy density $\rho$ inside a sphere of volume $R_0$, the actual field which 
energy density is $\rho$ there does not need to be confined within the sphere. It can actually 
extend much beyond and even overlap with over fields. The unic characteristic responsible for 
 this particular field to be a possible contributor to the gravific $\rho$ inside the sphere is 
that it is a spherical field which center of isotropy is the center of the sphere. Since the 
total energy of the field is not trapped inside the sphere, actually it's just a fraction of this 
energy that is sourcing gravity, the exact value of this fraction being determined by the value
of the gravitational constant G. 

So we can now consider that this field has a total energy which is not trapped inside the sphere 
of radius $R_0$ but rather by a spherical discontinuity at a radius $R$ much larger than $R_0$. 
Therefore the field can only be trapped in the form of a superposition of standing waves :
the fundamental with wavelength $4R$ and its harmonics.

For the time being, assume for instance that the field under consideration is a vector 
field which is not yet quantized. In the cavity and in its fundamental mode which amplitude 
takes the well known spherical standing wave form $A \frac{sin(\pi r/R)}{r}$, we know that this 
amplitude is rather constant for $r < R_0$ if $R_0 << R$, which justifies our treatment of the
 previous section where we assumed a constant $\rho$ inside the sphere of radius $R_0$.

The total integrated energy of our standing wave bounded by R is simply something like
\begin{equation}
m_0 = A^2 \int_0^R \frac{sin^2(\pi r/R)}{r^2} 4\pi r^2 dr 
\end{equation} 
This is the bubble rest energy or in other words, its mass $m_0$.
But while the wave is an extended object much beyond $R_0$ and hence so is its energy 
everywhere inside the bubble of radius R, 
we located and computed the gravific effect of $m_0$ only in a very smaller sphere 
of radius $R_0$ at the center of the bubble, almost point like.
So inside the volume delimited by the discontinuity 
our standing wave is now submitted 
to its own static gravitational field with gravific point mass $m_0$.
Assuming in a first order approximation that the wave function is not too much 
deformed by this potential, it is easy to compute the total self gravitational potential
 energy $E_p$ of this spherical standing wave in the bubble. 
 Delimited by the discontinuous potential barrier in its own central static 
gravitational field of its gravific point-mass $m_0$, it's $E_p$ is given by:

\begin{equation}
E_p=G m_0 A^2\int_0^R \frac{1}{r} \frac{sin^2(\pi r/R)}{r^2}4\pi r^2 dr
\end{equation} 
We can of course multiply and divide the latter expression by $m_0$ and get

\begin{equation}
E_p = m_0\frac{\int_0^R \frac{-G m_0 }{r} \frac{sin^2(\pi r/R)}{r^2}4\pi r^2 dr} {\int_0^R \frac{sin^2(\pi r/R)}{r^2} 4\pi r^2 dr}=-\frac{2.43 G m_0^2}{R}
\end{equation} 

With the exact expression for the gravitational
 potential $\frac{c^2}{4}(e^{-4\frac{G m_0}{rc^2}}-1)$ which in the weak field 
approximation gives back the Newtonian potential $-G m_0/r$ we have just
used to compute $E_p$, $E_p$ gets modified 
\begin{equation}
E_p = m_0\frac{\int_0^R \frac{c^2}{4}(e^{-4 \frac{R_0}{r} }-1) \frac{sin^2(\pi r/R)}{r^2}4\pi r^2 dr} {\int_0^R \frac{sin^2(\pi r/R)}{r^2} 4\pi r^2 dr}
\end{equation} 
We numerically compute
\begin{equation}
E_p/ (\frac{-G m_0^2}{R} ) = \frac{R}{R_0} 
\frac{\int_0^R \frac{1}{4} (1-e^{-4 \frac{R_0}{r}}) 
sin^2(\pi r/R) dr} 
{\int_0^R sin^2(\pi r/R) dr}
\end{equation} 
where we have introduced $R_0=\frac{G m_0}{c^2}$,
and find that this expression fastly converges to $2.43765$ when $R/R_0$ tends to infinity.
For R smaller than $R_0$ the expression would instead decrease as $R$. 

As a consequence, the self gravitational potential energy of our spherical wave in the bubble is
a quantum inversely proportionnal to the bubble radius R : $E = h_G c /4R$ provided we define
$h_G = 4 (2.43) \frac{G m_0^2}{c}$ : the gravitational Planck constant.
Thanks to $c/4R = c/\lambda  \approx \nu$, the frequency of our standing wave, we indeed obtain $E=h_G \nu$,
 the fundamental relation of Quantum Mechanics and by the way
establish a direct link between the gravitational Planck constant and $m_0$.

Of course $m_0$ is a fundamental mass of nature which 
is the same whatever the bubble radius R: in 
other words the bubble can expand or collapse at constant $m_0$. Only the gravitational self
 energy varies and vanishes for infinite d in which
 case for the corresponding center of isotropy the spin 1 field is spread all over the universe
 but is only observable as a gravific mass $m_0$ at r=0. This is actually just the point mass 
we postulated in the previous section where we did not contemplate self gravitating energies. 

Notice that $c/d \approx \nu$ is needed but $c/d = \nu$ is not possible otherwise the wave 
could not be trapped by the spherical discontinuity. Hence the spin 1 field we considered 
inside our cavity could not be an exactly massless field however its mass term is allowed to be 
very small if the discontinuity is a huge one.

Also notice that we take $m_0$ to be the gravific mass in order to compute the self gravitating 
energy $E_p$ but of course from the outside world point of view
the self gravitational potential energy does contribute to the gravific mass which is now 
$m_0+E_p$ and reduces to $m_0$ only in case the discontinuity is at infinity.

Now for any propagating wave packet of any kind of field (massive or not, charged or not, 
arbitratry spin, composite or not, for instance even the wave function of an as big molecule as 
a C60 fullerene 
for which the Quantum Mechanics predictions have been recently tested with flying colors) 
we can always consider its decomposition into spherical waves 
components.  Each of these spherical waves with frequency $\nu$ 
can be created or annihilated by the bubble state we have just described provided the bubble
 center is located at the center of isotropy of the spherical wave and of course provided 
other quantum numbers are conserved in the overall process. The important result is that a 
quantum of such isotropic centripetal wave with frequency  
$\nu$ can be absorbed (annihilated fom the point of view of the outside word) 
when a bubble which discontinuity was initially at infinity, instantaneously collapses 
and stabilizes at distance d allowing a standing wave frequency which is also 
$\nu$ inside the bubble. In this way a quantum of energy $E=h_G \nu$ from 
the initial wave is absorbed. If another identical quantum is absorbed the bubble 
will become twice smaller and so on.

For each quantum absorption or release, the increase or decrease of the
self gravitational potential energy of the bubble and its size follow accordingly and 
change the bubble gravific mass from $m_0$ to $m_0+E_p$ and back to $m_0$. As we 
anticipated in a previous section, such mechanism allows gravific mass to be defined
only near the considered 
center of isotropy as long as the wave is trapped. In between the network 
points where various fields are propagating, gravific mass is not defined.
This is our understanding of the link between $T^{\mu\nu}$ and $S^{\mu\nu}$ which also
provides an interesting picture to understand the origin of annihilation and creation operators
of Quantum Field Theory.

Notice however our new prediction: the Planck constant should no more be a constant 
but should start to decrease about the high energy threshold corresponding to a wavelength
 about $r_0$.

\subsection {Quantum gravity in the solar system}

Angular momentum must be quantized in the solar system according :
\[
mvr=n\hbar _G 
\]
Considering that the protoplanetary disk had constant surface density the mass found 
between radius r and r+dr is $m(r)\propto r$ while the fundamental relation of dynamics yields 
\[
v(r)\propto r^{-1/2}
\]
if the gravitational field of our central protostar was dominant.
Thus the disk must have fragmented and concentrated at quantified radii $r_n$:  
\[
r_{n}^{3/2}\propto n\hbar 
\]
The rotation period T is a quantity proportionnal to $r^{3/2}$ and must therefore be quantized. 
It was indeed shown by JM Souriau that solar system periods are not only approximate multiples of 30 days,
 but also occupy a Fibonacci suite of frequencies, in such a way that neighbour planets are 
minimally resonant.
After fragmentation of the primordial disk, protoplanetary masses grew
 in such a way that $m(r)\propto r$ was lost while radii and periods remained in the same proportions. 
If so $\hbar _G $ can be estimated. Given that the protoplanetary disk radius was approximately the present solar 
system radius r$_{Oort}$ and its mass $10^{-3}$ solar Mass, its initial surface density was
\[
\sigma =\frac{M_\odot }{1000\pi \left( {r_{Oort} } \right)^2}
\]
and
\[
\sigma 2\pi r_n (r_{n+1} -r_n )\frac{2\pi }{T_n }\left( {r_n } 
\right)^2=n\hbar _G 
\]
If 30 days is the fundamental period then Venus and Earth are at n=11 and n=12 respectively 
but may be the smaller self-rotation period in the solar system, ~10 hours for Jupiter, is a more likely fundamental one in which case
$n_{venus} \approx 990$ and $n_{earth} \approx 1080$ (L. Nottale formula has a 3 days fundamental period).  

\[
\begin{array}{l}
 \hbar _G \approx \frac{4\pi }{1000}\frac{M_\odot }{1000 T_{venus} 
}\frac{(r_{earth} -r_{venus} )\left( {r_{venus} } \right)^3}{\left( 
{r_{Oort} } \right)^2}=6.10^{29}Js \\ 
 \end{array}
\]
to be compared with the Planck constant of electromagnetism
\[
\hbar _q =10^{-34}Js
\]
It follows that
\[
m_{vacuum} = \sqrt {\frac{h_G c}{G }} \sim  10^{24} kg
\]
of the order of one fifth of the earth mass.
\[
d_{vacuum} \approx \frac{G m_{vacuum} }{c^2}1.7 \sim  1 mm
\]

Can vacuum effects related to this network of masses be tested?
All laboratory experiments involve masses moving at about 300 km/sec relative to our vacuum masses provided these are at rest with respect 
to the CMB frame. Being alternatively attracted and repelled a free mass test should vibrate or be submitted to deformations 
with typical frequencies of the order of 10 MHz in vacuum. Anyway it is probably difficult to extract such signal from the noise since it affects in almost the 
same way the experimental setup. No doubt that the optimal conditions are those of free motion 
with highly reduce noise i.e. free fall in space. Gravity Probe B is a free falling apparatus 
having  an extremely good control of deformation and motion of its gyroscopes (the most spherical ever man-made objects), 
rotating at 0.03 mm from their stator, and a read-out system highly torque sensitive
which should render it optimal for the detection of our  vacuum effects. It appears that indeed the experiment has discovered unexpected 
new phenomena among which resonance peaks in the drift rate of the gyroscopes axis.

If we were to adopt a more conservative point of view , we would have a single fundamental Planck constant for both gravity 
and electromagnetism and $m_{vacuum} = \sim 3. 10^{-8} kg$ and $d_{vacuum} \sim  3. 10^{-35} m$.

\subsection {Quantizing electromagnetism}

The same method used to quantize gravity allows us to quantize electromagnetism 
with another Planck constant $h_Q $ .
\begin{equation}
 \,\frac{h_q c}{d_e }=\frac{q^2}{4\pi \varepsilon _0 d_e }\Rightarrow h_q 
=\frac{q^2}{4\pi \varepsilon _0 c}=\frac{e^2}{c} 
 \end{equation}
assuming that in the same bubble we have in addition a genuine positronium electromagnetically bound state.
But now the vanishing energy condition should apply by compensating the electromagnetic potential energy of the pair by
its spin-spin magnetic energy.
 \begin{equation}
 (\frac{q \hbar_q}{2 m})^2\frac{1}{c^2 4\pi \varepsilon _0 d_e^3 }=\frac{q^2}{4\pi \varepsilon _0 d_e}
 \end{equation}
Yielding the bubble radius $d_e = \frac{\hbar_q}{2mc}=\frac{137}{2}r_e \approx 2 . 10^{-13}m$ where $r_e$ is the classical radius of the electron ($2.8$ $10^{-15}m$). 

Motivated by the approximate $z_{CMB}$ =1000 $\approx 2\pi .137$ in 
\[
h_q =z.\frac{e^2}{c}\approx 2\pi .137.\frac{e^2}{c}
\]
we suspect cosmological expansion to be responsible for a coevolution of $\alpha=\frac{e^2}{\hbar c}$ and the masses scale.

\subsection {Gravity, Quantum mechanics and Spirituality}

For several decades and in spite of theoreticians sustained efforts, quantizing gravity has raised the major issue, not solved to date,
 of the compatibility between the 
conceptual foundations of General Relativity and Quantum Mechanics, the two pilars of contemporary physics.
Indeed, these appear radically antinomic, the main obstacle at the origin of this incompatibility between MQ and GR certainly being
the inexistence of any privileged coordinate system and in particular the impossibility to define an absolute time in GR.
It is already encouraging to realize that such kind of obstacle immediately disappears within the framework of a theory as DG which
is built starting from an absolute and non dynamical flat space-time, a familiar framework for quantization. But well beyond, 
it is not only the simple perspective of unification between two ways of thinking
 nor merely a quantization program that opens up with DG. Indeed the theory, appears to generate quantization from its own principles 
 and does much more than throwing some new light on the well known interpretation issues of
QM: it solves them for the most part.

 In DG we find two cohabiting modes of gravity:
a continuous source one having propagating wave solutions and a discrete source one, a network of point masses  structuring the vacuum,
 each point being able to communicate via instantaneous gravity with all the others 
and having  in its neighborhood and being the center of isotropy of a stationary wave system oscillating inside 
a finite volume delimited by a gravity discontinuity. As we have already shown, each such system 
  can emit a spherical centrifugal wave or absorb a spherical centripetal wave thereby a paquet of energy (quantum) proportional
to the wave frequency. The fundamental relation of quantization linking energy and frequency of the absorbed/emitted wave, $E=h\nu$, 
is therefore a consequence of the theory. 
 On the other hand, the network points, while absorbing and emitting in a non local and concerted way a new system of spherical waves, is 
perfectly able to trigger the collapse of any QM wave paquet.

 We can now reconsider the most important interpretational issues of QM and explain which kind of solution DG offers in each case.
 We advise the reader to first read the first ten pages of "The transactionnal interpretation of Quantum Mechanics" by JG Cramer where the 
seven issues (Identity, Complexity, Collapse, Non locality, 
 Completeness, Predictivity and the uncertainty Principle) are introduced and discussed within the Copenhagen Interpretation. 

\begin{itemize}
\item  Identity 

What is the state vector (or wave function) of QM?  The wave paquet collapse is a so enigmatic and inacceptable process 
for most physicists that a positivist interpretation which does not take serious the physical reality of the wave function eventually standed out, 
 interpretation according to which it is no more than a tool for efficiently computing relations between observables. 
At the contrary we believe that QM waves are as real and on the same footing as classical electrodynamics wave solutions of Maxwell
equations. The state vector thus describes a purely wave phenomenum propagating in the continuous space-time of DG. 
Only when detection occurs (interaction or mesurement) and the wave paquet collapses, 
a very different physical process than mere propagation, does the more localised particle aspect manifests itself.
 This realistic way of thinking is not new actually: it is the a priori most obvious way first considered and studied by de Broglie and Heisenberg 
then criticized and unfortunately abandonned due to the non locality issues this approach raises.
\item  Non Locality

 The main reason why the collapse of the wave paquet is so disturbing is  that it is essentially non local.
 This is not only a prediction of the QM formalism but now an experimental fact
 after many historical experimental results (in particular the A Aspect one) have firmly established the existence of QM non local correlations in entangled systems. 
Thus one must recognize the strong physical reality of this process. DG allows to go one step beyond in the acceptation and visualisation
  of the process. The discrete mode of vacuum, the network of points, by annihilating or creating a system of spherical waves, can trigger
 the collapse of any wave paquet. This collapse is allowed to be non local since all points can communicate via DG instantaneous gravity. 
\item  The Wave Paquet Collapse (Why?, How?)

 The collapse has to be concerted ("decision taken in common by all involved network points") in order to respect the Born probability law:
 the energy of the vibration at each space-time point determines the probability for the wave paquet to collapse there. 
But there is no need for the transactionnal interpretation of J.G Cramer to justify this point. It simply results from the fact that 
for instance light intensity at each point
is according classical electrodynamics given by the signal energy there, i.e the mean of the squarred signal, 
i.e. the squarred modulus of the complex amplitudes sum that enter in the composition of this signal. 
For what concerns a light beam which photons are emitted one after the other, in between the emitter points and receiver points 
 we have nothing else but a wave paquet s(x,y,z,t), with energy at x,y,z always given by the temporal mean of $s^2(x,y,z,,t)$.
 But absorption or energy emission (involved in detecting a photon) can only occur in quantum paquets ($E=h\nu$) by the network points. Naturally 
we then expect that an energy absorption will be more probable when the available energy at a given point that the wave brings there is more important on the mean 
(at the particular time "chosen by the point networks" for the collapse during one period, if the instantaneous amplitude is not enough to provide the minimum energy at 
a given point the quantum will not be absorbed there).
Therefore it is not surprising that the energy available determines at least the mean probability that the wave paquet collapse takes place at that point. 
 Actually the collapse is not only possible but mandatory because of the discrete way network points can absorb or emit energy. 
The particle aspect is only manifested in the collapse: there is no more necessity 
  for the obscure if not paradoxical wave-particle duality of the Copenhaguen interpretation in our framework, since the wave and particle aspects 
are not really dual aspects of the same reality: they now just stand for the influence of the two independent DG modes of vacuum!
The wave paquet propagating and spreading in the continuous mode of vacuum space is just from time to time transformed into a new more
 localized wave-paquet by the non local and concerted action of the discrete mode of vacuum.

\item  Complexity

 Advanced waves of the transactional interpretation 
  ( positive energies going backward in time or negative energies going forward intime) are not available 
having been rejected from the formalism of modern Quantum Field Theory
 since at a second quantization level these are completely understood in terms of annihilation operators. The Born probability law
 thus can not be interpreted as a transaction 
 between a retarded field and its complex conjugate advanced field. By the way, let us recall that 
 antiparticles are not advanced waves since when they go backward in time, following the Feynmann point of view,
 they are negative energy objects (see www.darksideofgravity.com/antimatiere.htm). 
\item  Predictivity

 May be could we hope to be able to compute more than just a probability if we had access to more than 
a temporal mean of s(t), its instantaneous value or mean or integrated value on the reduced time interval where the collapse decision is taken by the points network. 
  It would allow hopefully to eliminate much of the indeterminism. If there is a hidden determinism that makes appeal to
 blind physical processes (a non spiritualist understanding), it is totally unknown and remains to be explored.
\item The Uncertainty Principle

 We know from Fourier analysis that the better the space-time localisation of a signal the poorer its localisation in the space of frequencies. 
The  time-frequency principle of uncertainty is therefore purely classical and not a mystery. Only when energy is susbstituted to frequency
 thanks to the quantization relation in the uncertainty principle do interpretational issues arise.
 In our framework, the time frequency uncertainty principle comes with the wave physics that takes place in the continuous mode of vacuum.
 The discrete mode of vacuum on the other hand establishes the link between energy and frequency, 
so that we can derive immediately the energy time uncertainty principle  that only deals with the detected quantum 
and the recreation of a new more localized wave paquet. The same for all other uncertainty relations. 
Interpretational issues most often related to the obscure concept of duality are avoided in this way (see previous paragraphs)
\end{itemize}

 Should the QM formalism evolve, leading to new possible testable effects? Certainly if more determinism is hidden.
 Even at the new level of understanding implied by DG, for the collapse to be possible by the discrete vacuum mode 
probably the usual formalism must already be modified since the spherical waves base 
  restricted to waves having as isotropic centers the network points is complete and considerably reduced 
 compared to the more usual one which isotropy centers scanned the whole continuum.

 There is still an important issue : what triggers and when the wave paquet collapse by the network ?
 We should not neglect the firmly spiritualist way of understanding motivated by a deep analogy that we find between 
the vacuum network and a more familiar one: the neural network of our branes. 
 In the same way as a correspondance exists between the activity states of billion neurons in a brain and mental or consciousness states, 
in the same way the states (vibration modes) of  all points in the vacuum cosmic network could represent the physical 
manifestation of a spirit or consciousness of the universe, the living mode of vacuum, the one which eventually triggers
the collapse of all wave paquets and by the way the periodic  reactualisation of the universe.
  The inter-subjective if not objective character of reality for all individual minds would be insured in this way. 
 The individual minds could be those of all living beings in the universe, may be
 as many components of the larger and encompassing cosmic one, the neural network being in interaction, 
in a way that remains to be studied, with the global vacuuum network in the volume occupied by our brane.

 Cerebral neural networks play the role of a fundamental interface if their activation is an essential step in the process that leads to the collapse 
of wave paquets 
 for instance if they provide the global network with the information (our brains would be the senses of the universe) 
necessary for it to decide the way the wave paquets should collapse may be by introducing in the process 
 a certain level of determinism hence ordering and favouring in a discreet but efficient way 
some states among those that the probability amplitude gives equiprobables.   
 Metaphysical outlooks are fascinating, particularly
 the idea that our states of consciousness are shared by the global spirit which in turn could enlarge our faculties 
(if we are willing to morally improve ourselves) and give our intuition an access to an infinite bank of knowledge through modified consciousness states.
 Mind survival to the brain death and integration into the universal mind is one of the most fascinating possibility in this perspective. 

\section{Conclusion}

We could settle down here the foundations of the new theory of gravitation on flat spacetime 
which necessarily imposes itself as soon as we give up the GR geometrical point of view.
Eventually, we find that this allows to solve many long lasting theoretical issues such as negative
 masses and stability, QFT vacuum divergences and the 
cosmological constant, negative energy and tachyonic representations of the Lorentz Group but also leads to 
very remarkable predictions:
 Locally, the disagreement with GR only arises at the PPN level in the preferred frame of our isometries, 
GR black holes disappear and gravitomagnetism can arise in an unusual way. 
Globally, an accelerating 
phase for a
 necessarily spatially flat universe in good agreement with the
 present data is easy to obtain.
The growing  of primordial fluctuations works well in this dark gravity theory without any need for dark energy 
nor dark matter components and the context is very promising to help solving the galaxy or cluster of galaxies 
missing mass issues. 
At last, we derived a gravitational wave solution leading to the observed decay of the binary pulsar orbital period.
More speculative but very promising developments have started to be explored in Ref \refcite{fhc2} and Ref \refcite{DGneut}.
Interesting ideas related at some level to the ones explored here can be found in Refs. 
\refcite{mof} \refcite{hoo} \refcite{bar} \refcite{chard} \refcite{ior}  \refcite{bog} (see also many references of interest 
therein).

\newpage
\begin{figure}[htbp]
\centerline{\includegraphics[width=6.5in,height=7.in]{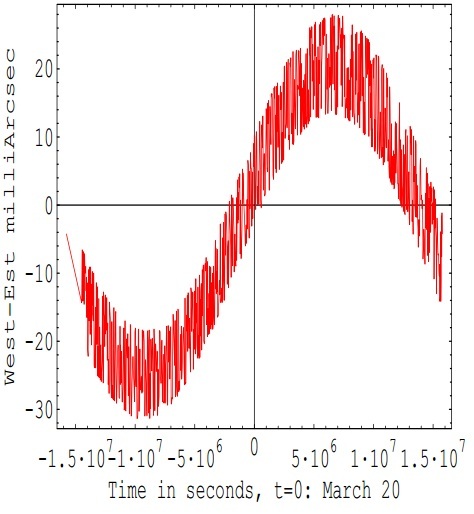}}
\label{fig1}
\end{figure}
\newpage
\begin{figure}[htbp]
\centerline{\includegraphics[width=6.5in,height=7.in]{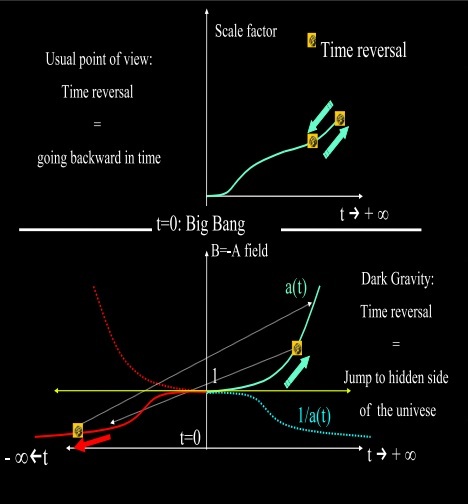}}
\label{fig2}
\end{figure}
\newpage
\begin{figure}[htbp]
\centerline{\includegraphics[width=4in,height=5in]{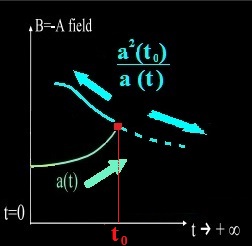}}
\label{fig3}
\end{figure}
\newpage
\begin{figure}[htbp]
\centerline{\includegraphics[width=6.5in,height=7.in]{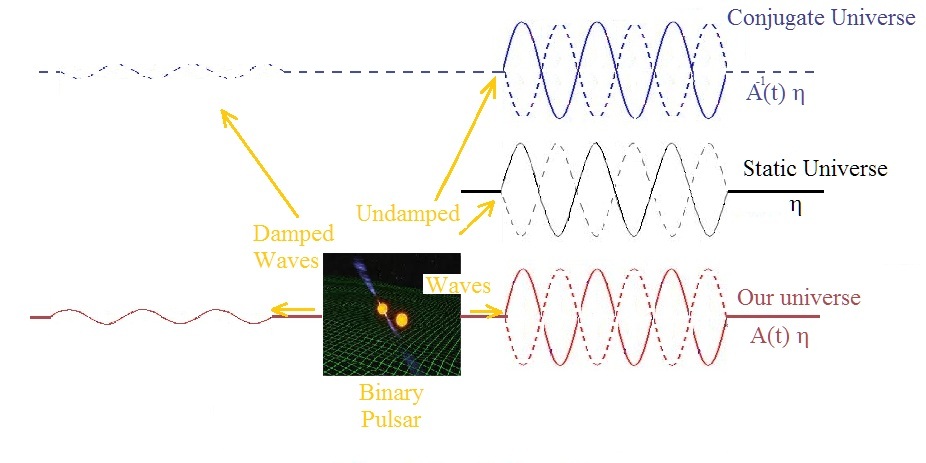}}
\label{fig4}
\end{figure}
\end{document}